\documentclass[pra,aps,10pt,twocolumn,amsmath,amssymb,floatfix]{revtex4-2}

\usepackage[english]{babel}
\usepackage{graphicx}
\usepackage{bm}
\usepackage{color}
\usepackage{slashed}
\usepackage{latexsym}
\usepackage{appendix}
\usepackage{dsfont}

\newcommand{\dd}{d}
\newcommand{\ii}{i}
\newcommand{\ee}{e}

\newcommand{\me}{m_\mathrm{e}} 
\newcommand{\op}[1]{\hat{#1}}

\newcommand{\invf}{\raisebox{0.5em}{\rotatebox{180}{{\scriptsize F}}}}

\newcommand{\vpedele}[1]{\bm{#1}_{-}}
\newcommand{\pedele}[1]{{#1}_{-}}
\newcommand{\spedele}[1]{\slashed{#1}_{-}}
\newcommand{\lamele}{\lambda_{-}}
\newcommand{\vpedpos}[1]{\bm{#1}_{+}}
\newcommand{\pedpos}[1]{{#1}_{+}}
\newcommand{\spedpos}[1]{\slashed{#1}_{+}}
\newcommand{\lampos}{\lambda_{+}}

\begin{document}
 
\title{Scattering matrix approach to dynamical Sauter-Schwinger process:\\ Spin- and helicity-resolved momentum distributions}
\author{M. M. Majczak} 
\author{K. Krajewska} \email{Katarzyna.Krajewska@fuw.edu.pl}
\author{J. Z. Kami\'nski}
\author{A. Bechler} \email{Adam.Bechler@fuw.edu.pl}

\affiliation{
Institute of Theoretical Physics, Faculty of Physics, University of Warsaw, Pasteura 5, 02-093 Warsaw, Poland}
\date{\today}

\begin{abstract}
Dynamical Sauter-Schwinger mechanism of electron-positron pair creation by a time-dependent electric field pulses is considered using the $S$-matrix approach and
reduction formulas. They lead to the development of framework based on the solutions of the Dirac equation with the Feynman- or
anti-Feynman boundary conditions. Their asymptotic properties are linked to the spin-resolved probability amplitudes of created pairs. The same
concerns the helicity-resolved amplitudes. Most importantly, the aforementioned spin- or helicity-resolved amplitudes,
when summed over spin or helicity configurations, reproduce the momentum distributions of created particles calculated with other
methods that are typically used in this context. This does validate the current approach. It also allows us to investigate the vortex structures 
in momentum distributions of produced particles, as the method provides an access to the phase of the probability amplitude. As we also
illustrate numerically, the method is applicable to arbitrary time-dependent electric fields with, in general, elliptical polarization. This
proves its great flexibility.
\end{abstract}

\maketitle

\section{Introduction}
\label{sec:introduction}

Control of physical processes by external electromagnetic fields is at the core of interests of modern physics. Note that, initially, these problems 
were analyzed using perturbative methods. However, with technological developments that allow to generate intense fields (e.g., in laser 
physics~\cite{RevModPhys.78.309} or in heavy ion collisions~\cite{Rafelski1978FermionsAB,Greiner1985QuantumElectrodynamics}) and observational 
methods confirming the existence of very strong fields in the Universe (e.g., strong magnetic fields in accretion disks, approaching the so-called 
Schwinger limit $\mathcal{B}_S\approx 4.4\times 10^9$~T~\cite{10.3389fphy.2014.00059,Ge2020AJL}), the shortcomings of perturbation theory have become
clear. This has led to the development of nonperturbative methods. The most prominent example is the generation of high-order harmonics resulting 
in the synthesis of short in atomic-scale and intense light pulses, which in turn has given rise to a new field of knowledge called 
attoscience~\cite{RevModPhys.81.163} (note that the generation of zeptosecond light pulses in Thomson or Compton scattering has also been 
investigated~\cite{PhysRevA.89.052123}). Another example is the construction of new types of accelerators that use the wake-wave phenomenon 
and permit charged particles to be accelerated to very high energies on relatively short distances~\cite{RevModPhys.81.1229,RevModPhys.85.751}. 
Accelerator physics, in turn, allows one to build generators of coherent radiation (Free-Electron Lasers, FELs) of frequencies in the ultraviolet 
and beyond~\cite{RevModPhys.88.015006}. Thus, we observe an increasing research effort aimed at developing new nonperturbative methods 
that permit, on one hand, to more accurately describe and interpret the already available observational data, and on other hand, to predict new 
physical effects that could be expected in such complex situations. 

One of the topics of such activity is to study the fundamental processes of Quantum Electrodynamics (QED) in the presence of strong 
electromagnetic fields~\cite{fradkin1991vacuumquantum}. Initially, these processes were studied in a monochromatic plane wave~\cite{RitusReview85}
or in a time-dependent electric field generated during the collision of heavy ions (cf.,~\cite{BAUR20071,PhysRevD.109.036008} and references therein).
However, with the development of computational techniques, the scope of these studies has significantly expanded, as it can be inferred from recent 
reviews~\cite{ehlotzky2009fundqed,ruffini2010pairastro,RevModPhys.84.1177,RevModPhys.94.045001,sun2022pair,fedetov2023qed,popruzhenko2023dynamics}.

In this context, one of the most important processes is the creation of real electron-positron ($e^-e^+$) pairs. The corresponding studies were initiated by 
the work of Breit and Wheeler~\cite{breit1934collision}, who considered the $e^-e^+$ pair creation in collisions of two photons with sufficiently high energies. 
Similar scenario but in the presence of a laser beam has been observed experimentally~\cite{PhysRevLett.76.3116,PhysRevLett.79.1626,PhysRevD.60.092004} 
as a result of laser field-induced Compton scattering (Compton process induced by short laser pulses was discussed, for instance, in 
Refs.~\cite{PhysRevD.3.1692,PhysRevA.80.053403,PhysRevA.83.022101,PhysRevA.83.032106,PhysRevA.85.062102}), generating high-energy photons, 
and the subsequent nonlinear Breit-Wheeler process (induced by short laser pulses Breit-Wheeler process was studied, for instance, in 
Refs.~\cite{Heinzl2010FiniteSE,PhysRevLett.108.240406,PhysRevA.86.052104,PhysRevD.91.013009,PhysRevD.93.085028,PhysRevLett.117.213201,PhysRevD.106.056003,PhysRevD.105.056018,PhysRevD.104.096019,PhysRevD.105.056018,PhysRevD.109.036030}).
In turn, the combination of the two-photon Breit-Wheeler process with the Dirac one~\cite{dirac1930theory,Dirac1930Annihilation} (i.e., annihilation 
of a pair into two photons) resulted in the fundamental works of Heisenberg and Euler~\cite{heisenberg1936diractheory}, and 
Schwinger~\cite{schwinger1951gaugeinvariance} on the effective Lagrangian of QED and the creation of pairs by a constant electric field. 
Let us mention that the phenomenon of pair creation from vacuum by a constant electric field was earlier investigated by Sauter~\cite{sauter1931pairdirac} 
using the tunneling theory. Therefore, the process is called either the Schwinger- or the Sauter-Schwinger effect.

The creation of $e^-e^+$ pairs by a constant electric field is significant if the field strength is close to the critical value, 
$\mathcal{E}_S\approx 1.3\times 10^{18}$~V/m (this value is also called the Schwinger limit for the electric field strength). For now, such fields 
are not attainable in laboratory settings, unless they are created in very short pulses, for instance, during heavy ion collisions.
In such circumstances, we typically talk about the dynamical Sauter-Schwinger effect~\cite{schutzhold2008schwinger,PhysRevA.97.022515,PhysRevD.100.116018,PhysRevD.97.116001}. 
Let us note, however, that there are equivalent phenomena which do not require such intense fields. Specifically, the Sauter-Schwinger scenario
can be realized experimentally in graphene, where the electron-hole pairs in an electric field can be produced~\cite{RevModPhys.81.109,PhysRevLett.124.110403}.

The currently dominant tools for nonperturbative analysis of the dynamical Sauter-Schwinger process within QED (note that schemes based on 
the worldline instantons~\cite{schubert2001perturbative,PhysRevD.72.105004,PhysRevD.107.056019} and other tunneling approaches, or the formalism 
called the computational quantum field theory~\cite{doi:10.1080/00107510903450559,PhysRevA.108.033112} are beyond the scope of this paper) 
are based on the Dirac-Heisenberg-Wigner (DHW) function~\cite{bialynicki1991diracvacuum} (equivalent to the bispinorial 
approach~\cite{bechler2023schwinger}) and the field-theoretical approach, which we call the spinorial one, because it reduces to the analysis
of an equation similar to the one describing the precession of spin 1/2 in a time-varying magnetic field 
(cf.,~\cite{PhysRevD.98.056009,PhysRevD.99.056006,PhysRevD.100.016013,bechler2023schwinger} and references therein). However, both methods 
do not allow for a full examination of the spin properties of the momentum distributions of created $e^-e^+$ pairs. In the case of the DHW function, 
the point being that the initial condition is set for values averaged over the initial spins~\cite{PhysRevD.99.096017}. This means that the spin correlations 
of created particles cannot be fully accounted for. In the spinorial approach, a linear polarization of the electric field 
and a special form of the particle spin states have to be imposed, which significantly restricts the applicability of this method. Note that the spin effects were partially taken into account in Refs.~\cite{PhysRevD.99.056006,PhysRevD.100.016013}.

The aim of this paper is to provide the method that allows to fully investigate the spin (helicity) effects in $e^-e^+$ pair creation and to determine 
the spin-resolved (helicity-resolved) momentum distributions. To achieve this goal, our starting point is the QED scattering 
matrix~\cite{dyson1949radiation,dyson1949s}, from which the Lehmann-Symanzik-Zimmermann (LSZ) reduction 
formula~\cite{lehmann1955formulierung,lehmann1957formulation,itzykson1980quantumfield,degli2023worldline} in the presence of an external 
electromagnetic field is derived. It connects the elements of the $S$-matrix with the propagator (or the Green's function) satisfying Feynman 
boundary conditions (Sec.~\ref{sec:reduction}). With the help of the reduction formula and the propagator, we then determine the so-called 
Feynman and anti-Feynman solutions of the Dirac equation and show how to relate their asymptotic properties in the far future and past with 
the spin-resolved momentum distributions of created pairs (Sec.~\ref{sec:FeynmanStates}). This is done for an arbitrary electromagnetic field, 
provided that its vector potential vanishes in the future and the past. Although this condition 
excludes the very often used Sauter pulse from further analysis, it is consistent with the form of the electric field generated in laser pulses. 
Let us mention, however, that our theoretical description can be generalized to the case when the electromagnetic potential takes 
different values in the past and in the future. 

Theoretical and numerical analysis of the dynamical Sauter-Schwinger effect in a general situation 
when the electric field changes rapidly in space and time (i.e., for distances and times comparable to the Compton wavelength $\lambda_C$ and 
the Compton time $\lambda_C/c$, respectively) is difficult and only recently some attempts have been made in this direction. The aim of our 
general analysis of the scattering matrix is to show that the problems of pair creation by an external electromagnetic field cannot be reduced 
to solving the Dirac equation with an arbitrarily chosen initial condition, because its form is correlated with the final states of the particles 
through the asymptotic Feynman conditions (if we determine electron distributions) or anti-Feynman conditions (if we determine positron distributions). 
In particular, their spin-momentum distributions significantly depend on the choice of the initial spin state. Therefore, to simplify further analysis, 
the process of $e^-e^+$ pair creation in a uniform in space electric pulse is discussed (Sec.~\ref{sec:IntegralEquation}). Additionally, in 
order to compare the results with those obtained by other methods, we define the momentum distributions summed over electron and positron spins, 
and introduce the additional normalization (Sec.~\ref{sec:momentumdistributions}). The latter, however, does not follow from the scattering matrix and 
reduction formulas, but is present in the method based on the DHW function (Sec.~\ref{sec:DHWfunction}) or in the spinorial approach 
(Sec.~\ref{sec:Spinorial}). Only after these additional steps, our results fully agree with the results obtained using other methods described 
in this work. However, it is the analysis based on the scattering matrix and reduction formulas with appropriate boundary conditions imposed on 
the solutions of the Dirac equation that allows to fully describe the spin (helicity) and momentum distributions of created particles. Numerical 
illustrations of the spin-resolved (helicity-resolved) momentum distributions are presented in Sec.~\ref{sec:Numer} both for linearly 
(Sec.~\ref{sec:LinPol}) and circularly (Sec.~\ref{sec:CirPol}) polarized electric field pulses. For both polarizations we explore the existence of quantized vortex 
structures in the momentum distributions by analyzing the phase of the corresponding probability amplitudes. Finally, in Sec.~\ref{sec:Conclusions} we draw some conclusions.

In numerical analysis, we use the relativistic units in which $\hbar=c=\me=|e|=1$ and $e=-|e|$, where $\me$ and $e$ are the electron rest 
mass and charge. The Schwinger value for the electric field strength, $\mathcal{E}_S$, is defined such that $\me c^2=|e|\mathcal{E}_S\lambdabar_C$, 
where $\lambdabar_C=\hbar/\me c$ is the reduced Compton wavelength. Additionally, the Compton time is defined as $t_C=\hbar/\me c^2$. For the 
$\gamma$ matrices we apply the Dirac representation, use the Feynman's notation $\slashed{a}=a_\mu\gamma^\mu$, and the metric $(+,-,-,-)$. 
In analytical formulas, we put only $\hbar=1$ and keep explicitly $\me$, $e$, and $c$.

\section{Reduction formulas for electron-positron pair creation}
\label{sec:reduction}

Consider the process of electron-positron pair creation from vacuum by a strong time-dependent external electromagnetic field, that 
is described by the vector potential $A_\mu(x)$, where $x=(x^0,{\bm x})=(ct,\bm{x})$. We assume that both the electric field and the vector potential 
vanish for $t\rightarrow -\infty$ and $t\rightarrow\infty$. The derivation below will be performed for a general time- and space-dependent
vector potential but the expressions for the probability amplitude of pair creation will be further specified for the case of a spatially homogeneous 
electric field. In the following we disregard mutual electromagnetic interactions between created particles.


The amplitude of pair creation can be written as
\begin{equation}\label{rf1}
{\cal A}(\vpedele{p},\lamele;\vpedpos{p},\lampos)=\frac{{\cal A}^\prime(\vpedele{p},\lamele;\vpedpos{p},\lampos)}{\langle 0;{\rm out}|0;{\rm in}\rangle},
\end{equation} 
where 
\begin{equation}\label{rf1a}
{\cal A}^\prime(\vpedele{p},\lamele;\vpedpos{p},\lampos)=\langle \vpedele{p},\lamele;\vpedpos{p},\lampos;{\rm out}|0;{\rm in}\rangle,
\end{equation} 
and
$\vpedele{p}$ and $\vpedpos{p}$ denote, respectively, the momentum of created electron and positron, whereas $\lamele$ and $\lampos$ are their spin quantum numbers. 
Here, we have introduced the in-vacuum and the out-vacuum states, $|0;{\rm in}\rangle$ and $|0;{\rm out}\rangle$, respectively.
Note that the vacuum-to-vacuum probability is not equal to unity since the time-dependent external field can create pairs, so that $|0;{\rm in}\rangle\neq |0;{\rm out}\rangle$. 
In the language of Feynman diagrams, dividing by the vacuum-to-vacuum amplitude in Eq.~\eqref{rf1} eliminates all vacuum diagrams not containing external fermion lines.
Finally, the $e^-e^+$ out-state $|\vpedele{p},\lamele;\vpedpos{p},\lampos;{\rm out}\rangle$ is obtained by acting with the electron and positron creation out-operators on $|0;{\rm out}\rangle$.

The fermion field operator $\op{\Psi}(x)$ fulfills the Dirac equation in external electromagnetic field, 
\begin{equation}\label{rf2}
[-\ii\slashed{\partial}+e\slashed{A}(x)+m_{\rm e}c]\op{\Psi}(x)=0.
\end{equation}
In the asymptotic regions in the remote future $(t\rightarrow +\infty)$ and the remote past $(t\rightarrow -\infty)$ the field operator fulfills 
the free Dirac equation and can be decomposed into free plane wave solutions with positive and negative frequencies. The asymptotic conditions look as
\begin{align}
&\lim_{x^0\rightarrow\infty}\op{\Psi}(x)=\op{\Psi}_{\rm out}(x)\label{rf3a}\\
&=\sum_{\lambda=\pm}\int\frac{d^3p}{(2\pi)^3}\Bigl[\op{c}^{(+)}_{{\rm out};\bm{p},\lambda}u^{(+)}_{\bm{p},\lambda}\ee^{-\ii p\cdot x}+\op{c}^{(-)\dag}_{{\rm out};\bm{p},\lambda}u^{(-)}_{\bm{p},\lambda}\ee^{\ii p\cdot x}\Bigr],\nonumber\\
&\lim_{x^0\rightarrow -\infty}\op{\Psi}(x)=\op{\Psi}_{\rm in}(x)\label{rf3b}\\
&=\sum_{\lambda=\pm}\int\frac{d^3p}{(2\pi)^3}\Bigl[\op{c}^{(+)}_{{\rm in};\bm{p},\lambda}u^{(+)}_{\bm{p},\lambda}\ee^{-\ii p\cdot x}+\op{c}^{(-)\dag}_{{\rm in};\bm{p},\lambda}u^{(-)}_{\bm{p},\lambda}\ee^{\ii p\cdot x}\Bigr],\nonumber
\end{align} 
where $\op{c}^{(+)}_{{\rm in};{\bm p},\lambda}$ and $\op{c}^{(-)}_{{\rm in};{\bm p},\lambda}$ are the annihilation operators for electron and positron, 
respectively, defining the in-vacuum state: $\op{c}^{(+)}_{{\rm in};{\bm p},\lambda}|0;{\rm in}\rangle=0$ and $\op{c}^{(-)}_{{\rm in};{\bm p},\lambda}|0;{\rm in}\rangle=0$. The same concerns the out-operators. The bispinors $u^{(\pm)}_{{\bm p},\lambda}$ fulfill the orthogonality and completeness relations,
\begin{equation}\label{rf4}
u^{(\beta)\dag}_{\beta\bm{p},\lambda}u^{(\beta')}_{\beta'\bm{p},\lambda'}=\delta_{\beta\beta'}\delta_{\lambda\lambda'},\quad
\sum_{\beta,\lambda}u^{(\beta)}_{\beta\bm{p},\lambda}u^{(\beta)\dag}_{\beta\bm{p},\lambda}=\mathds{I}_4,
\end{equation}
where $\beta=+$ stands for the electron (particle) whereas $\beta=-$ is for the positron (antiparticle), and $\mathds{I}_4$ is the unit $4\times 4$ matrix. 
In the standard representation of Dirac matrices, the bispinors $u^{(\pm)}_{{\bm p},\lambda}$ are given explicitly as
\begin{align}\label{rf5}
	u^{(+)}_{\bm{p},\lambda}&=\sqrt{\frac{p^0+m_{\rm e}c}{2p^0}}\left[\begin{array}{c}
	 \chi_\lambda \\  \frac{\bm{\sigma}\cdot\bm{p}}{p^0+m_{\rm e}c}\chi_\lambda\\  \end{array}\right],\nonumber\\
	 u^{(-)}_{\bm{p},\lambda}&=\sqrt{\frac{p^0+m_{\rm e}c}{2p^0}}\left[\begin{array}{c}
	 \frac{\bm{\sigma}\cdot\bm{p}}{p^0+m_{\rm e}c}\chi_\lambda\\ \chi_\lambda
	  \end{array}\right],
\end{align}
where $\chi_\lambda$ is the two component Pauli spinor describing spin state of the particle. As we will discuss it later, by replacing ${\bm p}\rightarrow -{\bm p}$ in the second bispinor and by properly choosing $\chi_\lambda$, the above bispinors also define the eigenstates of the helicity operator. With this normalization of the bispinors
$u^{(\pm)}_{{\bm p},\lambda}$  the nonvanishing anticommutators of the annihilation and creation operators have the form,
\begin{align}
		& \{\op{c}^{(\beta)}_{{\rm out};\bm{p},\lambda},\op{c}^{(\beta')\dag}_{{\rm out};\bm{p}',\lambda'}\}=(2\pi)^3\delta^{(3)}(\bm{p}-\bm{p}')\delta_{\beta\beta'}\delta_{\lambda\lambda'},\label{rf6a}\\
		& \{\op{c}^{(\beta)}_{{\rm in};\bm{p},\lambda},\op{c}^{(\beta')\dag}_{{\rm in};\bm{p}',\lambda'}\}=(2\pi)^3\delta^{(3)}(\bm{p}-\bm{p}')\delta_{\beta\beta'}\delta_{\lambda\lambda'}	\label{rf6b}.
\end{align}
Here, the symbol $\{\op{a},\op{b}\}=\op{a}\op{b}+\op{b}\op{a}$ means the anticommutator. Moreover, as it follows from Eqs.~\eqref{rf3a} and~\eqref{rf3b},
\begin{subequations}
\begin{align}
	& \op{c}^{(+)}_{{\rm out};\bm{p},\lambda}=\lim_{x^0\rightarrow\infty}\int d^3x\bar{\chi}^{(+)}_{\bm{p},\lambda}(x)\gamma^0\op{\Psi}(x)\label{rf7a},\\
	& \op{c}^{(-)}_{{\rm out};\bm{p},\lambda}=\lim_{x^0\rightarrow\infty}\int d^3x\bar{\op{\Psi}}(x)\gamma^0\chi^{(-)}_{\bm{p},\lambda}(x)\label{rf7b},\\
	& \op{c}^{(+)}_{{\rm in};\bm{p},\lambda}=\lim_{x^0\rightarrow -\infty}\int d^3x\bar{\chi}^{(+)}_{\bm{p},\lambda}(x)\gamma^0\op{\Psi}(x)\label{rf7c},\\
	& \op{c}^{(-)}_{{\rm in};\bm{p},\lambda}=\lim_{x^0\rightarrow -\infty}\int d^3x\bar{\op{\Psi}}(x)\gamma^0\chi^{(-)}_{\bm{p},\lambda}(x)\label{rf7d},
\end{align}
\end{subequations}
where
\begin{align}\label{rf8}
	\chi^{(\beta)}_{\bm{p},\lambda}(x)=u^{(\beta)}_{\bm{p},\lambda}\ee^{- \ii\beta p\cdot x}
\end{align}
are the free electron $\chi^{(+)}_{\bm{p},\lambda}(x)$ and positron $\chi^{(-)}_{\bm{p},\lambda}(x)$ states. 

To perform reduction of the electron in the matrix element~\eqref{rf1a} we note that
\begin{align}\label{rf9}
&{\cal A}^\prime (\vpedele{p},\lamele;\vpedpos{p},\lampos)\\
&\, =\lim_{x^0\rightarrow\infty}\int d^3x\,\bar{\chi}^{(+)}_{\vpedele{p},\lamele}(x)\gamma^0
		\langle \vpedpos{p},\lampos;{\rm out}|\op{\Psi}(x)|0;{\rm in}\rangle,\nonumber
\end{align}
where Eq.~\eqref{rf7a} was used. Proceeding in the standard way, we write Eq.~\eqref{rf9} as
\begin{align}\label{rf10}
&{\cal A}^\prime (\vpedele{p},\lamele;\vpedpos{p},\lampos)\\
&\, = \lim_{x^0\rightarrow -\infty}\int d^3x\,\bar{\chi}^{(+)}_{\vpedele{p},\lamele}(x)\gamma^0
	\langle \vpedpos{p},\lampos;{\rm out}|\hat{\Psi}(x)|0;{\rm in}\rangle \nonumber\\
&+\int_{-\infty}^{\infty}dx^0\,\frac{\partial}{\partial x^0}\int d^3x\,\bar{\chi}^{(+)}_{\vpedele{p},\lamele}(x)\gamma^0 \langle \vpedpos{p},\lampos;{\rm out}|\op{\Psi}(x)|0;{\rm in}\rangle. \nonumber
\end{align}
The first term vanishes since, according to Eq.~\eqref{rf7c}, it contains electron annihilation in-operator acting on the in-vacuum state. Using further
\begin{align}\label{rf11}
	-\ii\partial_0\bar{\chi}^{(+)}_{\vpedele{p},\lamele}(x)\gamma^0=\bar{\chi}^{(+)}_{\vpedele{p},\lamele}(x)(\ii\gamma^j\overleftarrow{\partial}_j+m_{\rm e}c),
\end{align}
and performing integration by parts over $\bm{x}$ we get
\begin{align}\label{rf12}
&{\cal A}^\prime (\vpedele{p},\lamele;\vpedpos{p},\lampos)\\
&\quad	=\ii\int d^4x\,\bar{\chi}^{(+)}_{\vpedele{p},\lamele}(x)D_x\langle \vpedpos{p},\lampos;{\rm out}|\op{\Psi}(x)|0;{\rm in}\rangle, \nonumber
\end{align}
where 
\begin{align}\label{rf13}
	D_x=-\ii\gamma^\mu\frac{\partial}{\partial x^\mu}+m_{\rm e}c.
\end{align}
Reduction of the positron proceeds in a similar way. Using Eq.~\eqref{rf7b}, we can write Eq.~\eqref{rf12} as
\begin{align}\label{rf14}
&{\cal A}^\prime (\vpedele{p},\lamele;\vpedpos{p},\lampos)=\ii\lim_{y^0\rightarrow\infty}\int d^4x\int d^3y \\
	&\quad \times \bar{\chi}^{(+)}_{\vpedele{p},\lamele}(x) D_x\langle 0;{\rm out}|\bar{\op{\Psi}}(y)\gamma^0\chi^{(-)}_{\vpedpos{p},\lampos}(y)\op{\Psi}(x)|0;{\rm in}\rangle. \nonumber
\end{align}
Due to the limit $y^0\rightarrow\infty$, this expression can be written as
\begin{align}\label{rf15}
&{\cal A}^\prime (\vpedele{p},\lamele;\vpedpos{p},\lampos)=-\ii\lim_{y^0\rightarrow\infty}\int d^4x\int d^3y\\
	&\times\bar{\chi}^{(+)}_{\vpedele{p},\lamele}(x)D_x\langle 0;{\rm out}|\op{T}[\op{\Psi}(x)\bar{\op{\Psi}}(y)]|0;{\rm in}\rangle\gamma^0\chi^{(-)}_{\vpedpos{p},\lampos}(y),\nonumber
\end{align}
where $\op{T}$ denotes the time ordering operator. Note change of the overall sign due to reversed order of fermion field operators 
in Eq.~\eqref{rf15} as compared to Eq.~\eqref{rf14}. Applying now the same trick as in Eq.~\eqref{rf10} leads to
\begin{align}\label{rf16}
&{\cal A}^\prime (\vpedele{p},\lamele;\vpedpos{p},\lampos)=-\ii\int d^4x\, \bar{\chi}^{(+)}_{\vpedele{p},\lamele}(x)D_x\int d^4y \nonumber\\
	&\times\frac{\partial}{\partial y^0}\{\langle 0;{\rm out}|\op{T}[\op{\Psi}(x)\bar{\op{\Psi}}(y)]|0;{\rm in}\rangle\gamma^0\chi^{(-)}_{\vpedpos{p},\lampos}(y)\},
\end{align} 
where the boundary term with $y^0\rightarrow -\infty$ vanishes due to action of positron annihilation in-operator on $|0;{\rm in}\rangle$. 
Using further the free Dirac equation fulfilled by $\chi^{(-)}_{\vpedpos{p},\lampos}(y)$ and performing integration by parts over $\bm{y}$ we obtain
\begin{align}\label{rf17}
&{\cal A}^\prime (\vpedele{p},\lamele;\vpedpos{p},\lampos)=-\int d^4x\int d^4y\, \bar{\chi}^{(+)}_{\vpedele{p},\lamele}(x) \nonumber\\
	& \times D_x\langle 0;{\rm out}|\op{T}[\Psi(x)\bar{\Psi}(y)]|0;{\rm in}\rangle\overleftarrow{D}_y\chi^{(-)}_{\vpedpos{p},\lampos}(y),
\end{align}
where
\begin{align}\label{rf18}
	\overleftarrow{D}_y=\ii\gamma^\mu\frac{\overleftarrow{\partial}}{\partial y^\mu}+m_{\rm e}c,
\end{align}
with the $y$-differentiation acting to the left. After dividing by the vacuum-to-vacuum amplitude according to Eq.~\eqref{rf1} the pair creation 
amplitude takes the form
\begin{align}\label{rf19}
&{\cal A}(\vpedele{p},\lamele;\vpedpos{p},\lampos) \\
&	=\ii\int d^4x\int d^4y \bar{\chi}^{(+)}_{\vpedele{p},\lamele}(x)D_xK_\mathrm{F}(x,y)\overleftarrow{D}_y\chi^{(-)}_{\vpedpos{p},\lampos}(y). \nonumber
\end{align}
Here, the Feynman propagator in the external field has been defined,
\begin{align}\label{rf20}
	K_\mathrm{F}(x,y)=\ii\frac{\langle 0;{\rm out}|\op{T}[\op{\Psi}(x)\bar{\op{\Psi}}(y)]|0;{\rm in}\rangle}{\langle 0;{\rm out}|0;{\rm in}\rangle},
\end{align}
which fulfills the equation,
\begin{align}\label{rf21}
	[-\ii\slashed{\partial}+e\slashed{A}(x)+m_{\rm e}c]K_\mathrm{F}(x,y)=\delta^{(4)}(x-y),
\end{align}
with proper boundary conditions that are analyzed next.

\section{Feynman and anti-Feynman boundary conditions}
\label{sec:FeynmanStates}

The reduction formula~\eqref{rf19} represents the amplitude of $e^-e^+$ pair creation in an external space- and time-dependent electromagnetic field. It can be further transformed in two ways. First, from Eq.~\eqref{rf21} one has
\begin{align}\label{rf22}
	D_xK_\mathrm{F}(x,y)=\delta^{(4)}(x-y)-e\slashed{A}(x)K_\mathrm{F}(x,y),
\end{align} 
which, after substituting into Eq.~\eqref{rf19}, results in
\begin{align}\label{rf23}
&{\cal A}(\vpedele{p},\lamele;\vpedpos{p},\lampos)=-\ii\int d^4x\int d^4y\\
	&\qquad \times\bar{\chi}^{(+)}_{\vpedele{p},\lamele}(x)e\slashed{A}(x)K_\mathrm{F}(x,y)
	\overleftarrow{D}_y\chi^{(-)}_{\vpedpos{p},\lampos}(y).\nonumber
\end{align}
Note that the $\delta$-function term in Eq.~\eqref{rf22} does not contribute here.  Using further
\begin{align}\label{rf24}
	K_\mathrm{F}(x,y)\overleftarrow{D}_y=\delta^{(4)}(x-y)-K_\mathrm{F}(x,y)e\slashed{A}(y),
\end{align}
one can write Eq.~\eqref{rf23} as
\begin{align}\label{rf25}
&{\cal A}(\vpedele{p},\lamele;\vpedpos{p},\lampos)=-\ii\int d^4x\bar{\chi}^{(+)}_{\vpedele{p},\lamele}(x)e\slashed{A}(x)\nonumber\\
	&\times\left[\chi^{(-)}_{\vpedpos{p},\lampos}(x)-\int d^4yK_\mathrm{F}(x,y)e\slashed{A}(y)\chi^{(-)}_{\vpedpos{p},\lampos}(y)\right]\nonumber\\
	&=-\ii\int d^4x\, \bar{\chi}^{(+)}_{\vpedele{p},\lamele}(x)e\slashed{A}(x)\psi^{(-)}_{\mathrm{F};\vpedpos{p},\lampos}(x),
\end{align}
where the Feynman-type wave function $\psi^{(-)}_{\mathrm{F};\vpedpos{p},\lampos}(x)$ has been defined as
\begin{align}\label{rf26}
\psi^{(-)}_{\mathrm{F};\vpedpos{p},\lampos}(x)&=\chi^{(-)}_{\vpedpos{p},\lampos}(x) \\
&-\int d^4yK_\mathrm{F}(x,y)e\slashed{A}(y)\chi^{(-)}_{\vpedpos{p},\lampos}(y). \nonumber
\end{align}
It fulfills the integral equation,
\begin{align}\label{rf27}
	\psi^{(-)}_{\mathrm{F};\vpedpos{p},\lampos}(x)&=\chi^{(-)}_{\vpedpos{p},\lampos}(x) \\
	&-\int d^4y\, S_\mathrm{F}(x-y)e\slashed{A}(y)\psi^{(-)}_{\mathrm{F};\vpedpos{p},\lampos}(y), \nonumber
\end{align}
where the free-particle Feynman propagator $S_\mathrm{F}(x-y)$ is equal to~\cite{itzykson1980quantumfield,bjorken1964relquantum}
\begin{align}\label{rf27a}
S_\mathrm{F}(x-y)=-\int\frac{d^4p}{(2\pi)^4}\frac{\slashed{p}+m_{\bf e}c}{p^2-(m_{\rm e}c)^2+\ii\epsilon}\ee^{-\ii p\cdot (x-y)},
\end{align}
with infinitesimally small and positive $\epsilon$. Note the minus sign in front of the integral, which is due to the definition of the Feynman propagator \eqref{rf21} (cf., Ref.~\cite{bialynicki1975quantumelectro}).

The boundary conditions satisfied by $\psi^{(-)}_{\mathrm{F};\vpedpos{p},\lampos}(x)$ look as follows: in the remote past, i.e., for $x^0\rightarrow -\infty$, it contains only negative frequencies, whereas in the remote future, i.e., for $x^0\rightarrow\infty$, it contains both the negative and positive frequencies with the negative frequency contribution given exactly by $\chi^{(-)}_{\vpedpos{p},\lampos}(x)$. Writing it explicitly, the so-called Feynman state \cite{bialynicki1975quantumelectro}, $\psi^{(-)}_{\mathrm{F};\vpedpos{p},\lampos}(x)$, fulfills the following boundary conditions,
\begin{align}\label{rf27b}
&\lim_{x^0\rightarrow -\infty}\psi^{(-)}_{\mathrm{F};\vpedpos{p},\lampos}(x) 
=\psi^{(-)}_{\mathrm{in};\vpedpos{p},\lampos}(x) \\
&\quad =\sum_{\lambda=\pm}\int\frac{d^3p}{(2\pi)^3} C^{(-)}_{\mathrm{F};\vpedpos{p},\lampos}(\bm{p},\lambda)\chi^{(-)}_{\bm{p},\lambda}(x)
\nonumber
\end{align}
and
\begin{align}\label{rf27c}
&\lim_{x^0\rightarrow \infty}\psi^{(-)}_{\mathrm{F};\vpedpos{p},\lampos}(x)
=\chi^{(-)}_{\vpedpos{p},\lampos}(x) \\
&\quad +\sum_{\lambda=\pm}\int\frac{d^3p}{(2\pi)^3} C^{(+)}_{\mathrm{F};\vpedpos{p},\lampos}(\bm{p},\lambda)\chi^{(+)}_{\bm{p},\lambda}(x),
\nonumber
\end{align}
with some complex functions, $C^{(\pm)}_{\mathrm{F};\vpedpos{p},\lampos}(\bm{p},\lambda)$, that determine the momentum and spin distributions 
of created pairs. To show this, let us note that the Feynman state satisfies the Dirac equation \eqref{rf2}, meaning that
\begin{align}\label{rf28}
	-e\slashed{A}(x)\psi^{(-)}_{\mathrm{F};\vpedpos{p},\lampos}(x)=D_x\psi^{(-)}_{\mathrm{F};\vpedpos{p},\lampos}(x).
\end{align} 
Substituting the right-hand-side of this equation into Eq.~\eqref{rf25}, performing integration by parts, and taking into account that 
$\bar{\chi}^{(+)}_{\vpedpos{p},\lampos}(x)\overleftarrow{D}_x=0$, one arrives at
\begin{align}\label{rf29}
&	{\cal A}(\vpedele{p},\lamele;\vpedpos{p},\lampos) \\
&\quad =\int_{-\infty}^\infty dx^0\, \frac{\partial}{\partial x^0}\int d^3x\bar{\chi}^{(+)}_{\vpedele{p},\lamele}(x)\gamma^0\psi^{(-)}_{\mathrm{F};\vpedpos{p},\lampos}(x)\nonumber\\
&\quad =\lim_{x^0\rightarrow\infty}\int d^3x\, \bar{\chi}^{(+)}_{\vpedele{p},\lamele}(x)\gamma^0\psi^{(-)}_{\mathrm{F};\vpedpos{p},\lampos}(x). \nonumber
\end{align}
Note that, due to boundary condition imposed on $\psi^{(-)}_{\mathrm{F};\vpedpos{p},\lampos}(x)$, the lower integration limit $x^0\rightarrow-\infty$ does not 
contribute here. Finally, applying the orthogonality condition for free-particle solutions of the Dirac equation, we obtain
\begin{align}\label{rf30}
{\cal A}(\vpedele{p},\lamele;\vpedpos{p},\lampos) 
= C^{(+)}_{\mathrm{F};\vpedpos{p},\lampos}(\vpedele{p},\lamele).
\end{align}
Therefore, in this formulation, we interpret the amplitude introduced in Eq.~\eqref{rf1} as the conditional probability amplitude of detecting 
electron of the momentum $\vpedele{p}$ and spin polarization $\lamele$ provided that positron has the momentum $\vpedpos{p}$ and spin polarization 
$\lampos$. Moreover, by normalizing the Feynman state in the remote past to unity we obtain the momentum and spin distribution of created 
electrons,
\begin{align}\label{rf30a}
F^{(+)}_{\vpedpos{p},\lampos}(\vpedele{p},\lamele)=|A^{(+)}_{\vpedpos{p},\lampos}(\vpedele{p},\lamele)|^2,
\end{align}
where
\begin{align}\label{rf30b}
A^{(+)}_{\vpedpos{p},\lampos}(\vpedele{p},\lamele)=\frac{C^{(+)}_{\mathrm{F};\vpedpos{p},\lampos}(\vpedele{p},\lamele)}{\mathcal{N}_{\mathrm{F};\vpedpos{p},\lampos}}
\end{align}
and
\begin{align}\label{rf30c}
\mathcal{N}^2_{\mathrm{F};\vpedpos{p},\lampos}=\sum_{\lambda=\pm}\int\frac{d^3p}{(2\pi)^3} |C^{(-)}_{\mathrm{F};\vpedpos{p},\lampos}(\bm{p},\lambda)|^2.
\end{align}
Note that the electron distribution~\eqref{rf30a} is defined provided that positrons are detected with momenta $\vpedpos{p}$ and spins 
$\lampos$, similarly to the famous Bothe's coincidence experiment, originally performed for Compton scattering 
(see, e.g., Refs.~\cite{bothe1926kopplung,garrison2008quantumopt}). Hence, in order to define the electron momentum and spin distribution 
irrespective of the positron parameters, we have to integrate over momenta and sum over spins of created positrons such that
\begin{align}\label{rf30aa}
F^{(+)}(\vpedele{p},\lamele)=\sum_{\lampos=\pm}\int\frac{d^3\pedpos{p}}{(2\pi)^3}F^{(+)}_{\vpedpos{p},\lampos}(\vpedele{p},\lamele).
\end{align}
Additionally, $C^{(-)}_{\mathrm{F};\vpedpos{p},\lampos}(\bm{p},\lambda)$ fixes the initial profile of the positron wave packet that leads to the Feynman asymptotic conditions. This profile is \textit{a priori} not known as it is determined by the solution of the Dirac equation in the future.

It is well-known that in quantum theory  the scattering amplitude can be represented in two equivalent forms in which the exact scattering state 
appears either with the out-going or in-coming spherical waves~\cite{goldberger1964collision,newton1982scattering}. We meet a similar situation 
in the case of pair production. Indeed, by rearranging Eq.~\eqref{rf25} we arrive at
\begin{align}\label{rf32}
&{\cal A}(\vpedele{p},\lamele;\vpedpos{p},\lampos) \\
&\quad	=-\ii\int d^4x\bar{\psi}^{(+)}_{{\invf};\vpedele{p},\lamele}(x)e\slashed{A}(x)\chi^{(-)}_{\vpedpos{p},\lampos}(x).
	\nonumber
\end{align}
This expression for the probability amplitude contains Dirac conjugation of the so-called anti-Feynman wave function~\cite{bialynicki1975quantumelectro} 
$\psi^{(+)}_{{\invf};\vpedele{p},\lamele}(x)$ given by
\begin{align}\label{rf33}
	\psi^{(+)}_{\invf;\vpedele{p},\lamele}(x)&=\chi^{(+)}_{\vpedele{p},\lamele}(x) \\
	&-\int d^4y\, K_{\invf}(x,y)e\slashed{A}(y)\chi^{(+)}_{\vpedele{p},\lamele}(y).
	\nonumber
\end{align}
Here, the anti-Feynman propagator of the Dirac equation, $K_{\invf}(x,y)$, is defined as
\begin{align}\label{rf35}
	K_{\invf}(x,y)=\gamma^0K^\dag_\mathrm{F}(y,x)\gamma^0,
\end{align}
or
\begin{align}\label{rf34}
		K_{\invf}(x,y)=-\ii\frac{\langle 0;{\rm in}|\op{\bar{T}}[\op{\Psi}(x)\bar{\op{\Psi}}(y)]|0;{\rm out}\rangle}{\langle 0;{\rm in}|0;{\rm out}\rangle},
\end{align}
with $\op{\bar{T}}$ denoting anti-chronological ordering. Now, the anti-Feynman wave function fulfills the integral equation 
[here, $S_{{\invf}}(x-y)=\gamma^0S^\dag_\mathrm{F}(y-x)\gamma^0$, similarly to Eq.~\eqref{rf35}],
\begin{align}\label{rf36}
	\psi^{(+)}_{{\invf};\vpedele{p},\lamele}(x)&=\chi^{(+)}_{\vpedele{p},\lamele}(x) \\
	& -\int d^4y\, S_{{\invf}}(x-y)e\slashed{A}(y)
	\psi^{(+)}_{{\invf};\vpedele{p},\lamele}(y),
	\nonumber
\end{align}
along with the boundary conditions: in the remote past it contains only solutions of the Dirac equation with positive frequencies, 
whereas in the remote future it contains both positive and negative frequencies with the positive frequency part exactly equal to $\chi^{(+)}_{\vpedele{p},\lamele}(x)$. More explicitly,
\begin{align}\label{rf35a}
&\lim_{x^0\rightarrow -\infty}\psi^{(+)}_{{\invf};\vpedele{p},\lamele}(x) 
=\psi^{(+)}_{\mathrm{in};\vpedele{p},\lamele}(x) \\
&\quad =\sum_{\lambda=\pm}\int\frac{d^3p}{(2\pi)^3} C^{(+)}_{{\invf};\vpedele{p},\lamele}(\bm{p},\lambda)\chi^{(+)}_{\bm{p},\lambda}(x)
\nonumber
\end{align}
and
\begin{align}\label{rf35b}
&\lim_{x^0\rightarrow \infty}\psi^{(+)}_{{\invf};\vpedele{p},\lamele}(x)
=\chi^{(+)}_{\vpedpos{p},\lampos}(x) \\
&\quad +\sum_{\lambda=\pm}\int\frac{d^3p}{(2\pi)^3} C^{(-)}_{{\invf};\vpedele{p},\lamele}(\bm{p},\lambda)\chi^{(-)}_{\bm{p},\lambda}(x).
\nonumber
\end{align}
Next, using the relation
\begin{align}\label{rf37}
	-\bar{\psi}^{(+)}_{{\invf};\vpedele{p},\lamele}(x)e\slashed{A}(x)=\bar{\psi}^{(+)}_{{\invf};\vpedele{p},\lamele}(x)\overleftarrow{D}_x,
\end{align} 
and performing similar steps as above for the Feynman state, we obtain another equivalent expression for the amplitude of pair creation in the form of a scalar product in remote future,
\begin{align}\label{rf38}
&
{\cal A}(\vpedele{p},\lamele;\vpedpos{p},\lampos) 
\\ &\quad 
=-\lim_{x^0\rightarrow\infty}\int d^3x\, \bar{\psi}^{(+)}_{{\invf};\vpedele{p},\lamele}(x)\gamma^0\chi^{(-)}_{\vpedpos{p},\lampos}(x) 
\nonumber\\ &\quad 
=-C^{(-)}_{{\invf};\vpedele{p},\lamele}(\vpedpos{p},\lampos).
\nonumber
\end{align}
This time, the amplitude ${\cal A}(\vpedele{p},\lamele;\vpedpos{p},\lampos)$ can be interpreted as the conditional probability amplitude 
for the creation of positrons of the momentum $\vpedpos{p}$ and spin $\lampos$ provided that electrons are created with the momentum $\vpedele{p}$ 
and spin $\lamele$. Similarly, by normalizing to unity the initial state $\psi^{(+)}_{\mathrm{in};\vpedele{p},\lamele}(x)$ in Eq.~\eqref{rf35a}, 
we define the conditional momentum and spin distribution for created positrons,
\begin{align}\label{rf37a}
F^{(-)}_{\vpedele{p},\lamele}(\vpedpos{p},\lampos)=|A^{(-)}_{\vpedele{p},\lamele}(\vpedpos{p},\lampos)|^2,
\end{align}
where
\begin{align}\label{rf37b}
A^{(-)}_{\vpedele{p},\lamele}(\vpedpos{p},\lampos)=-\frac{C^{(-)}_{{\invf};\vpedele{p},\lamele}(\vpedpos{p},\lampos)}{\mathcal{N}_{{\invf};\vpedele{p},\lamele}}
\end{align}
and
\begin{align}\label{rf37c}
\mathcal{N}^2_{{\invf};\vpedele{p},\lamele}=\sum_{\lambda=\pm}\int\frac{d^3p}{(2\pi)^3} |C^{(+)}_{{\invf};\vpedele{p},\lamele}(\bm{p},\lambda)|^2.
\end{align}
Similar to Eq.~\eqref{rf30aa}, we can also define the positron distribution,
\begin{align}\label{rf37aa}
F^{(-)}(\vpedpos{p},\lampos)=\sum_{\lamele=\pm}\int\frac{d^3\pedele{p}}{(2\pi)^3}F^{(-)}_{\vpedele{p},\lamele}(\vpedpos{p},\lampos),
\end{align}
that is irrespective of the electron spin and momentum characteristics. Moreover, $C^{(+)}_{{\invf};\vpedele{p},\lamele}(\bm{p},\lambda)$ fixes 
the profile of the initial electron wave packet, which is also \textit{a priori} unknown, as it has to be chosen such that the solution evolves 
to the anti-Feynman asymptotic conditions in the future.

While the above considerations are very general, in the following sections we focus on the case when the $e^-e^+$ pair creation occurs in a spatially homogeneous electric field.

\section{Integral equations}
\label{sec:IntegralEquation}

In this section, we analyze in detail the integral equations~\eqref{rf27} and~\eqref{rf36} in the case of a spatially homogeneous
electromagnetic field, i.e., assuming that the vector potential depends only on time, $A(x^0)=(0,\bm{A}(x^0))$.
For the spatially homogeneous vector potential, the wave function $\psi^{(-)}_{\mathrm{F};\vpedpos{p},\lampos}(x)$ can be written as
\begin{equation}\label{rf2-1}
	\psi^{(-)}_{\mathrm{F};\vpedpos{p},\lampos}(x)=\ee^{-\ii\vpedpos{p}\cdot\bm{x}}\Phi^{(-)}_{\mathrm{F};\vpedpos{p},\lampos}(x^0).
\end{equation}
Using further Eq.~\eqref{rf8} one can write Eq.~\eqref{rf27} as
\begin{align}\label{rf2-2}
&
	\Phi^{(-)}_{\mathrm{F};\vpedpos{p},\lampos}(x^0)=\ee^{\ii\pedpos{p}^0 x^0}u^{(-)}_{\vpedpos{p},\lampos}
\\ & \quad
	-\int d^4yS_\mathrm{F}(x-y)e\slashed{A}(y^0)\ee^{\ii\vpedpos{p}\cdot(\bm{x}-\bm{y})}\Phi^{(-)}_{\mathrm{F};\vpedpos{p},\lampos}(y^0).
\nonumber
\end{align}
Substituting here the integral representation of the free Feynman propagator,
\begin{align}\label{rf2-3}
&
S_\mathrm{F}(x-y)
\\ & \quad
=\ii\theta(x^0-y^0)\int\frac{d^3k}{(2\pi)^3}\frac{\slashed{k}+m_{\rm e}c}{2k^0}\ee^{-\ii k\cdot (x-y)}
\nonumber \\ & \quad
-\ii\theta(y^0-x^0)\int\frac{d^3k}{(2\pi)^3}\frac{\slashed{k}-m_{\rm e}c}{2k^0}\ee^{\ii k\cdot (x-y)},
\nonumber
\end{align}
where $k=(k^0,\,\bm{k})$ with $k^0=\sqrt{\bm{k}^2+(m_{\rm e}c)^2}$, we obtain the integral equation for the Feynman-type function
\begin{align}\label{rf2-4}
& 
\Phi^{(-)}_{\mathrm{F};\vpedpos{p},\lampos}(x^0)=\ee^{\ii \pedpos{p}^0x^0}u^{(-)}_{\vpedpos{p},\lampos}
\\ & \quad
-\ii \ee^{-\ii\pedpos{p}^0x^0}\frac{\spedpos{p}'+m_{\rm e}c}{2\pedpos{p}^0}\int_{-\infty}^{x^0}dy^0
\ee^{\ii\pedpos{p}^0y^0}e\slashed{A}(y^0)\Phi^{(-)}_{\mathrm{F};\vpedpos{p},\lampos}(y^0) 
\nonumber\\ & \quad
+\ii \ee^{\ii\pedpos{p}^0x^0}\frac{\spedpos{p}-m_{\rm e}x}{2\pedpos{p}^0}\int_{x^0}^{\infty}dy^0
\ee^{-\ii\pedpos{p}^0y^0}e\slashed{A}(y^0)\Phi^{(-)}_{\mathrm{F};\vpedpos{p},\lampos}(y^0),
\nonumber
\end{align}
where $\pedpos{p}=(\pedpos{p}^0,\vpedpos{p})$ and $\pedpos{p}'=(\pedpos{p}^0,-\vpedpos{p})$.

For the anti-Feynman wave function $\psi^{(+)}_{{\invf};\vpedele{p},\lamele}(x)$, we take
\begin{equation}\label{rf2-5}
	\psi^{(+)}_{{\invf};\vpedele{p},\lamele}(x)=\ee^{\ii\vpedele{p}\cdot\bm{x}}\Phi^{(+)}_{{\invf};\vpedele{p},\lamele}(x^0),
\end{equation}
where $\vpedele{p}$ is now interpreted as the electron momentum. Moreover, in the integral equation~\eqref{rf36}  we use
\begin{align}\label{rf2-6}
& 
		S_{{\invf}}(x-y)=\gamma^0S^\dagger_\textrm{F}(y-x)\gamma^0
\\ & \quad
		=\ii\theta(x^0-y^0)\int\frac{d^3k}{(2\pi)^3}
		\frac{\slashed{k}-m_{\rm e}c}{2k^0}\ee^{\ii k\cdot (x-y)}
\nonumber\\ & \quad
		-\ii\theta(y^0-x^0)\int\frac{d^3k}{(2\pi)^3}
	\frac{\slashed{k}+m_{\rm e}c}{2k^0}\ee^{-\ii k\cdot (x-y)}.
\nonumber
\end{align}
After performing in Eq.~\eqref{rf36} the integrals over $\bm{y}$ and $\bm{k}$, for the anti-Feynman function we obtain that
\begin{align}\label{rf2-7}
& 
\Phi^{(+)}_{{\invf};\vpedele{p},\lamele}(x^0)=\ee^{-\ii p^0_-x^0}u^{(+)}_{\vpedele{p},\lamele}
\\ & \quad
-\ii\ee^{\ii \pedele{p}^0x^0}\frac{\spedele{p}'-m_{\rm e}c}{2\pedele{p}^0}\int_{-\infty}^{x^0}dy^0
\ee^{-\ii \pedele{p}^0y^0}e\slashed{A}(y^0)\Phi^{(+)}_{{\invf};\vpedele{p},\lamele}(y^0) 
\nonumber\\ & \quad
+\ii\ee^{-\ii \pedele{p}^0x^0}\frac{\spedele{p}+m_{\rm e}c}{2\pedele{p}^0}\int_{x^0}^{\infty}dy^0
\ee^{\ii \pedele{p}^0y^0}e\slashed{A}(y^0)\Phi^{(+)}_{{\invf};\vpedele{p},\lamele}(y^0),
\nonumber
\end{align}
where $\pedele{p}=(\pedele{p}^0,\vpedele{p})$ and $\pedele{p}'=(\pedele{p}^0,-\vpedele{p})$.

Note that the boundary conditions satisfied by the Feynman and anti-Feynman wave functions can be deduced from Eqs.~\eqref{rf2-4} and~\eqref{rf2-7}. 
The Feynman-type wave function $\Phi_{{\rm F};\vpedpos{p},\lampos}^{(-)}(x^0)$ in the remote past takes the form,
\begin{align}\label{rf2-8}
& 
\Phi^{(-)}_{{\rm F};\vpedpos{p},\lampos}(x^0)\big\lvert_{x^0\rightarrow -\infty}\sim \ee^{\ii\pedpos{p}^0x^0}u^{(-)}_{\vpedpos{p},\lampos}
\\ & \quad
+\ii \mathrm{e}^{\ii\pedpos{p}^0x^0}\frac{\spedpos{p}-m_{\rm e}c}{2\pedpos{p}^0}\int_{-\infty}^{\infty}dy^0{\rm e}^{-\ii\pedpos{p}^0y^0}e\slashed{A}(y^0)\Phi^{(-)}_{{\rm F};\vpedpos{p},\lampos}(y^0),
\nonumber
\end{align} 
i.e., it contains only solutions of the free Dirac equation with negative frequencies. On the other hand, in the far future, we have
\begin{align}\label{rf2-9}
& 
\Phi^{(-)}_{{\rm F};\vpedpos{p},\lampos}(x^0)\big\lvert_{x^0\rightarrow\infty}\sim \ee^{\ii\pedpos{p}^0x^0}u^{(-)}_{\vpedpos{p},\lampos}
\\ & \quad
-\ii \ee^{-\ii\pedpos{p}^0x^0}\frac{\spedpos{p}'+m_{\rm e}c}{2\pedpos{p}^0}\int_{-\infty}^{\infty}dy^0\ee^{\ii\pedpos{p}^0y^0}e
\slashed{A}(y^0)\Phi^{(-)}_{{\rm F};\vpedpos{p},\lampos}(y^0).
\nonumber
\end{align}
Thus, the Feynman-type wave function is a combination of the solutions of the free Dirac equation with negative and positive frequencies and with prescribed negative 
frequency part given by $\ee^{\ii\pedpos{p}^0x^0}u^{(-)}_{\vpedpos{p},\lampos}$; the latter corresponding to the created positron with the given momentum and spin. 
The anti-Feynman wave function $\Phi^{(+)}_{{\invf};\vpedele{p},\lamele}(x^0)$ in the remote past contains only positive frequency part,
\begin{align}\label{rf2-10}
& 
\Phi^{(+)}_{{\invf};\vpedele{p},\lamele}(x^0)\big\vert_{x^0\rightarrow -\infty}\sim \ee^{-\ii\pedele{p}^0x^0}u^{(+)}_{\vpedele{p},\lamele}
\\ & \quad
+\ii\ee^{-\ii\pedele{p}^0x^0}\frac{\spedele{p}+m_{\rm e}c}{2\pedele{p}^0}\int_{-\infty}^{\infty}d y^0\ee^{\ii\pedele{p}^0y^0}e\slashed{A}(y^0)
\Phi^{(+)}_{{\invf};\vpedele{p},\lamele}(y^0).
\nonumber
\end{align}
In the remote future, it contains both positive and negative frequency parts with the positive frequency contribution given by 
$\ee^{-\ii\pedele{p}^0x^0}u^{(+)}_{\vpedele{p},\lamele}$, corresponding to the given momentum and spin of the created electron,
\begin{align}\label{rf2-11}
& 
\Phi^{(+)}_{{\invf};\vpedele{p},\lamele}(x^0)\big\vert_{x^0\rightarrow \infty}\sim \ee^{-\ii\pedele{p}^0x^0}u^{(+)}_{\vpedele{p},\lamele}
\\ & \quad
-\ii\ee^{\ii\pedele{p}^0x^0}\frac{\spedele{p}'-m_{\rm e}c}{2\pedele{p}^0}\int_{-\infty}^{\infty}d y^0\ee^{-\ii\pedele{p}^0y^0}e\slashed{A}(y^0)\Phi^{(+)}_{{\invf};\vpedele{p},\lamele}(y^0).
\nonumber
\end{align}
As we will demonstrate next, the above asymptotic conditions for the Feynman and anti-Feynman states allow us to define 
the momentum and spin distributions of created particles.

\section{Pair creation amplitudes and momentum distributions}
\label{sec:momentumdistributions}

The Feynman-type wave function $\Phi^{(-)}_{{\rm F};\vpedpos{p},\lampos}(x^0)$ fulfills the Dirac equation in the form obtained after separating 
the coordinate dependent factor $\ee^{-\ii\vpedpos{p}\cdot\bm{x}}$,
\begin{align}\label{rf3-1}
	[-\ii\gamma^0\partial_0-\bm{\gamma}\cdot\vpedpos{p}+e\slashed{A}(x^0)+m_{\rm e}c]\Phi^{(-)}_{F;\vpedpos{p},\lampos}(x^0)=0.
\end{align}
For the purpose of detailed analysis of the time evolution, the wave function is decomposed into free solutions of the Dirac equation,
\begin{align}\label{rf3-2}
\Phi^{(-)}_{{\rm F};\vpedpos{p},\lampos}(x^0)=\sum_{\lambda=\pm}
&
\bigl[C^{(+)}_{{\rm F};\vpedpos{p},\lampos,\lambda}(x^0)u^{(+)}_{-\vpedpos{p},\lambda}\ee^{-\ii\pedpos{p}^0x^0}
\\ & 
+C^{(-)}_{{\rm F};\vpedpos{p},\lampos,\lambda}(x^0)u^{(-)}_{\vpedpos{p},\lambda}\ee^{\ii\pedpos{p}^0x^0}\bigr].
\nonumber
\end{align}
Momentum of the Dirac bispinors corresponding to components with positive frequency has been chosen as $-\vpedpos{p}$ since $\vpedpos{p}$ is interpreted as the positron momentum so that the electron momentum is equal to $\vpedele{p}=-\vpedpos{p}$. It follows from the asymptotic conditions analyzed in the previous section, Eqs.~\eqref{rf2-8} and~\eqref{rf2-9}, that
\begin{align}\label{rf3-3}
C^{(+)}_{{\rm F};\vpedpos{p},\lampos,\lambda}(-\infty)=0,\, C^{(-)}_{{\rm F};\vpedpos{p},\lampos,\lambda}(\infty)=\delta_{\lampos\lambda}.
\end{align}
Pair creation amplitude can be now obtained from the formula \eqref{rf29} by substituting Eq.~\eqref{rf2-1} and using the explicit form of the free solution $\chi^{(+)}_{\vpedele{p},\lamele}(x)$. This gives
\begin{align}\label{rf3-4}
&
{\cal A}(\vpedele{p},\lamele;\vpedpos{p},\lampos)=(2\pi)^3\delta^{(3)}(\vpedele{p}+\vpedpos{p})
\\ & \quad \nonumber 
\times \ee^{\ii\pedele{p}^0x^0}\bar{u}^{(+)}_{\vpedele{p},\lamele}\gamma^0\Phi^{(-)}_{{\rm F};\vpedpos{p},\lampos}(x^0)\big\lvert_{x^0\rightarrow \infty},
\end{align}
where the momentum conservation delta is due to spatial integration in Eq.~\eqref{rf29}. In this version of the amplitude, momentum and polarization of the positron are fixed and we determine polarization of the electron (its momentum is opposite to the positron momentum). From the asymptotic condition in remote future~\eqref{rf3-3} we have
\begin{align}\label{rf3-6}
&
\Phi^{(-)}_{{\rm F};\vpedpos{p},\lampos}(x^0)\big\lvert_{x^0\rightarrow\infty}=u^{(-)}_{\vpedpos{p},\lampos}\ee^{\ii\pedpos{p}^0x^0}
\\ & \quad \nonumber 
+\sum_{\lambda=\pm}C^{(+)}_{{\rm F};\vpedpos{p},\lampos,\lambda}(x^0)\big\lvert_{x^0\rightarrow\infty}u^{(+)}_{-\vpedpos{p},\lambda}{\rm e}^{-\ii\pedpos{p}^0x^0}.
\end{align}
After substituting Eq.~\eqref{rf3-6} into Eq.~\eqref{rf3-4}, we obtain
\begin{align}\label{rf3-7}
{\cal A}(\vpedele{p},\lamele;\vpedpos{p},\lampos)
=(2\pi)^3\delta^{(3)}(\vpedele{p}+\vpedpos{p})C^{(+)}_{{\rm F};\vpedpos{p},\lampos,\lamele}(\infty).
\end{align}
This is the analogue of Eq.~\eqref{rf30} for the time-dependent electric field. Further, the time-evolution of the coefficients $C^{(+)}_{{\rm F};\vpedpos{p},\lampos,\lambda}(x^0)$ and $C^{(-)}_{{\rm F};\vpedpos{p},\lampos,\lambda}(x^0)$ is analyzed by substituting Eq.~\eqref{rf3-2} into Eq.~\eqref{rf3-1} and solving the resulting system of equations with the asymptotic conditions~\eqref{rf3-3}.

In order to define the momentum distributions of created electrons per unit volume let us introduce the norm of the initial positron state,  
\begin{align}\label{rf3-7a}
\tilde{\mathcal{N}}^2_{{\rm F};\vpedpos{p},\lampos} =\sum_{\lambda=\pm}|C^{(-)}_{{\rm F};\vpedpos{p},\lampos,\lambda}(-\infty)|^2.
\end{align}
Introducing now the `renormalized' amplitude, 
\begin{align}\label{rf3-7b}
A^{(+)}_{\lampos}(\vpedele{p},\lamele)=\frac{1}{\tilde{\mathcal{N}}_{{\rm F};-\vpedele{p},\lampos}}
C^{(+)}_{{\rm F};-\vpedele{p},\lampos,\lamele}(\infty),
\end{align}
with the replacement $\vpedpos{p}=-\vpedele{p}$, we obtain the momentum and spin distribution for created electrons per unit volume, $f^{(+)}_{\lampos}(\vpedele{p},\lamele)$, provided that positrons are created with the spin $\lampos$,
\begin{align}\label{rf3-7c}
f^{(+)}_{\lampos}(\vpedele{p},\lamele)
&
=|A^{(+)}_{\lampos}(\vpedele{p},\lamele)|^2
\\ & \nonumber
=\int\frac{d^3\pedpos{p}}{(2\pi)^3}\frac{|{\cal A}(\vpedele{p},\lamele;\vpedpos{p},\lampos)|^2}{V\tilde{\mathcal{N}}^2_{{\rm F};\vpedpos{p},\lampos}}.
\end{align}
Here, the volume $V$ has been introduced by applying the standard procedure,
\begin{align}\label{rf3-7d}
[(2\pi)^3\delta^{(3)}(\vpedpos{p}+\vpedele{p})]^2\longrightarrow V(2\pi)^3\delta^{(3)}(\vpedpos{p}+\vpedele{p}),
\end{align}
which is well-known in scattering theory. In addition, we introduce the spin-independent momentum distribution for created electrons,
\begin{align}\label{rf3-7e}
f^{(+)}(\vpedele{p})=\sum_{\lampos,\lamele=\pm}f^{(+)}_{\lampos}(\vpedele{p},\lamele).
\end{align}
Note that the above normalization of probability distributions does not follow from the $S$-matrix theory. We have introduced it exclusively 
to provide a meaningful comparison with the distributions calculated by other methods, which will be presented in the next sections.


For the anti-Feynman states we proceed in the same way. The function $\Phi^{(+)}_{{\invf};\vpedele{p},\lamele}(x^0)$
is presented in the form
\begin{align}\label{rf3-7f}
\Phi^{(+)}_{{\invf};\vpedele{p},\lamele}(x^0)=\sum_{\lambda=\pm}
&
\bigl[C^{(+)}_{{\invf};\vpedele{p},\lamele,\lambda}(x^0)u^{(+)}_{\vpedele{p},\lambda}\ee^{-\ii\pedele{p}^0x^0}
\\ & 
+C^{(-)}_{{\invf};\vpedele{p},\lamele,\lambda}(x^0)u^{(-)}_{-\vpedele{p},\lambda}\ee^{\ii\pedele{p}^0x^0}\bigr],
\nonumber
\end{align}
but now with the asymptotic conditions,
\begin{align}\label{rf3-7g}
C^{(-)}_{{\invf};\vpedele{p},\lamele,\lambda}(-\infty)=0,\, C^{(+)}_{{\invf};\vpedele{p},\lamele,\lambda}(\infty)=\delta_{\lamele\lambda}.
\end{align}
Introducing the normalization constant for the state in the remote past,
\begin{align}\label{rf3-7h}
\tilde{\mathcal{N}}^2_{{\invf};\vpedele{p},\lamele} =\sum_{\lambda=\pm}|C^{(+)}_{{\invf};\vpedele{p},\lamele,\lambda}(-\infty)|^2,
\end{align}
we define the `renormalized' amplitude,
\begin{align}\label{rf3-7i}
A^{(-)}_{\lamele}(\vpedpos{p},\lampos)=-\frac{1}{\tilde{\mathcal{N}}_{{\invf};-\vpedpos{p},\lamele}}
C^{(-)}_{{\invf};-\vpedpos{p},\lamele,\lampos}(\infty).
\end{align}
Hence, we arrive at the momentum and spin distribution for created positrons per unit volume, $f^{(-)}_{\lamele}(\vpedpos{p},\lampos)$, provided that electrons are created with the spin $\lamele$,
\begin{align}\label{rf3-7j}
f^{(-)}_{\lamele}(\vpedpos{p},\lampos)=|A^{(-)}_{\lamele}(\vpedpos{p},\lampos)|^2.
\end{align}
Further, by summing up over the final spin states of electrons and positrons we obtain the momentum distribution of created positrons,
\begin{align}\label{rf3-7k}
f^{(-)}(\vpedpos{p})=\sum_{\lamele,\lampos=\pm}f^{(-)}_{\lamele}(\vpedpos{p},\lampos).
\end{align}
Note that for space-homogeneous electric fields it holds that $f^{(-)}(\bm{p})=f^{(+)}(-\bm{p})$.

In summary of theoretical analysis presented in Secs.~\ref{sec:reduction}-\ref{sec:momentumdistributions}, we note that, although the momentum 
and spin distributions of created $e^-e^+$ pairs are determined by the solutions of the Dirac equation, they are not related to the 
`\textit{initial value problem}'. In the considered case, they have to satisfy the asymptotic conditions both in the remote past and future, 
that are imposed either on electrons in the past and positrons in the future (Feynman states), or on positrons in the past and electrons 
in the future (anti-Feynman states). Therefore, determination of the momentum and spin distributions in the dynamical Sauter-Schwinger 
process cannot be reduced to the solution of the Dirac equation with some initial conditions. In particular, the analysis carried out in these 
sections shows that, if in the future we have positrons with the precisely defined momentum and spin state, this determines unambiguously the 
positron state in the past. It also means that the initial positron state cannot be chosen arbitrarily (the same is true for electrons). Therefore, it can be concluded that in the dynamical Sauter-Schwinger process '\textit{the future determines the past}' (see also the discussion presented in the book \cite{bialynicki1975quantumelectro}, cf. Sec.~17, and in the original Feynman's papers \cite{feymann1949qed,feymann1949positrons}). This severely complicates the numerical analysis of the pair creation by an arbitrary space- and time-dependent electromagnetic fields. For this reason other approaches have been developed that try to surpass these difficulties, and which so far have been successfully applied only for time-dependent and homogeneous electric fields. We mean here the Dirac-Heisenberg-Wigner (DHW) function approach (or, equivalently, the bispinorial one \cite{bechler2023schwinger}), originally introduced in Ref.~\cite{bialynicki1991diracvacuum} for the Dirac equation (for the nonrelativistic Schr\"odinger equation, see Ref.~\cite{bialynicki1991TheoryOfQuanta}). Or, the earlier developed method that we call the spinorial method (see, e.g., Refs.~\cite{PhysRevD.83.065020,PhysRevD.98.056009}), which is equivalent to the one based on the quantum Vlasov equation~\cite{PhysRevD.88.045017,PhysRevA.100.012104}. Both are shortly discussed in the following two sections.


\section{The DHW-function approach}
\label{sec:DHWfunction}

The DHW-function for the fermion field $\op{\Psi}(t,{\bm x})$ is defined as
\begin{align}\label{0.1}
W_{\mu\nu}&(t,\bm{x},\bm{p})=-\frac{1}{2}\int d^3s\, \ee^{-\ii\bm{p}\cdot\bm{s}}\\
&\times\langle 0|\mathcal{U}(t,\bm{s},\bm{x})[\op{\Psi}_{\mu}(t,\bm{x}+\bm{s}/2),\op{\Psi}^\dag_\nu(t,\bm{x}-\bm{s}/2)]|0\rangle,\nonumber
\end{align}
where
\begin{align}\label{0.1a}
\mathcal{U}(t,\bm{s},\bm{x})=\exp\Bigl[-\ii e\int_{-1/2}^{1/2}d\xi\, \bm{s}\cdot\bm{A}(t,\bm{x}+\xi\bm{s}) \Bigr] .
\end{align}
It can be decomposed in terms of the Dirac matrices (for the definition of the Dirac matrices $\gamma^\mu$ and the related ones which are used here, 
see Ref.~\cite{bechler2023schwinger}),
\begin{align}\label{0.5}
 W(t,\bm{x},\bm{p})&=\frac{1}{4}(f_0+\gamma_5f_1-\ii\gamma^0\gamma_5f_2+\gamma^0f_3\\
 &+\bm{\Sigma}\cdot\bm{g}_0+\bm{\alpha}\cdot\bm{g}_1
 -\ii\bm{\gamma}\cdot\bm{g}_2+\gamma^0\bm{\Sigma}\cdot\bm{g}_3),\nonumber
\end{align}
with sixteen dimensionless expansion coefficients. For a spatially homogeneous electric field only ten functions: 
$f_3,\,\bm{g}_0,\,\bm{g}_1, \,\bm{g}_2$ survive. Denoting by $W$ the ten-dimensional column vector (the superscript $T$ below means the transposition),
\begin{align}\label{0.6}
W(t,\bm{p})=\left[\begin{array}{c}f_3(t,\bm{p}),\, \bm{g}_0(t,\bm{p}),\, \bm{g}_1(t,\bm{p}),\, \bm{g}_2(t,\bm{p})\end{array}\right]^T,
\end{align}
we can write equations fulfilled by the expansion coefficients in the compact matrix form
\begin{align}\label{0.7}
\dot{W}(t,\bm{p}(t))=M(\bm{p}(t))W(t,\bm{p}(t)),
\end{align}
where $\bm{p}(t)=\bm{p}-e\bm{A}(t)$ and dot denotes complete time derivative. The matrix $M$ has the following block structure,
\begin{equation}\label{0.8}
M(\bm{p})=\left[\begin{array}{cccc}
		0 & \bm{0}^T & \bm{0}^T & 2c\bm{p}^T\\
		\bm{0} & \mathds{O}_3 & 2c\bm{p}\times & \mathds{O}_3\\
		\bm{0} & 2c\bm{p}\times & \mathds{O}_3 & -2m_{\mathrm{e}}c^2 \mathds{I}_3\\
		-2c\bm{p} & \mathds{O}_3 & 2m_{\mathrm{e}}c^2 \mathds{I}_3 & \mathds{O}_3
	\end{array}\right],
\end{equation}
where the symbol $\bm{p}\times$ means the vector product with the corresponding $\bm{g}_0$ or $\bm{g}_1$ vector functions.

Note that the ten-dimensional vector $W$ [Eq.~\eqref{0.6}] can be also constructed by means of so-called bispinorial representation of DHW function~\cite{bechler2023schwinger}
\begin{align}\label{eq1}
{\cal V}(x^0,\bm{p})=\left[ \begin{array}{c}
\mathrm{Tr}(\gamma^0A_{\bm{p}}(x^0))\\\mathrm{Tr}(\bm{\Sigma}A_{\bm{p}}(x^0))\\\mathrm{Tr}(\bm{\alpha}A_{\bm{p}}(x^0))\\\mathrm{Tr}(-\ii\bm{\gamma}A_{\bm{p}}(x^0))
\end{array}\right],
\end{align}
where $A_{\bm p}$ is the following $4\times 4$ matrix,
\begin{align}\label{eq2}
A_{\bm{p}}(x^0)=\sum_\lambda\Phi_{\bm{p},\lambda}(x^0)\Phi^\dag_{\bm{p},\lambda}(x^0),
\end{align}
and $\Phi_{\bm{p},\lambda}(x^0)$ is implicitly defined in Eq.~\eqref{rf2-1} for the Feynman- and in Eq.~\eqref{rf2-5} for the anti-Feynman boundary condition.
It has been shown in Ref.~\cite{bechler2023schwinger} that the vector ${\cal V}$ fulfills the same system of differential equations as~\eqref{0.7}. Moreover,  
its form for $x^0\rightarrow -\infty$ is the same as the vacuum value of $W$~\cite{bialynicki1991diracvacuum},
\begin{align}\label{eq3}
W^{\rm vac}(\bm{p})=\frac{2}{p^0}\left[\begin{array}{c}
-m_ec\\  \bm{0}\\-\bm{p}\\ \bm{0}
\end{array}
\right],
\end{align}
provided that
\begin{align}\label{eq4}
A_{\bm{p}}(x^0)|_{x^0\rightarrow -\infty}&=\sum_\lambda\Phi^{(-)}_{F;\bm{p},\lambda}(x^0)\Phi^{(-)\dag}_{F;\bm{p},\lambda}(x^0)\bigr|_{x^0\rightarrow -\infty}\nonumber\\
&=\sum_\lambda u^{(-)}_{-\bm{p},\lambda}u^{(-)\dag}_{-\bm{p},\lambda}.
\end{align} 
The latter corresponding to the Feynman-type boundary condition in the past. Since $W$ and ${\cal V}$ fulfill the same system of differential equations and,
with the Feynman-type boundary condition in the past, these two vectors have the same initial values, therefore ${\cal V}=W$. This shows explicitly the connection
of the DHW-function approach with the boundary conditions exploited in this paper.

Basically, Eq.~\eqref{0.7} could be used for numerical analysis, e.g., in the case of circularly or elliptically polarized field. However, it has been argued in Ref.~\cite{blinne2014pair} that there is a serious precision problem and reformulation of the equations is required. The idea is to decompose $W$ in the following manner
\begin{align}\label{0.9}
	W=(e_1\cdot W)e_1+\mathcal{P}W,
\end{align}
where  $e_1$ is the four-vector proportional to the vacuum value of $W$,
\begin{align}\label{0.10}
	e_1=-\frac{1}{2}W^\mathrm{vac}=\dfrac{1}{p^0}\left[\begin{array}{c}
m_{\mathrm{e}}c,\, \bm{0},\, \bm{p},\, \bm{0}
\end{array}\right]^T,
\end{align}
$\mathcal{P}$ is  the projection matrix onto the nine-dimensional subspace orthogonal to $e_1$, and $p^0=(\bm{p}^2+m_{\mathrm{e}}^2c^2)^{1/2}$. Next, we express the scalar product in Eq.~\eqref{0.9} by one-particle distribution function per unit volume, $f^{(+)}_{\rm W}(t,\bm{p})$,
\begin{align}\label{0.11}
		e_1\cdot W=\frac{m_{\mathrm{e}}cf_3}{p^0}+\frac{\bm{p}\cdot\bm{g}_1}{p^0}=2(f_{\rm W}^{(+)}-1).
\end{align}
Also, following Ref.~\cite{blinne2014pair}, we write part of $W$ orthogonal to $e_1$ as
\begin{align}\label{0.12}
	\mathcal{P}W=\mathcal{T}w_9,
\end{align}
where $\mathcal{T}$ is the $10\times 9$ matrix such that $e_1^T\mathcal{T}=0_9^T$ ($0_9$ denotes nine-dimensional column null vector). 
Here, the vector $w_9$ has nine components,
\begin{align}\label{0.13}
	w_9=\left[\begin{array}{c}
\bm{v}_1,\, \bm{v}_2,\, \bm{v}_3	
 \end{array}\right]^T.
\end{align}
As a result, Eq.~\eqref{0.9} becomes
\begin{align}\label{0.14}
	W=2(f_{\mathrm{W}}^{(+)}-1)e_1+\mathcal{T}w_9.
\end{align}
Note that the choice of matrix $\mathcal{T}$ fulfilling condition $e_1^T\mathcal{T}=0_9^T$ is not unique. Here, we use its form in analogy with Ref.~\cite{blinne2014pair}, adjusted to our ordering of the DHW-coefficients,
\begin{align}\label{0.16}
	\mathcal{T}=\left[\begin{array}{ccc}
		-\dfrac{\bm{p}^T}{m_{\mathrm{e}}c} & \bm{0}^T & \bm{0}^T  \\ 
		\mathds{O}_3 & \mathds{I}_3 & \mathds{O}_3\\ 
			\mathds{I}_3 & \mathds{O}_3 & \mathds{O}_3\\ 
				\mathds{O}_3 & \mathds{O}_3 & \mathds{I}_3
		\end{array}
	\right],
\end{align}
where $\mathds{O}_3$ and $\mathds{I}_3$ denote respectively the $3\times 3$ null and identity matrices. After some algebra it leads to the following system of differential equations,
\begin{align}\label{0.22}
	\begin{split}
	&\dot{f}_{\rm W}^{(+)}=\frac{e\bm{\mathcal{E}}\cdot\bm{v}_1}{2p^0},\\
	&\dot{\bm{v}}_1=\frac{2e}{(p^0)^3}[\bm{p}(\bm{p}\cdot\bm{\mathcal{E}})-(p^0)^2\bm{\mathcal{E}}](f_{\mathrm{W}}^{(+)}-1) \\
	&\quad -\frac{e\bm{p}(\bm{\mathcal{E}}\cdot\bm{v}_1)}{(p^0)^2}+2c\bm{p}\times\bm{v}_2-2m_{\mathrm{e}}c^2\bm{v}_3,\\
	&\dot{\bm{v}}_2=2c\bm{p}\times\bm{v}_1,\\
	&\dot{\bm{v}}_3=\frac{2}{m_{\mathrm{e}}}[\bm{p}(\bm{p}\cdot\bm{v}_1)+(m_{\mathrm{e}}c)^2\bm{v}_1],
	\end{split}
\end{align} 
where $\bm{p}$ has to be replaced by $\bm{p}(t)$ and $\dot{\bm{p}}(t)=-e\dot{\bm{A}}(t)=e\bm{\mathcal{E}}(t)$ was used. The initial conditions for $t\rightarrow -\infty$ are
\begin{align}\label{0.23}
	f_{\mathrm{W}}^{\mathrm{in}}=0,\quad \bm{v}_1^{\mathrm{in}}=\bm{v}_2^{\mathrm{in}}=\bm{v}_3^{\mathrm{in}}=\bm{0},
\end{align}
where the vector potential vanishes and $\bm{p}(t)\rightarrow\bm{p}$, with $\bm{p}$ being the momentum of the created electron. These equations are the same as those derived and discussed in Refs.~\cite{blinne2014pair,blinne2016wigner,li2017pairvortex,li2023pairspiralorginal}.

It has to be stressed that Eqs.~\eqref{0.22} have been derived under the assumption of Feynman-type boundary conditions. For anti-Feynman boundary conditions, the Dirac wave function in the past contains only positive frequencies, i.e.,
\begin{align}\label{0.26}
	\psi(t,\bm{x})|_{t\rightarrow -\infty}=\ee^{-\ii cp^0 t+\ii\bm{p}\cdot\bm{x}}u^{(+)}_{\bm{p},\lambda}.
\end{align}
Hence,
\begin{align}\label{0.27}
	W^{\mathrm{vac}}=\sum_\lambda u^{(+)\dag}_{\bm{p},\lambda}
	\left[\begin{array}{c}
		\gamma^0\\ \bm{\Sigma}\\ \bm{\alpha}\\  -\ii\bm{\gamma}
	\end{array}\right]u^{(+)}_{\bm{p},\lambda} 
	=\frac{2}{p^0}\left[
	\begin{array}{c}
		m_{\mathrm{e}}c\\ \bm{0}\\ \bm{p}\\ \bm{0}
	\end{array}\right],
\end{align}
which is a negative "Feynman-type" vacuum condition. The one-particle distribution function per unit volume, $f_{\mathrm{W}}^{(-)}(t,\bm{p})$, is now given by
\begin{align}\label{0.28}
	f_{\mathrm{W}}^{(-)}=\frac{1}{2}(e_1\cdot W-e_1\cdot W^{\mathrm{vac}})=\frac{m_{\mathrm{e}}cf_3+\bm{p}\cdot\bm{g}_1}{2p^0}-1,
\end{align}
whereas for the "Feynman-type" vacuum conditions we had $+1$, since in that case $(1/2)e_1\cdot W^{\mathrm{vac}}=-1$. Due to this change of sign, Eq.~\eqref{0.14} has to be replaced by
\begin{align}\label{0.29}
	W=2(f_{\mathrm{W}}^{(-)}+1)e_1+\mathcal{T}w_9.
\end{align}
This leads to the system of differential equations,
\begin{align}\label{0.30}
	\begin{split}
	&\dot{f}_{\rm W}^{(-)}=\frac{e\bm{\mathcal{E}}\cdot\bm{v}_1}{2p^0},\\
	&\dot{\bm{v}}_1=\frac{2e}{(p^0)^3}[\bm{p}(\bm{p}\cdot\bm{\mathcal{E}})-(p^0)^2\bm{\mathcal{E}}](f_{\mathrm{W}}^{(-)}+1) \\
	&\quad -\frac{e\bm{p}(\bm{\mathcal{E}}\cdot\bm{v}_1)}{(p^0)^2}+2c\bm{p}\times\bm{v}_2-2m_{\mathrm{e}}c^2\bm{v}_3,\\
	&\dot{\bm{v}}_2=2c\bm{p}\times\bm{v}_1,\\
	&\dot{\bm{v}}_3=\frac{2}{m_{\mathrm{e}}}[\bm{p}(\bm{p}\cdot\bm{v}_1)+(m_{\mathrm{e}}c)^2\bm{v}_1].
	\end{split}
\end{align} 
with the same initial conditions~\eqref{0.23}, but now $-\bm{p}$ is the momentum of the created positron.

The functions $f_{\mathrm{W}}^{(\pm)}(t,\bm{p})$ and $f^{(\pm)}(\bm{p}_{\mp})$, the latter defined by Eqs.~\eqref{rf3-7e} and~\eqref{rf3-7k}, are related with each other. Namely,
\begin{align}\label{0.31}
\lim_{t\rightarrow\infty}f_{\mathrm{W}}^{(\pm)}(t,\pm\bm{p})=f^{(\pm)}(\bm{p}),
\end{align} 
which is going to be the topic of our numerical analysis. Moreover, identical equations can be derived by using the bispinorial approach~\cite{bechler2023schwinger}.

Concluding this section, we note that only for the space homogeneous electric field the determination of the spin summed up momentum distributions of electrons or positrons can be reduced to the initial value problem, provided that extra normalizations \eqref{rf3-7b} or \eqref{rf3-7i} are applied. However, already for the spin-resolved distributions one has to account for the Feynman or anti-Feynman asymptotic conditions that couple `the future to the past', i.e., depending on spins and momenta of created pairs the initial state has to be prepared in a particular well-defined spin state.

\section{The spinorial approach}
\label{sec:Spinorial}

Eqs.~\eqref{0.22} or~\eqref{0.30} can be further simplified for the linearly polarized electric field, $\bm{\mathcal{E}}(t)=\mathcal{E}(t)\bm{n}$, 
where $\bm{n}$ is a constant unit vector. Then, one can show that instead of solving the system of ten differential equations the problem can be 
reduced to the solution of two equations for complex amplitudes. This system of equations has the form similar to the precession of spin 1/2 
in a time-dependent magnetic field,
\begin{align}\label{spinor1}
	\ii\frac{d}{dt}\left[\begin{array}{c}
		c_{\bm{p}}^{(+)}(t)\\c_{\bm{p}}^{(-)}(t)\end{array}\right]=\left[\begin{array}{cc}
			\omega_{\bm{p}}(t) & \ii\Omega_{\bm{p}}(t)\\-\ii\Omega_{\bm{p}}(t) & -\omega_{\bm{p}}(t)
	\end{array}\right]\left[\begin{array}{c}
	c_{\bm{p}}^{(+)}(t)\\c_{\bm{p}}^{(-)}(t)\end{array}\right],
\end{align} 
where
\begin{align}\label{spinor2}
	\omega_{\bm p}(t)=c\sqrt{\bm{p}_\perp^2+[p_{\|}-eA(t)]^2+m_{\mathrm{e}}^2c^2} 
\end{align} 
and
\begin{align}\label{spinor3}
\Omega_{\bm p}(t)=\frac{c\epsilon_\perp e\mathcal{E}(t)}{2[\omega_{\bm p}(t)]^2},
\end{align} 
with $\epsilon_\perp=c\sqrt{\bm{p}_\perp^2+m^2c^2}$, $p_{\|}=\bm{p}\cdot\bm{n}$, and $\bm{p}_\perp=\bm{p}-p_{\|}\bm{n}$. 

By solving the system of equations \eqref{spinor1} with the initial conditions $c^{(+)}_{\bm{p}}(-\infty)=1$ and $c^{(-)}_{\bm{p}}(-\infty)=0$ (analogue of the anti-Feynman boundary conditions) we obtain the momentum distribution for created positrons,
\begin{align}\label{spinor4}
f^{(-)}(\bm{p})=2|c^{(-)}_{-\bm{p}}(\infty)|^2.
\end{align}
On the other hand, by choosing the initial conditions as $c^{(+)}_{\bm{p}}(-\infty)=0$ and $c^{(-)}_{\bm{p}}(-\infty)=1$ (analogue of the Feynman boundary conditions) we arrive at the momentum distribution for created electrons,
\begin{align}\label{spinor5}
f^{(+)}(\bm{p})=2|c^{(+)}_{\bm{p}}(\infty)|^2.
\end{align} 
In closing this section, we recall that the spinorial approach has been developed much earlier than the DHW-function method (see, Refs.~\cite{grib1972pair,mostepanenko1974pair,bagrov1975pair,grib1994vacuumquantum,gavrilov1996vacuum}). Moreover, a similar system of equations can be also obtained for the Klein-Gordon equation, for which the system is pseudo-hermitian \cite{PhysRevA.100.062116}. Finally, let us remark that the equivalent approach to the one presented in this section has been developed by Avetissian and co-workers \cite{avetissian2002pair} and extensively used in Refs.~\cite{jansen2013strongpair,villalbachavez2019schwinger,folkerts2023pair}.


\section{Numerical illustrations}
\label{sec:Numer}

The aim of a numerical analysis presented in this section is, among others, to compare three approaches discussed above. We start our discussion with linearly polarized electric field pulses, for which all three methods can be applied.
Next, we shall consider circularly polarized pulses, for which only the S-matrix and the DHW-function approaches can be used. In both cases, we have to define the time-dependent pulses of the electric field, $\bm{\mathcal{E}}(t)$, that satisfy the condition
\begin{equation}\label{num1}
\int_{-\infty}^{\infty} \dd t \,\bm{\mathcal{E}}(t)=\bm{0}.
\end{equation} 
Due to this equation, the vector potential
\begin{equation}\label{num2}
\bm{A}(t)=\int_t^{\infty} \dd t' \,\bm{\mathcal{E}}(t'),
\end{equation} 
fulfills the conditions $\bm{A}(\pm\infty)=\bm{0}$, under which our theoretical analysis has been carried out.

\subsection{Linear polarization}
\label{sec:LinPol}

\subsubsection{Sauter-like electric field pulse}
\label{sauter-pulse}

For the Sauter-Schwinger process in a linearly polarized electric field, the standard choice of the pulse shape is the Sauter one. Its time dependence 
is given by the function $1/\cosh^2(t/\tau_0)$, where the parameter $\tau_0$ determines the pulse duration. Although exact analytical solutions 
of the spinorial equations~\eqref{spinor1} can be constructed in this case, the pulse itself does not meet the condition~\eqref{num1}. For this 
reason, we introduce a function $F(t)$,
\begin{equation}\label{num3}
F(t)=\frac{\tau_0}{\cosh(t/\tau(t))}, \, \tau(t)=\tau_0[1+\sigma\tanh(t/t_0)],
\end{equation} 
and define the electric field and vector potential as
\begin{equation}\label{num4}
\bm{\mathcal{E}}(t)={\cal E}(t)\bm{\varepsilon}=-\mathcal{E}_0F'(t)\bm{\varepsilon}, \quad \bm{A}(t)=A(t)\bm{\varepsilon}= \mathcal{E}_0F(t)\bm{\varepsilon}.
\end{equation} 
Here, $\bm{\varepsilon}$ is a constant unit vector determining the polarization of the electric field and $\mathcal{E}_0$ fixes its strength. 
${\cal E}(t)$ and $A(t)$ are referred to as shape functions for the electric field and the vector potential, respectively, and are defined 
by Eqs.~\eqref{num3} and~\eqref{num4}. The pulse additionally depends on three parameters, $\tau_0>0$, $-1<\sigma<1$, and $t_0>0$,
of which $\tau_0$ controls the duration of the pulse, whereas $\sigma$ relates to the rate of its growth and decay. In particular, if $\sigma$ is 
close to unity, the pulse builds up quickly and decays slowly. The opposite situation takes place when $\sigma$ is close to $-1$. The third parameter, $t_0$, plays a minor role and in our numerical analysis is put equal to $\tau_0$.

\begin{figure}
\includegraphics[width=6cm]{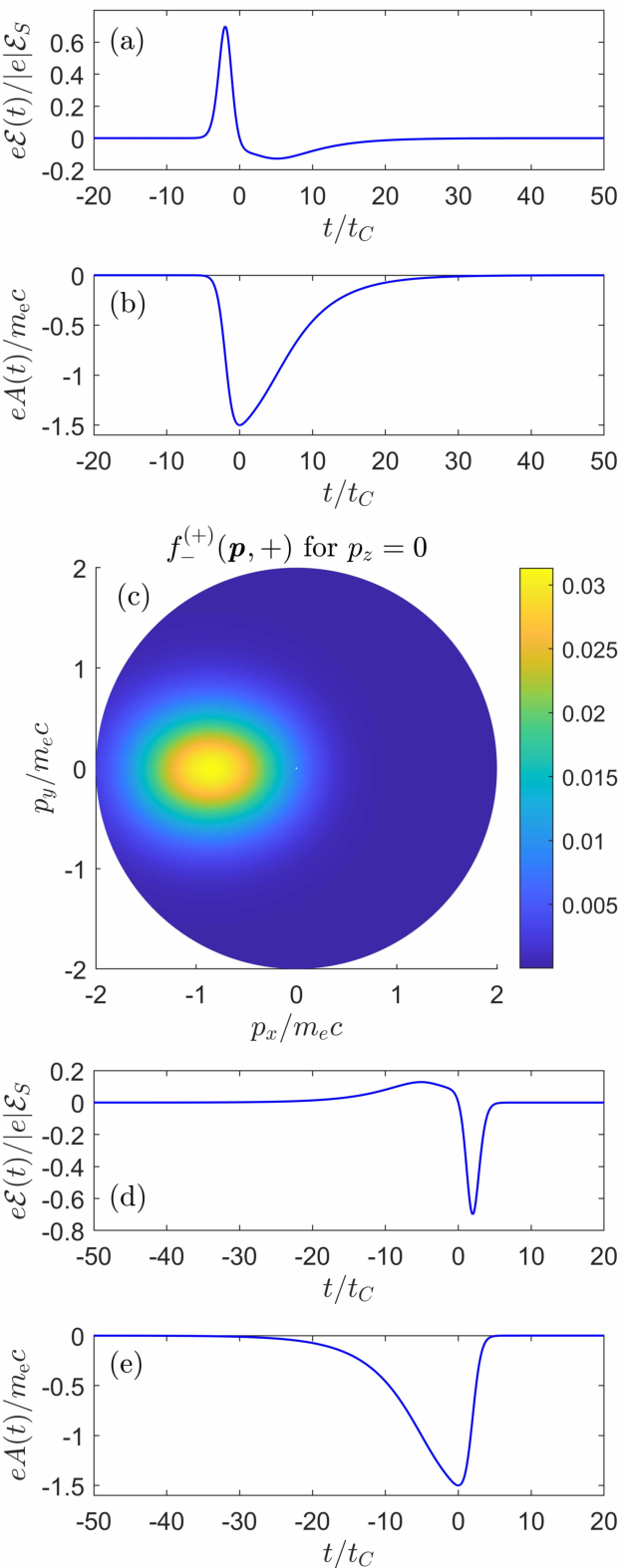}
\caption{Shape functions for (a) the electric field, ${\cal E}(t)$, and (b) the vector potential, $A(t)$, (both multiplied by the electron charge 
$e=-|e|$), defined by Eqs.~\eqref{num3} and~\eqref{num4}. Here, $\bm{\varepsilon}=\bm{e}_x$ and $\mathcal{E}_0=0.5\mathcal{E}_S$, $\tau_0=t_0=3t_C$, 
$\sigma=0.8$. Panels (d) and (e) show the same but for $\sigma=-0.8$. In both cases the electric field has the dominant extremum at 
$t_+\approx -2(\sigma/|\sigma|)t_C$, at which $|\mathcal{E}(t_+)|\approx 0.7\mathcal{E}_S$ and $eA(t_+)\approx -0.8m_\mathrm{e}c$. Panel (c) 
demonstrates the spin-resolved electron momentum distribution $f^{(+)}_-(\bm{p},+)$ for the spin quantization axis along the $z$-direction. Note, that $f^{(+)}_+(\bm{p},+)=0$. Relativistic units are used.
}
\label{fpolafasymlin1and2aK}
\end{figure}

Examples of such pulses are illustrated in Fig.~\ref{fpolafasymlin1and2aK}. As we see, for $|\sigma|=0.8$ the electric field has the dominant single peak for time $t_+\approx -2(\sigma/|\sigma|)t_C$, at which, according to the tunneling theory (see, e.g., \cite{brezin1970pair,VladimirSPopov2004} and references therein), the pairs are mostly created with the vanishing momenta. Since, due to the classical Newton equation for electrons,
\begin{equation}\label{num5}
\frac{\dd}{\dd t}[\bm{p}(t)+e\bm{A}(t)]=0,
\end{equation} 
and accounting for the facts that $\bm{p}(t_+)=\bm{0}$ and $\bm{A}(\infty)=\bm{0}$, we get the final electron momentum of created pairs,
\begin{equation}\label{num6}
\bm{p}=\bm{p}(\infty)=e\bm{A}(t_+)\approx -0.8\me c\bm{\varepsilon},
\end{equation} 
at which the electron momentum distribution reaches its maximum. This rough estimation agrees very well with the distribution $f^{(+)}_-(\bm{p},+)$ for $p_z=0$ and $\bm{\varepsilon}=\bm{e}_x$, presented in Fig.~\ref{fpolafasymlin1and2aK}(c). In this analysis the spin quantization axis has been chosen as the $z$-axis. Moreover, our numerical analysis shows that $f^{(+)}_+(\bm{p},-)=f^{(+)}_-(\bm{p},+)$ and $f^{(+)}_+(\bm{p},+)=f^{(+)}_-(\bm{p},-)=0$. Thus, the spin summed up electron distribution equals $f^{(+)}(\bm{p})=2f^{(+)}_-(\bm{p},+)$, which agrees with the DHW-function (or bispinorial) and spinorial approaches discussed above.

\begin{figure}
\includegraphics[width=6cm]{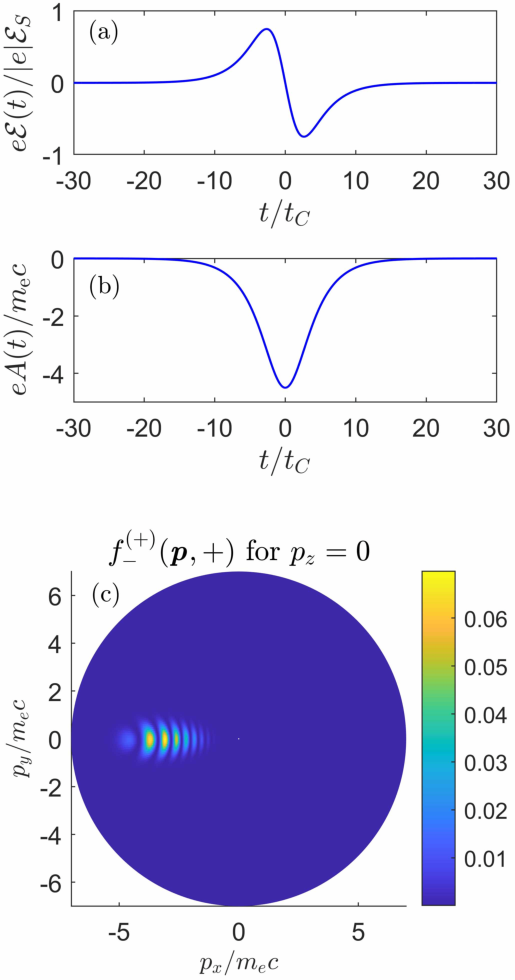}
\caption{
Shape functions for (a) the electric field, ${\cal E}(t)$, and (b) the vector potential, $A(t)$, (both multiplied by the electron charge 
$e=-|e|$), defined by Eqs.~\eqref{num3} and~\eqref{num4} for $\bm{\varepsilon}=\bm{e}_x$, $\mathcal{E}_0=1.5\mathcal{E}_S$, $\tau_0=t_0=3t_C$, 
$\sigma=0$. Panel (c) presents the spin-resolved electron momentum distribution $f^{(+)}_-(\bm{p},+)$ for the spin quantization axis along the 
$z$-direction, in which we observe the interference pattern due to the existence of two extrema for the electric field at times 
$t_\pm\approx \pm 2.64t_C$, for which $|\mathcal{E}(t_\pm)|\approx 0.75\mathcal{E}_S$ and $eA(t_\pm)\approx -3.2m_\mathrm{e}c$.
Relativistic units are used.
}
\label{fpolafsymlin0K}
\end{figure}

The pattern is changed for $\sigma=0$, as illustrated in Fig.~\ref{fpolafsymlin0K}. Now we have two extrema for the electric field strength at times $t_\pm\approx \pm 2.64t_C$, at which pairs are created predominantly. According to the classical analysis, the final momenta of electrons created at these times are the same and equal to $e\bm{A}(t_\pm)\approx -3.2m_\mathrm{e}c\bm{e}_x$. Therefore, as a result of the superposition of two amplitudes, we observe the interference pattern in the momentum distribution. The center of this interference pattern is well reproduced by the classical analysis. Similarly to the previous case, $f^{(+)}_+(\bm{p},-)=f^{(+)}_-(\bm{p},+)$, $f^{(+)}_+(\bm{p},+)=f^{(+)}_-(\bm{p},-)=0$ and the spin summed up distribution $f^{(+)}(\bm{p})$ is the same by applying all three methods described above.

\subsubsection{Helicity-resolved distributions}
\label{sec:LinPolHel}

In Sec.~\ref{sauter-pulse}, we have investigated the spin-resolved momentum distributions for $p_z=0$ and for the spin-quantization axis chosen 
in the $z$-direction. In this case, distributions for the same spin projections of electrons and positrons are zero. For other choices of 
quantization axis the situation becomes more complex and, in general, all spin-resolved distributions are nonzero. Still, the total spin summed up
distribution stays unchanged. To illustrate this, we consider now the case when the pairs are created in their helicity states, 
for which the free bispinors have the form [cf., Eq.~\eqref{rf5}],
\begin{align}\label{rf5a}
u^{(+)}_{\bm{p},\lambda}&=\sqrt{\frac{p^0+m_{\rm e}c}{2p^0}}\left[\begin{array}{c}
\chi_\lambda(\bm{p}) \\  \frac{\bm{\sigma}\cdot\bm{p}}{p^0+m_{\rm e}c}\chi_\lambda(\bm{p})\\  \end{array}\right],\nonumber\\
u^{(-)}_{-\bm{p},\lambda}&=\sqrt{\frac{p^0+m_{\rm e}c}{2p^0}}\left[\begin{array}{c}
-\frac{\bm{\sigma}\cdot\bm{p}}{p^0+m_{\rm e}c}\chi_\lambda(\bm{p})\\ \chi_\lambda(\bm{p})
\end{array}\right],
\end{align}
for $\lambda=\pm$, and
\begin{align}\label{rf5b}
\chi_+(\bm{p})=&\left[\begin{array}{c}\ee^{-\ii\varphi_{\bm{p}}/2}\cos(\theta_{\bm{p}}/2) \\ \ee^{\ii\varphi_{\bm{p}}/2}\sin(\theta_{\bm{p}}/2) \end{array}\right],
\, \nonumber\\
\chi_-(\bm{p})=&\left[\begin{array}{c}-\ee^{-\ii\varphi_{\bm{p}}/2}\sin(\theta_{\bm{p}}/2) \\ \ee^{\ii\varphi_{\bm{p}}/2}\cos(\theta_{\bm{p}}/2) \end{array}\right].
\end{align}
Here, $\theta_{\bm{p}}$ and $\varphi_{\bm{p}}$ denote the polar and azimuthal angles of the particle momentum $\bm{p}$, respectively,
\begin{equation}\label{rf5c}
\frac{\bm{p}}{|\bm{p}|}=\bm{n}_{\bm{p}}=[\sin\theta_{\bm{p}}\cos\varphi_{\bm{p}},\sin\theta_{\bm{p}}\sin\varphi_{\bm{p}},\cos\theta_{\bm{p}}]^T,
\end{equation}
and the spinors $\chi_\lambda(\bm{p})$ fulfill the eigenvalue equation,
\begin{equation}\label{rf5d}
(\bm{\sigma}\cdot\bm{n}_{\bm{p}})\chi_\lambda(\bm{p})=\lambda\chi_\lambda(\bm{p}).
\end{equation}
Hence, the helicity states are the eigenstates of both the free Dirac Hamiltonian and the helicity operator (see, e.g., Ref.~\cite{bialynicki1975quantumelectro}).

\begin{figure}
\includegraphics[width=6cm]{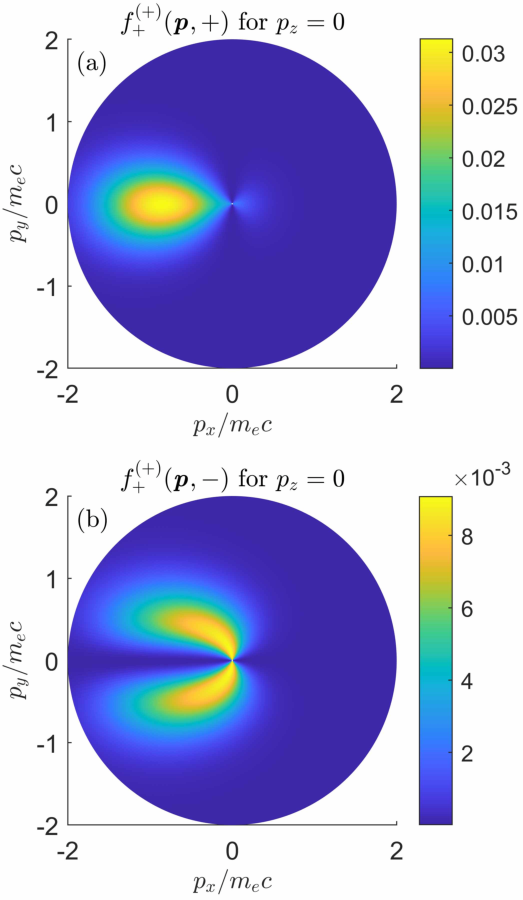}
\caption{
Helicity-resolved momentum distributions for created electrons if positrons and electrons are generated in their helicity eigenstates. 
The parameters are the same is in Fig.~\ref{fpolafasymlin1and2aK}. In this case, all distributions (in relativistic units) are nonzero 
with $f^{(+)}_+(\bm{p},+)=f^{(+)}_-(\bm{p},-)$ and $f^{(+)}_+(\bm{p},-)=f^{(+)}_-(\bm{p},+)$.
}
\label{fpolafasymlin1h1sK}
\end{figure}

In Fig.~\ref{fpolafasymlin1h1sK}, we present the helicity-resolved momentum distributions for created electrons. For this particular geometry and 
kinematics of the Sauter-Schwinger process we find that $f^{(+)}_+(\bm{p},+)=f^{(+)}_-(\bm{p},-)$ and $f^{(+)}_+(\bm{p},-)=f^{(+)}_-(\bm{p},+)$. 
Moreover, the sum $f^{(+)}_+(\bm{p},+)+f^{(+)}_+(\bm{p},-)$ exactly reproduces the distribution presented in Fig.~\ref{fpolafasymlin1and2aK}. 
As we see, the electrons and positrons with the same helicities are generated along the electric field polarization direction and their distributions 
are well reproduced by the tunneling theory. This is not longer valid for electrons and positrons of the opposite helicities which are rather created
in the perpendicular direction with respect to the electric field polarization vector. However, for the chosen parameters  
the dominant contribution to the $e^-e^+$ pair creation comes from processes where the created particles have the same helicity. In this case, the maximum of the momentum distribution 
is more than three times larger then for the processes that result in production of particles with the opposite helicities. This is, however, not the universal property, as it is going to be discussed in the following section.

\subsubsection{Oscillating electric field pulse}
\label{sec:LinPolOsc}

In this section, we study the Sauter-Schwinger process by linearly polarized, oscillating in time electric field pulses. 
For this purpose, we consider the vector potential [Eq.~\eqref{num4}] with the shape function,
\begin{equation}\label{num7}
A(t)=F(t)\cos(\omega t+\chi),
\end{equation} 
where $\omega$ is the carrier wave frequency whereas $\chi$ denotes the carrier envelope phase. Moreover, the pulse envelope is defined by 
Eq.~\eqref{num3}. As before, due to Eq.~\eqref{num4}, the condition~\eqref{num1} is satisfied.


\begin{figure}
\includegraphics[width=6.2cm]{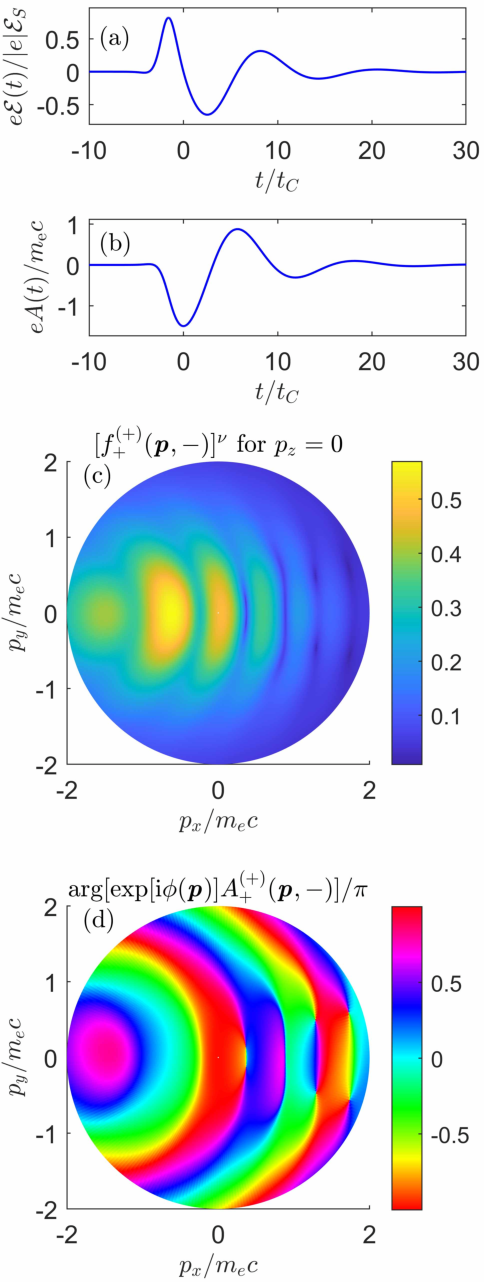}
\caption{
In panels (a) and (b) we demonstrate the electric field, ${\mathcal{E}}(t)$, and the vector potential, $A(t)$, shape functions (both multiplied by 
the electron charge $e=-|e|$), as defined by Eq.~\eqref{num4}, for $\bm{\varepsilon}=\bm{e}_x$, $\mathcal{E}_0=0.5\mathcal{E}_S$, 
$\tau_0=t_0=3t_C$, $\sigma=0.8$, $\omega=0.5\me c^2$, and $\chi=0$. Panel (c) exhibits the spin-resolved electron momentum distribution 
$f^{(+)}_-(\bm{p},+)$ [expressed in relativistic units and raised to the power $\nu=1/4$ for visual purposes] for the spin quantization axis ${\bm e}_z$. 
Note that $f^{(+)}_+(\bm{p},-)=f^{(+)}_-(\bm{p},+)$ and $f^{(+)}_-(\bm{p},-)=f^{(+)}_+(\bm{p},+)=0$. In panel (d) the phase of the 
amplitude $A^{(+)}_+(\bm{p},-)$ with an extracted phase factor, $\ee^{-\ii\phi(\bm p)}$, is shown [cf. Eq.~\eqref{rf3-7b}]. More details on
the phase factor $\ee^{\ii\phi(\bm p)}$ is given in the text.
}
\label{fpolafasymlinosclow0bK}
\end{figure}

First, we consider the low-frequency case, when $\omega < \me c^2$. Under these circumstances, the tunneling theory is applicable and we expect 
the electron momentum distribution to be concentrated along the polarization vector of the electric field, $\bm{\varepsilon}$. This is indeed the case, as presented in 
Fig.~\ref{fpolafasymlinosclow0bK}. Moreover, from the analysis of Figs.~\ref{fpolafasymlinosclow0bK}(a) and~\ref{fpolafasymlinosclow0bK}(b) we 
learn that the electric field has local extrema for $t_1=-1.6t_C$, $t_2=2.5t_C$, and $t_3=8.2t_C$, at which $e\mathcal{E}(t_1)=0.82\mathcal{E}_S$, 
$e\mathcal{E}(t_2)=-0.65\mathcal{E}_S$, and $e\mathcal{E}(t_3)=0.315\mathcal{E}_S$. Applying now the classical analysis discussed above we expect 
to have maxima of the distribution for momenta $e\bm{A}(t_1)=-0.74\me c\bm{\varepsilon}$, $e\bm{A}(t_2)=-0.4\me c\bm{\varepsilon}$, and 
$e\bm{A}(t_3)=0.37\me c\bm{\varepsilon}$. This quite well agrees with the results presented in Fig.~\ref{fpolafasymlinosclow0bK}(c), except for $t_2$, for which the maximum appears for the smaller momentum. We suspect that this is due to the interference effects.

The inspection of Fig.~\ref{fpolafasymlinosclow0bK}(c) also indicates the existence of points at which the distribution is very small. In order to 
check if these are the zeros of $f^{(+)}_+(\bm{p},-)$, we have to analyze the phase of the amplitude $A^{(+)}_+(\bm{p},-)$. The reason being
that, at these points, the phase is not defined whereas in their vicinity it acquires all possible values (see, e.g., the original discussion of 
this problem presented in Ref.~\cite{dirac1931quantised}). However, the total phase of $A^{(+)}_+(\bm{p},-)$ makes such analysis difficult, as it is 
the sum of the regular part (called the dynamical phase), that is continuous for all momenta, and the singular part (called the geometrical 
or Berry phase), that is not defined for zeros of the amplitude. Unfortunately, the regular phase can change rapidly with momentum, which makes it 
difficult to trace the singular points from the phase image. The solution to this problem could be to extract from the amplitude the regular
phase factor, so that the remaining phase is a slowly changing function of momentum beyond singularities. Therefore, we consider the modified amplitude 
$\ee^{\ii\phi(\bm{p})}A^{(+)}_+(\bm{p},-)$, where $\phi(\bm{p})$ is a regular function of ${\bm p}$. \textit{A priori} the function $\phi(\bm{p})$ is not known and, in fact, 
it is not uniquely defined. One can only guess its approximate form by analyzing the phase image. In doing so, we have found that its possible form, 
suitable for our numerical investigations, could be
\begin{equation}\label{num9}
\phi(\bm{p})=\eta c\sqrt{\bm{p}^2+(\me c)^2}(t_\mathrm{f}-t_\mathrm{i}),\, \eta=1.8,
\end{equation} 
where $t_\mathrm{i}$ and $t_\mathrm{f}$ are the initial and final times in numerical integration of the Dirac equation. For our choice of pulses 
and for $\sigma=0.8$ these are: $t_\mathrm{i}=-20t_C$ and $t_\mathrm{f}=200t_C$. Of course, by changing a little the factor $\eta$ one can even 
significantly modify the phase image, but such that the positions of the singular points are not changed. We have applied this form of the phase factor in all numerical illustrations presented in this paper.

In Fig.~\ref{fpolafasymlinosclow0bK}(d), we demonstrate the phase image of the amplitude, calculated in this way. Indeed, we observe singular points
at which the phase is not defined and which exactly relate to zeros of the momentum distribution [Fig.~\ref{fpolafasymlinosclow0bK}(c)]. They are located for positive values 
of $p_x$ and are symmetrically distributed with respect to the electric field polarization vector. The winding numbers (called also topological charges),
representing number of times when the phase changes by $2\pi$ counterclockwise around a singularity,
are opposite for symmetrically located singular points, which is the indication that they belong to the same circular {\it vortex line} (see, e.g., 
Ref.~\cite{bechler2023schwinger}, and a general discussion presented in Ref.~\cite{PhysRevA.61.032110}).

\begin{figure}
\includegraphics[width=6cm]{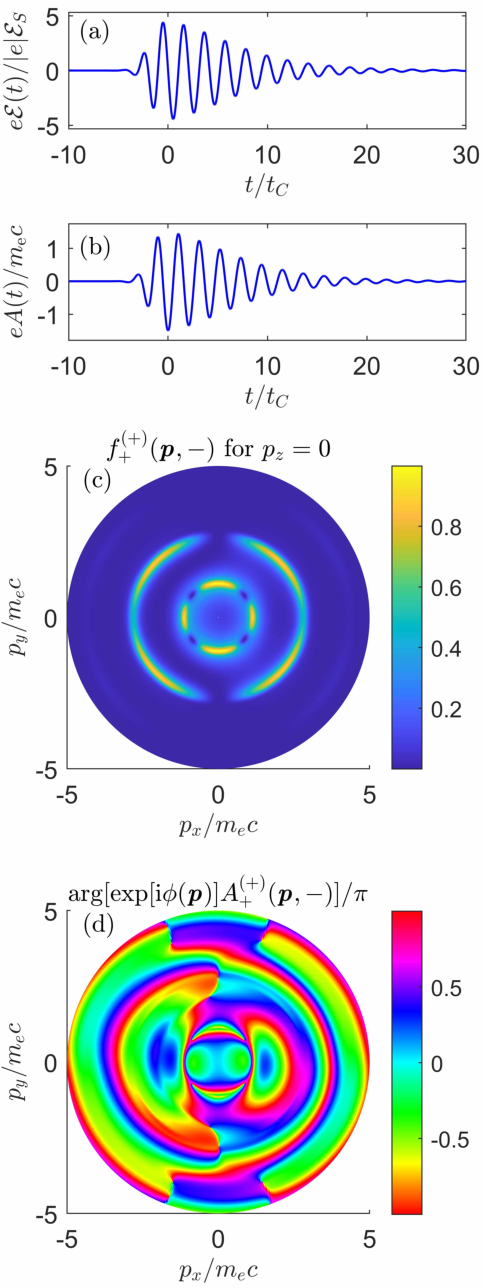}
\caption{
The same is in Fig.~\ref{fpolafasymlinosclow0bK} but for a larger electric field frequency, $\omega=3\me c^2$.
}
\label{fpolafasymlinoschigh0bK}
\end{figure}

The picture presented above changes significantly for electric field frequencies larger than the pair-creation threshold, equal to $2\me c^2$. 
In Fig.~\ref{fpolafasymlinoschigh0bK}, we analyze the electron momentum distribution and the phase image of the corresponding amplitude for 
$\omega=3\me c^2$. The momentum distribution shown in Fig.~\ref{fpolafasymlinoschigh0bK}(c) consists of two rings (in fact, also the third ring 
is nearly visible). A tempting interpretation of such a distribution may be based on ideas taken from atomic physics where the ionization process 
is perceived as the result of multiphoton absorption. Pair creation is formally similar to ionization. Therefore, the appearance of rings in the 
momentum distribution can be interpreted as the result of the absorption of one or more `photons' (or energy quanta that are multiples of $\omega$) 
from the electric field. Indeed, for the $n$-th ring we have $2c\sqrt{p^2+(\me c)^2}=n\omega$, from which we get the radius of the first ring equal 
to $p_1=\sqrt{1.25}\me c\approx 1.1\me c$, and the radius of the second one, $p_2=2\sqrt{2}\me c\approx 2.8\me c$. These values suit very well to our data.

The analysis of the phase image presented in Fig.~\ref{fpolafasymlinoschigh0bK}(d) indicates that the first ring there exhibits four zero-points at which the phase is singular. As the topological charges of the points located symmetrically with respect to the polarization vector are opposite, therefore, similarly to the low-frequency case we can presume that these points belong to the same circular vortex line. The same conclusion can be drawn with respect to the two singular points embedded in the second ring for nearly vanishing $p_x$ momenta.

\begin{figure}
\includegraphics[width=8.5cm]{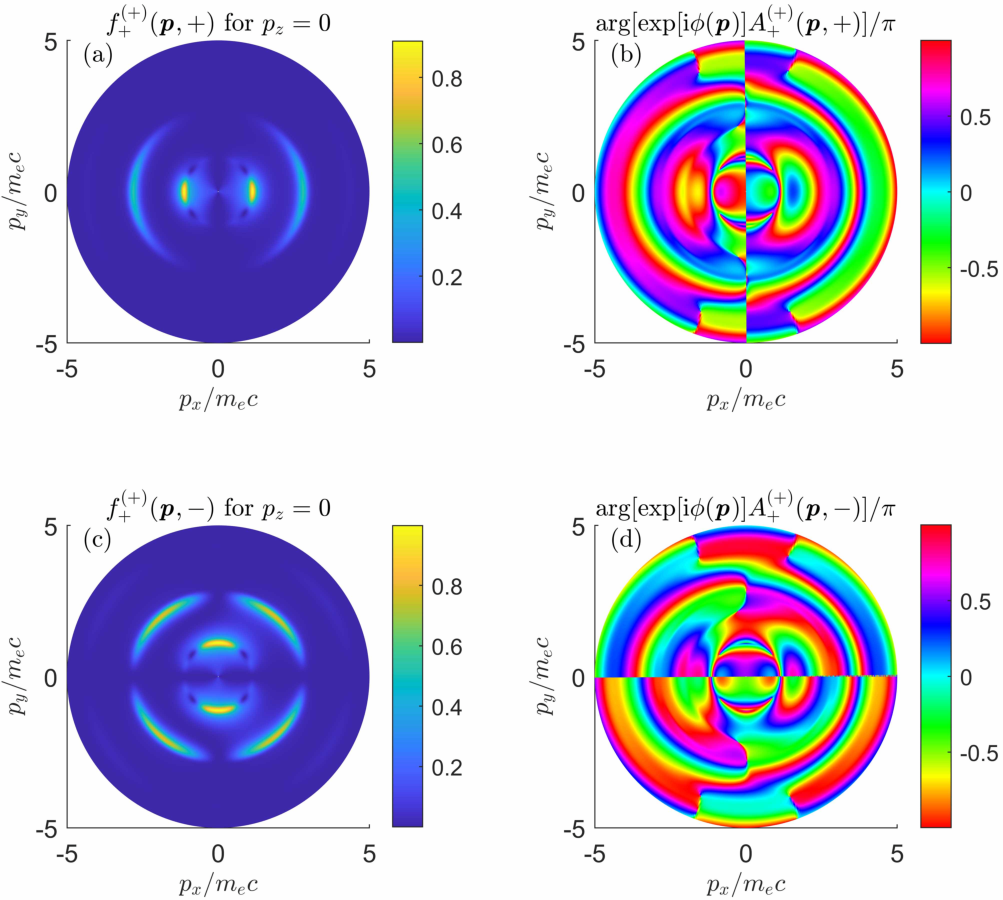}
\caption{
The same as in Fig.~\ref{fpolafasymlinoschigh0bK} but for the helicity-resolved distributions. Note that in this case, $f^{(+)}_+(\bm{p},+)=f^{(+)}_-(\bm{p},-)$ and $f^{(+)}_+(\bm{p},-)=f^{(+)}_-(\bm{p},+)$. The momentum distributions are given in relativistic units.
}
\label{fpolafasymlinoschighbK}
\end{figure}

A more detailed insight into the vortex structures exhibited by the probability amplitude of pair creation can be obtained from the helicity-resolved 
distributions, presented in Fig.~\ref{fpolafasymlinoschighbK} along with their corresponding phase images. As one can see, the momentum distributions 
depend significantly on the mutual helicities of generated positrons and electrons. For the same helicities, similarly to the low-frequency 
case and at least for the inner ring, pair creation occurs mainly along the electric field [cf., Fig.~\ref{fpolafasymlinoschighbK}(a)]. However,
the pairs of particles with the opposite helicities are mostly generated in the direction perpendicular to the electric field [cf., Fig.~\ref{fpolafasymlinoschighbK}(c), again at least for the inner ring]. In both cases, apart from the singular points corresponding to the vortex lines, we also observe lines along which the distributions vanish and the phase jumps by $\pi$ [the vertical line in Fig.~\ref{fpolafasymlinoschighbK}(b) and the horizontal one in Fig.~\ref{fpolafasymlinoschighbK}(d)]. However, they should not be confused with vortex lines, as they are part of the so-called nodal surfaces. We can verify this by analyzing, for instance, the distributions in the plane $(p_y,p_z)$ for $p_x=0$.

Note that similar vortex and nodal structures in momentum distributions of photoelectrons are observed in multiphoton ionization~\cite{Larionov2018Perturbation}
or detachment~\cite{PhysRevA.102.043102,PhysRevA.102.043117,PhysRevA.104.033111,PhysRevA.104.043116,majczak2022vorticesphotodetachment}. Here, we 
reveal another (i.e., apart from multiphoton absorption and tunneling) similarity between those processes and the dynamical Sauter-Schwinger pair production; all of them being threshold-related phenomena.

To summarize this section, devoted to the dynamical Sauter-Schwinger pair creation in linearly polarized electric fields, we have shown that the 
spin or helicity summed up momentum distributions of created particles agree for all three approaches considered in this paper, provided that normalizations,
Eqs.~\eqref{rf3-7b} or~\eqref{rf3-7i}, are applied. Note, however, that the extra normalization does not follow from the definition of the QED amplitude 
for the pair creation [Eq.~\eqref{rf1}] and have been introduced \textit{only} to compare results with other approaches. For weak electric fields, the 
normalization constants are very close to one; thus, the differences between these approaches are rather marginal. However, for sufficiently strong fields 
they could be significant. Independently of this, the advantage of using the Feynman or anti-Feynman boundary 
conditions while solving the Dirac equation is such that it allows us to investigate the spin- or helicity-resolved momentum distributions. As we have 
also demonstrated, it gives us access to the phase properties of the conditional probability amplitudes of produced particles, 
which allows to study vortex structures in Sauter-Schwinger process. Although the last topic can be also investigated by applying the spinorial 
approach~\cite{bechler2023schwinger}, we need to stress that it is applicable only for linearly polarized electric fields, for a particularly chosen spin quantization 
axis, and not for the helicity states. In this respect, the current method is more flexible which we shall demonstrate below for pair production
by circularly polarized electric fields.

\begin{figure}
\includegraphics[width=8.5cm]{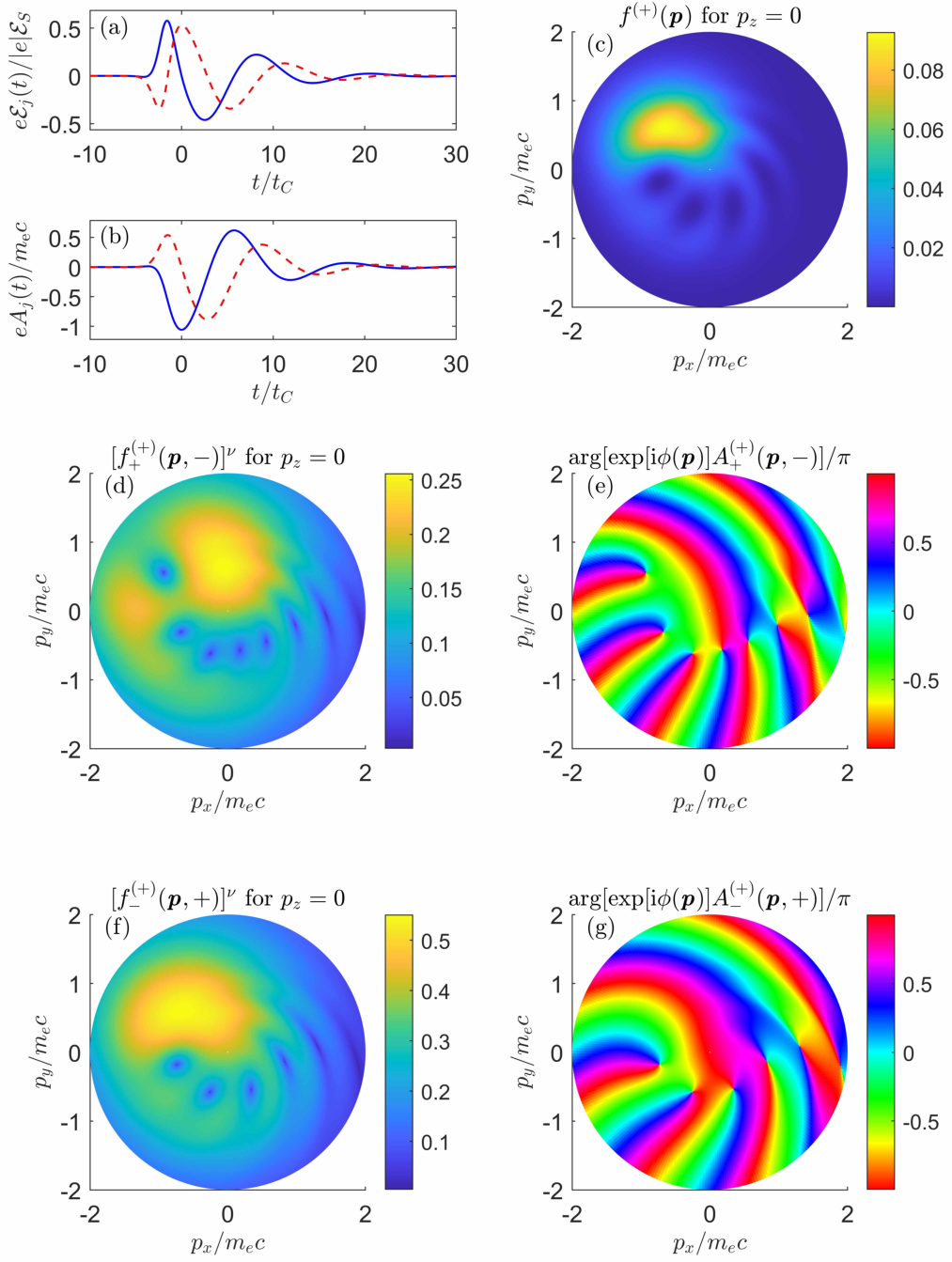}
\caption{
In panels (a) and (b) the components (solid line for $j=1$ and dashed line for $j=2$) of the electric field, $\bm{\mathcal{E}}(t)$, and the vector 
potential, $\bm{A}(t)$, (both multiplied by the electron charge $e=-|e|$) are presented for $\bm{\varepsilon}_1=\bm{e}_x$, 
$\bm{\varepsilon}_2=\bm{e}_y$, $\delta=\pi/4$ and for $\mathcal{E}_0=0.5\mathcal{E}_S$, $\tau_0=t_0=3t_C$, $\sigma=0.8$, $\omega=0.5\me c^2$, 
$\chi=0$ [see, Eq.~\eqref{num11} and Eq.~\eqref{num3} for the definition of the envelope $F(t)$]. Panel (c) exhibits the total electron momentum 
distribution $f^{(+)}(\bm{p})$. The spin-resolved distributions are presented in panels (d) and (f) [for the visual purpose, raised to the power 
$\nu=1/4$] for the spin quantization axis along the $z$-direction. Note that $f^{(+)}_+(\bm{p},+)=f^{(+)}_-(\bm{p},-)=0$ and all distributions
are given in relativistic units. In panels (e) and (g) the phase images of the corresponding amplitudes are shown [cf., Eq.~\eqref{rf3-7b}]. 
The meaning of the extra phase factor, $\ee^{\ii\phi(\bm{p})}$, is discussed in the text.
}
\label{fpolafasymcirosc4K}
\end{figure}

\subsection{Circular polarization}
\label{sec:CirPol}

In the most general case of an elliptically polarized pulse we define two shape functions,
\begin{align}\label{num10}
A_1(t)=&F(t)\cos(\omega t+\chi)\cos\delta, \nonumber \\
A_2(t)=&F(t)\sin(\omega t+\chi)\sin\delta,
\end{align} 
and the vector potential,
\begin{equation}\label{num11}
\bm{A}(t)=A_1(t)\bm{\varepsilon}_1+A_2(t)\bm{\varepsilon}_2,
\end{equation} 
where $\bm{\varepsilon}_j$ ($j=1,\, 2$) are two orthogonal real vectors, normalized to unity, whereas the angle $\delta$ controls the pulse ellipticity.
Specifically, for $\delta=0$ we recover the linearly polarized pulse discussed above, and for $\delta=\pi/4$ we obtain the circularly polarized pulse 
investigated in this section. The remaining parameters are the same as before and the pulse envelope $F(t)$ is defined by Eq.~\eqref{num3}. 
Then, the electric field is equal to $\bm{\mathcal{E}}(t)=-\dot{\bm{A}}(t)$.

The nonvanishing spin-resolved distributions together with the spin summed up distribution are presented in Fig.~\ref{fpolafasymcirosc4K} for the spin
polarization axis in the $z$-direction. Note that now $f^{(+)}_-(\bm{p},+)\neq f^{(+)}_+(\bm{p},-)$, with the dominant contribution coming from $f^{(+)}_-(\bm{p},+)$. Contrary to the linearly polarized field we observe the singular points of only positive topological charges (i.e., during the traversing around such a point in the counterclockwise direction the phase monotonically increases, as follows from the phase images). As in all other cases considered in this paper the spin (or helicity) summed up momentum distributions agree within the numerical error with the predictions of the DHW-function or bispinorial approaches discussed above. However, contrary to these methods we now have the access to the full information about the spin- or helicity-resolved distributions for arbitrary time-dependent electric field pulses.

\begin{figure}
\includegraphics[width=8.5cm]{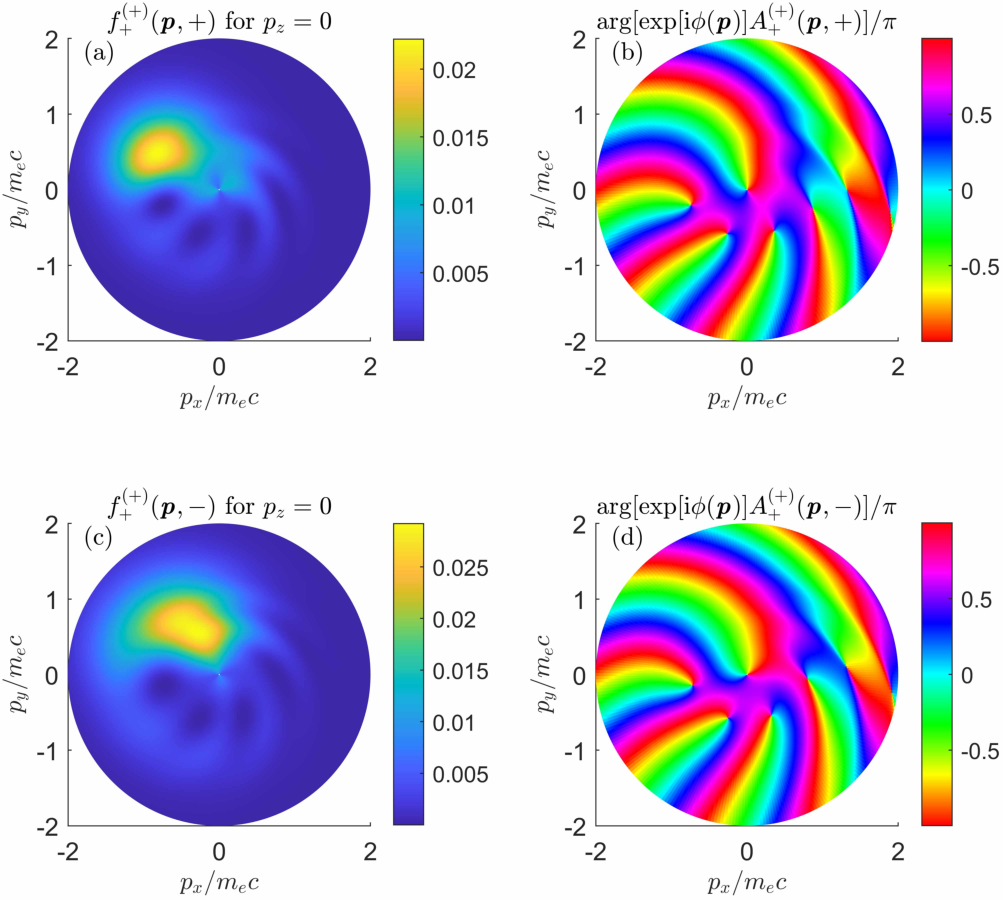}
\caption{
The same as in Fig.~\ref{fpolafasymcirosc4K}, but for the helicity-resolved electron momentum distributions.
In this case, $f^{(+)}_+(\bm{p},+)=f^{(+)}_-(\bm{p},-)$ and $f^{(+)}_+(\bm{p},-)=f^{(+)}_-(\bm{p},+)$. The momentum distributions are given in relativistic units.
}
\label{fpolafasymciroscbsK}
\end{figure}

For completeness of our analysis, in Fig.~\ref{fpolafasymciroscbsK} we present the helicity-resolved electron momentum distributions for the case 
discussed in this section. In contrast to the distributions with a fixed spin quantization axis (see, Fig.~\ref{fpolafasymcirosc4K}), in the phase 
images presented here we see an additional singular point for $\bm{p}=\bm{0}$, even though at this point the distributions are not vanishing. The reason for this peculiarity is that the helicity operator is not uniquely defined for $\bm{p}=\bm{0}$. Hence, we observe an isolated phase-singular point at ${\bm p}={\bm 0}$ in the helicity-resolved momentum distributions, while the vortex points form lines (close or open) in three-dimensional momentum space. In contrast, for a linearly polarized electric field, the singular point at ${\bm p}={\bm 0}$ was masked by the line where the probability amplitude phase jumps by $\pi$ (see, Fig.~\ref{fpolafasymlinoschighbK}).

\section{Conclusions}
\label{sec:Conclusions}

The main motivation of this work was to present an analysis of the dynamical Sauter-Schwinger process based on the standard QED formalism, i.e., 
on the scattering matrix and reduction formulas. We have demonstrated that this formalism provides a way to determine the spin-momentum distributions 
of created $e^-e^+$ pairs and the corresponding probability amplitudes. Knowledge of the amplitudes and their phases allows us to study the interference 
effects occurring in this process, in particular the vortex structures. These distributions result from certain solutions of the Dirac equation that 
satisfy either the Feynman asymptotic conditions (for the distributions of particles, i.e., electrons) or the anti-Feynman ones (for the distributions 
of anti-particles, i.e., positrons). Both are not the initial conditions for the Dirac equation, but rather the boundary ones. However, the numerical 
analysis of the Sauter-Schwinger effect presented here shows that for the situation when the electric field pulse depends only on time and for 
the momentum distributions summed over the electron and positron spins, the approach presented in this paper gives results identical to those obtained 
by methods using initial conditions, such as the DHW function formalism. We need to stress that this agreement is achieved only after additional normalization 
of the Dirac solution in the far past. The normalizing factor appearing here is close to unity for electric field strengths not approaching 
the Schwinger value $\mathcal{E}_S$, as in this case the maximum values of momentum distributions are much smaller than one. However, for stronger fields the results obtained by using the formalism presented here differ from those that follow, for example, from the DHW function, if additional normalization is not applied.

We have shown that the Sauter-Schwinger process can be studied by solving the Dirac equation with the appropriate boundary conditions. 
It seems that the discussed in this paper approach has a significant advantage over other approaches used in this context, as it allows one 
to determine the spin or helicity distributions of probability amplitudes. The issue of creating pairs by space- and time-dependent electric
field pulses remains, however, an open problem. In such situations, the formalism of the DHW function leads to a system of nonlocal equations 
(i.e., containing spatial derivatives of arbitrarily high order), the numerical analysis of which has only begun to be developed 
recently~\cite{hebenstreit2010schwinger,kohlfurst2020magneticpair,brodin2021plasmadynamics,kohlfurst2022collidinglaser}. It turns out that the approach presented
here is equally complex, as it requires to solve the Dirac equation with appropriate boundary conditions, imposed both in the past and the future.


\section*{Acknowledgements}

This work is supported by the National Science Centre (Poland) under Grant No. 2018/31/B/ST2/01251.

\bibliography{sf1biblio}

\begin{thebibliography}{108}%
\makeatletter
\providecommand \@ifxundefined [1]{%
 \@ifx{#1\undefined}
}%
\providecommand \@ifnum [1]{%
 \ifnum #1\expandafter \@firstoftwo
 \else \expandafter \@secondoftwo
 \fi
}%
\providecommand \@ifx [1]{%
 \ifx #1\expandafter \@firstoftwo
 \else \expandafter \@secondoftwo
 \fi
}%
\providecommand \natexlab [1]{#1}%
\providecommand \enquote  [1]{``#1''}%
\providecommand \bibnamefont  [1]{#1}%
\providecommand \bibfnamefont [1]{#1}%
\providecommand \citenamefont [1]{#1}%
\providecommand \href@noop [0]{\@secondoftwo}%
\providecommand \href [0]{\begingroup \@sanitize@url \@href}%
\providecommand \@href[1]{\@@startlink{#1}\@@href}%
\providecommand \@@href[1]{\endgroup#1\@@endlink}%
\providecommand \@sanitize@url [0]{\catcode `\\12\catcode `\$12\catcode
  `\&12\catcode `\#12\catcode `\^12\catcode `\_12\catcode `\%12\relax}%
\providecommand \@@startlink[1]{}%
\providecommand \@@endlink[0]{}%
\providecommand \url  [0]{\begingroup\@sanitize@url \@url }%
\providecommand \@url [1]{\endgroup\@href {#1}{\urlprefix }}%
\providecommand \urlprefix  [0]{URL }%
\providecommand \Eprint [0]{\href }%
\providecommand \doibase [0]{https://doi.org/}%
\providecommand \selectlanguage [0]{\@gobble}%
\providecommand \bibinfo  [0]{\@secondoftwo}%
\providecommand \bibfield  [0]{\@secondoftwo}%
\providecommand \translation [1]{[#1]}%
\providecommand \BibitemOpen [0]{}%
\providecommand \bibitemStop [0]{}%
\providecommand \bibitemNoStop [0]{.\EOS\space}%
\providecommand \EOS [0]{\spacefactor3000\relax}%
\providecommand \BibitemShut  [1]{\csname bibitem#1\endcsname}%
\let\auto@bib@innerbib\@empty
\bibitem [{\citenamefont {Mourou}\ \emph {et~al.}(2006)\citenamefont {Mourou},
  \citenamefont {Tajima},\ and\ \citenamefont {Bulanov}}]{RevModPhys.78.309}%
  \BibitemOpen
  \bibfield  {author} {\bibinfo {author} {\bibfnamefont {G.~A.}\ \bibnamefont
  {Mourou}}, \bibinfo {author} {\bibfnamefont {T.}~\bibnamefont {Tajima}},\
  and\ \bibinfo {author} {\bibfnamefont {S.~V.}\ \bibnamefont {Bulanov}},\
  }\bibfield  {title} {\bibinfo {title} {{Optics in the relativistic regime}},\
  }\href {https://doi.org/10.1103/RevModPhys.78.309} {\bibfield  {journal}
  {\bibinfo  {journal} {Rev. Mod. Phys.}\ }\textbf {\bibinfo {volume} {78}},\
  \bibinfo {pages} {309} (\bibinfo {year} {2006})}\BibitemShut {NoStop}%
\bibitem [{\citenamefont {Rafelski}\ \emph {et~al.}(1978)\citenamefont
  {Rafelski}, \citenamefont {Fulcher},\ and\ \citenamefont
  {Klein}}]{Rafelski1978FermionsAB}%
  \BibitemOpen
  \bibfield  {author} {\bibinfo {author} {\bibfnamefont {J.}~\bibnamefont
  {Rafelski}}, \bibinfo {author} {\bibfnamefont {L.~P.}\ \bibnamefont
  {Fulcher}},\ and\ \bibinfo {author} {\bibfnamefont {A.}~\bibnamefont
  {Klein}},\ }\bibfield  {title} {\bibinfo {title} {{Fermions and bosons
  interacting with arbitrarily strong external fields}},\ }\href
  {https://api.semanticscholar.org/CorpusID:120359985} {\bibfield  {journal}
  {\bibinfo  {journal} {Phys. Rep.}\ }\textbf {\bibinfo {volume} {38}},\
  \bibinfo {pages} {227} (\bibinfo {year} {1978})}\BibitemShut {NoStop}%
\bibitem [{\citenamefont {Greiner}\ \emph {et~al.}(1980)\citenamefont
  {Greiner}, \citenamefont {M\"uller},\ and\ \citenamefont
  {Rafelski}}]{Greiner1985QuantumElectrodynamics}%
  \BibitemOpen
  \bibfield  {author} {\bibinfo {author} {\bibfnamefont {W.}~\bibnamefont
  {Greiner}}, \bibinfo {author} {\bibfnamefont {B.}~\bibnamefont {M\"uller}},\
  and\ \bibinfo {author} {\bibfnamefont {J.}~\bibnamefont {Rafelski}},\
  }\href@noop {} {\emph {\bibinfo {title} {{Quantum Electrodynamics of Strong
  Field}}}}\ (\bibinfo  {publisher} {Springer},\ \bibinfo {address} {Berlin},\
  \bibinfo {year} {1980})\BibitemShut {NoStop}%
\bibitem [{\citenamefont {Treumann}\ \emph {et~al.}(2014)\citenamefont
  {Treumann}, \citenamefont {Baumjohann},\ and\ \citenamefont
  {Balogh}}]{10.3389fphy.2014.00059}%
  \BibitemOpen
  \bibfield  {author} {\bibinfo {author} {\bibfnamefont {R.~A.}\ \bibnamefont
  {Treumann}}, \bibinfo {author} {\bibfnamefont {W.}~\bibnamefont
  {Baumjohann}},\ and\ \bibinfo {author} {\bibfnamefont {A.}~\bibnamefont
  {Balogh}},\ }\bibfield  {title} {\bibinfo {title} {{The strongest magnetic
  fields in the universe: how strong can they become?}},\ }\href
  {https://doi.org/10.3389/fphy.2014.00059} {\bibfield  {journal} {\bibinfo
  {journal} {Front. Phys.}\ }\textbf {\bibinfo {volume} {2}},\ \bibinfo {pages}
  {59} (\bibinfo {year} {2014})}\BibitemShut {NoStop}%
\bibitem [{\citenamefont {Ge}\ \emph {et~al.}(2020)\citenamefont {Ge},
  \citenamefont {Ji}, \citenamefont {Zhang},\ and\ \citenamefont {\textit{et
  al.}}}]{Ge2020AJL}%
  \BibitemOpen
  \bibfield  {author} {\bibinfo {author} {\bibfnamefont {M.~Y.}\ \bibnamefont
  {Ge}}, \bibinfo {author} {\bibfnamefont {L.}~\bibnamefont {Ji}}, \bibinfo
  {author} {\bibfnamefont {S.~N.}\ \bibnamefont {Zhang}},\ and\ \bibinfo
  {author} {\bibnamefont {\textit{et al.}}},\ }\bibfield  {title} {\bibinfo
  {title} {{Insight-HXMT Firm Detection of the Highest-energy Fundamental
  Cyclotron Resonance Scattering Feature in the Spectrum of GRO J1008-57}},\
  }\href {https://doi.org/10.3847/2041-8213/abac05} {\bibfield  {journal}
  {\bibinfo  {journal} {Astrophys. J. Lett.}\ }\textbf {\bibinfo {volume}
  {899}},\ \bibinfo {pages} {L19} (\bibinfo {year} {2020})}\BibitemShut
  {NoStop}%
\bibitem [{\citenamefont {Krausz}\ and\ \citenamefont
  {Ivanov}(2009)}]{RevModPhys.81.163}%
  \BibitemOpen
  \bibfield  {author} {\bibinfo {author} {\bibfnamefont {F.}~\bibnamefont
  {Krausz}}\ and\ \bibinfo {author} {\bibfnamefont {M.}~\bibnamefont
  {Ivanov}},\ }\bibfield  {title} {\bibinfo {title} {{Attosecond physics}},\
  }\href {https://doi.org/10.1103/RevModPhys.81.163} {\bibfield  {journal}
  {\bibinfo  {journal} {Rev. Mod. Phys.}\ }\textbf {\bibinfo {volume} {81}},\
  \bibinfo {pages} {163} (\bibinfo {year} {2009})}\BibitemShut {NoStop}%
\bibitem [{\citenamefont {Krajewska}\ \emph {et~al.}(2014)\citenamefont
  {Krajewska}, \citenamefont {Twardy},\ and\ \citenamefont
  {Kami\ifmmode~\acute{n}\else \'{n}\fi{}ski}}]{PhysRevA.89.052123}%
  \BibitemOpen
  \bibfield  {author} {\bibinfo {author} {\bibfnamefont {K.}~\bibnamefont
  {Krajewska}}, \bibinfo {author} {\bibfnamefont {M.}~\bibnamefont {Twardy}},\
  and\ \bibinfo {author} {\bibfnamefont {J.~Z.}\ \bibnamefont
  {Kami\ifmmode~\acute{n}\else \'{n}\fi{}ski}},\ }\bibfield  {title} {\bibinfo
  {title} {{Global phase and frequency comb structures in nonlinear Compton and
  Thomson scattering}},\ }\href {https://doi.org/10.1103/PhysRevA.89.052123}
  {\bibfield  {journal} {\bibinfo  {journal} {Phys. Rev. A}\ }\textbf {\bibinfo
  {volume} {89}},\ \bibinfo {pages} {052123} (\bibinfo {year}
  {2014})}\BibitemShut {NoStop}%
\bibitem [{\citenamefont {Esarey}\ \emph {et~al.}(2009)\citenamefont {Esarey},
  \citenamefont {Schroeder},\ and\ \citenamefont
  {Leemans}}]{RevModPhys.81.1229}%
  \BibitemOpen
  \bibfield  {author} {\bibinfo {author} {\bibfnamefont {E.}~\bibnamefont
  {Esarey}}, \bibinfo {author} {\bibfnamefont {C.~B.}\ \bibnamefont
  {Schroeder}},\ and\ \bibinfo {author} {\bibfnamefont {W.~P.}\ \bibnamefont
  {Leemans}},\ }\bibfield  {title} {\bibinfo {title} {{Physics of laser-driven
  plasma-based electron accelerators}},\ }\href
  {https://doi.org/10.1103/RevModPhys.81.1229} {\bibfield  {journal} {\bibinfo
  {journal} {Rev. Mod. Phys.}\ }\textbf {\bibinfo {volume} {81}},\ \bibinfo
  {pages} {1229} (\bibinfo {year} {2009})}\BibitemShut {NoStop}%
\bibitem [{\citenamefont {Macchi}\ \emph {et~al.}(2013)\citenamefont {Macchi},
  \citenamefont {Borghesi},\ and\ \citenamefont {Passoni}}]{RevModPhys.85.751}%
  \BibitemOpen
  \bibfield  {author} {\bibinfo {author} {\bibfnamefont {A.}~\bibnamefont
  {Macchi}}, \bibinfo {author} {\bibfnamefont {M.}~\bibnamefont {Borghesi}},\
  and\ \bibinfo {author} {\bibfnamefont {M.}~\bibnamefont {Passoni}},\
  }\bibfield  {title} {\bibinfo {title} {{Ion acceleration by superintense
  laser-plasma interaction}},\ }\href
  {https://doi.org/10.1103/RevModPhys.85.751} {\bibfield  {journal} {\bibinfo
  {journal} {Rev. Mod. Phys.}\ }\textbf {\bibinfo {volume} {85}},\ \bibinfo
  {pages} {751} (\bibinfo {year} {2013})}\BibitemShut {NoStop}%
\bibitem [{\citenamefont {Pellegrini}\ \emph {et~al.}(2016)\citenamefont
  {Pellegrini}, \citenamefont {Marinelli},\ and\ \citenamefont
  {Reiche}}]{RevModPhys.88.015006}%
  \BibitemOpen
  \bibfield  {author} {\bibinfo {author} {\bibfnamefont {C.}~\bibnamefont
  {Pellegrini}}, \bibinfo {author} {\bibfnamefont {A.}~\bibnamefont
  {Marinelli}},\ and\ \bibinfo {author} {\bibfnamefont {S.}~\bibnamefont
  {Reiche}},\ }\bibfield  {title} {\bibinfo {title} {{The physics of x-ray
  free-electron lasers}},\ }\href
  {https://doi.org/10.1103/RevModPhys.88.015006} {\bibfield  {journal}
  {\bibinfo  {journal} {Rev. Mod. Phys.}\ }\textbf {\bibinfo {volume} {88}},\
  \bibinfo {pages} {015006} (\bibinfo {year} {2016})}\BibitemShut {NoStop}%
\bibitem [{\citenamefont {Fradkin}\ \emph {et~al.}(1991)\citenamefont
  {Fradkin}, \citenamefont {Gitman},\ and\ \citenamefont
  {Shvartsman}}]{fradkin1991vacuumquantum}%
  \BibitemOpen
  \bibfield  {author} {\bibinfo {author} {\bibfnamefont {E.}~\bibnamefont
  {Fradkin}}, \bibinfo {author} {\bibfnamefont {D.}~\bibnamefont {Gitman}},\
  and\ \bibinfo {author} {\bibfnamefont {S.}~\bibnamefont {Shvartsman}},\
  }\href@noop {} {\emph {\bibinfo {title} {{Quantum Electrodynamics with
  Unstable Vacuum}}}}\ (\bibinfo  {publisher} {Springer},\ \bibinfo {address}
  {Berlin},\ \bibinfo {year} {1991})\BibitemShut {NoStop}%
\bibitem [{\citenamefont {Ritus}(1985)}]{RitusReview85}%
  \BibitemOpen
  \bibfield  {author} {\bibinfo {author} {\bibfnamefont {V.}~\bibnamefont
  {Ritus}},\ }\bibfield  {title} {\bibinfo {title} {{Quantum effects of the
  interaction of elementary particles with an intense electromagnetic field}},\
  }\href {https://doi.org/10.1007/BF01120220} {\bibfield  {journal} {\bibinfo
  {journal} {J. Sov. Laser Res.}\ }\textbf {\bibinfo {volume} {6}},\ \bibinfo
  {pages} {497} (\bibinfo {year} {1985})}\BibitemShut {NoStop}%
\bibitem [{\citenamefont {Baur}\ \emph {et~al.}(2007)\citenamefont {Baur},
  \citenamefont {Hencken},\ and\ \citenamefont {Trautmann}}]{BAUR20071}%
  \BibitemOpen
  \bibfield  {author} {\bibinfo {author} {\bibfnamefont {G.}~\bibnamefont
  {Baur}}, \bibinfo {author} {\bibfnamefont {K.}~\bibnamefont {Hencken}},\ and\
  \bibinfo {author} {\bibfnamefont {D.}~\bibnamefont {Trautmann}},\ }\bibfield
  {title} {\bibinfo {title} {{Electron–positron pair production in
  ultrarelativistic heavy ion collisions}},\ }\href
  {https://doi.org/https://doi.org/10.1016/j.physrep.2007.09.002} {\bibfield
  {journal} {\bibinfo  {journal} {Phys. Rep.}\ }\textbf {\bibinfo {volume}
  {453}},\ \bibinfo {pages} {1} (\bibinfo {year} {2007})}\BibitemShut {NoStop}%
\bibitem [{\citenamefont {Dulaev}\ \emph {et~al.}(2024)\citenamefont {Dulaev},
  \citenamefont {Telnov}, \citenamefont {Shabaev}, \citenamefont {Kozhedub},
  \citenamefont {Maltsev}, \citenamefont {Popov},\ and\ \citenamefont
  {Tupitsyn}}]{PhysRevD.109.036008}%
  \BibitemOpen
  \bibfield  {author} {\bibinfo {author} {\bibfnamefont {N.~K.}\ \bibnamefont
  {Dulaev}}, \bibinfo {author} {\bibfnamefont {D.~A.}\ \bibnamefont {Telnov}},
  \bibinfo {author} {\bibfnamefont {V.~M.}\ \bibnamefont {Shabaev}}, \bibinfo
  {author} {\bibfnamefont {Y.~S.}\ \bibnamefont {Kozhedub}}, \bibinfo {author}
  {\bibfnamefont {I.~A.}\ \bibnamefont {Maltsev}}, \bibinfo {author}
  {\bibfnamefont {R.~V.}\ \bibnamefont {Popov}},\ and\ \bibinfo {author}
  {\bibfnamefont {I.~I.}\ \bibnamefont {Tupitsyn}},\ }\bibfield  {title}
  {\bibinfo {title} {{Angular and energy distributions of positrons created in
  subcritical and supercritical slow collisions of heavy nuclei}},\ }\href
  {https://doi.org/10.1103/PhysRevD.109.036008} {\bibfield  {journal} {\bibinfo
   {journal} {Phys. Rev. D}\ }\textbf {\bibinfo {volume} {109}},\ \bibinfo
  {pages} {036008} (\bibinfo {year} {2024})}\BibitemShut {NoStop}%
\bibitem [{\citenamefont {Ehlotzky}\ \emph {et~al.}(2009)\citenamefont
  {Ehlotzky}, \citenamefont {Krajewska},\ and\ \citenamefont
  {Kamiński}}]{ehlotzky2009fundqed}%
  \BibitemOpen
  \bibfield  {author} {\bibinfo {author} {\bibfnamefont {F.}~\bibnamefont
  {Ehlotzky}}, \bibinfo {author} {\bibfnamefont {K.}~\bibnamefont
  {Krajewska}},\ and\ \bibinfo {author} {\bibfnamefont {J.~Z.}\ \bibnamefont
  {Kamiński}},\ }\bibfield  {title} {\bibinfo {title} {{Fundamental processes
  of quantum electrodynamics in laser fields of relativistic power}},\ }\href
  {https://doi.org/10.1088/0034-4885/72/4/046401} {\bibfield  {journal}
  {\bibinfo  {journal} {Rep. Prog. Phys.}\ }\textbf {\bibinfo {volume} {72}},\
  \bibinfo {pages} {046401} (\bibinfo {year} {2009})}\BibitemShut {NoStop}%
\bibitem [{\citenamefont {Ruffini}\ \emph {et~al.}(2010)\citenamefont
  {Ruffini}, \citenamefont {Vereshchagin},\ and\ \citenamefont
  {Xue}}]{ruffini2010pairastro}%
  \BibitemOpen
  \bibfield  {author} {\bibinfo {author} {\bibfnamefont {R.}~\bibnamefont
  {Ruffini}}, \bibinfo {author} {\bibfnamefont {G.}~\bibnamefont
  {Vereshchagin}},\ and\ \bibinfo {author} {\bibfnamefont {S.-S.}\ \bibnamefont
  {Xue}},\ }\bibfield  {title} {\bibinfo {title} {{Electron–positron pairs in
  physics and astrophysics: From heavy nuclei to black holes}},\ }\href
  {https://doi.org/https://doi.org/10.1016/j.physrep.2009.10.004} {\bibfield
  {journal} {\bibinfo  {journal} {Phys. Rep.}\ }\textbf {\bibinfo {volume}
  {487}},\ \bibinfo {pages} {1} (\bibinfo {year} {2010})}\BibitemShut {NoStop}%
\bibitem [{\citenamefont {Di~Piazza}\ \emph {et~al.}(2012)\citenamefont
  {Di~Piazza}, \citenamefont {M\"uller}, \citenamefont {Hatsagortsyan},\ and\
  \citenamefont {Keitel}}]{RevModPhys.84.1177}%
  \BibitemOpen
  \bibfield  {author} {\bibinfo {author} {\bibfnamefont {A.}~\bibnamefont
  {Di~Piazza}}, \bibinfo {author} {\bibfnamefont {C.}~\bibnamefont {M\"uller}},
  \bibinfo {author} {\bibfnamefont {K.~Z.}\ \bibnamefont {Hatsagortsyan}},\
  and\ \bibinfo {author} {\bibfnamefont {C.~H.}\ \bibnamefont {Keitel}},\
  }\bibfield  {title} {\bibinfo {title} {{Extremely high-intensity laser
  interactions with fundamental quantum systems}},\ }\href
  {https://doi.org/10.1103/RevModPhys.84.1177} {\bibfield  {journal} {\bibinfo
  {journal} {Rev. Mod. Phys.}\ }\textbf {\bibinfo {volume} {84}},\ \bibinfo
  {pages} {1177} (\bibinfo {year} {2012})}\BibitemShut {NoStop}%
\bibitem [{\citenamefont {Gonoskov}\ \emph {et~al.}(2022)\citenamefont
  {Gonoskov}, \citenamefont {Blackburn}, \citenamefont {Marklund},\ and\
  \citenamefont {Bulanov}}]{RevModPhys.94.045001}%
  \BibitemOpen
  \bibfield  {author} {\bibinfo {author} {\bibfnamefont {A.}~\bibnamefont
  {Gonoskov}}, \bibinfo {author} {\bibfnamefont {T.~G.}\ \bibnamefont
  {Blackburn}}, \bibinfo {author} {\bibfnamefont {M.}~\bibnamefont
  {Marklund}},\ and\ \bibinfo {author} {\bibfnamefont {S.~S.}\ \bibnamefont
  {Bulanov}},\ }\bibfield  {title} {\bibinfo {title} {{Charged particle motion
  and radiation in strong electromagnetic fields}},\ }\href
  {https://doi.org/10.1103/RevModPhys.94.045001} {\bibfield  {journal}
  {\bibinfo  {journal} {Rev. Mod. Phys.}\ }\textbf {\bibinfo {volume} {94}},\
  \bibinfo {pages} {045001} (\bibinfo {year} {2022})}\BibitemShut {NoStop}%
\bibitem [{\citenamefont {Sun}\ \emph {et~al.}(2022)\citenamefont {Sun},
  \citenamefont {Zhao}, \citenamefont {Xue}, \citenamefont {Lu}, \citenamefont
  {Ji}, \citenamefont {Wan}, \citenamefont {Wang}, \citenamefont {Salamin},\
  and\ \citenamefont {Li}}]{sun2022pair}%
  \BibitemOpen
  \bibfield  {author} {\bibinfo {author} {\bibfnamefont {T.}~\bibnamefont
  {Sun}}, \bibinfo {author} {\bibfnamefont {Q.}~\bibnamefont {Zhao}}, \bibinfo
  {author} {\bibfnamefont {K.}~\bibnamefont {Xue}}, \bibinfo {author}
  {\bibfnamefont {Z.-W.}\ \bibnamefont {Lu}}, \bibinfo {author} {\bibfnamefont
  {L.-L.}\ \bibnamefont {Ji}}, \bibinfo {author} {\bibfnamefont
  {F.}~\bibnamefont {Wan}}, \bibinfo {author} {\bibfnamefont {Y.}~\bibnamefont
  {Wang}}, \bibinfo {author} {\bibfnamefont {Y.~I.}\ \bibnamefont {Salamin}},\
  and\ \bibinfo {author} {\bibfnamefont {J.-X.}\ \bibnamefont {Li}},\
  }\bibfield  {title} {\bibinfo {title} {{Production of polarized particle
  beams via ultraintense laser pulses}},\ }\href@noop {} {\bibfield  {journal}
  {\bibinfo  {journal} {Rev. Mod. Plasma Phys.}\ }\textbf {\bibinfo {volume}
  {6}},\ \bibinfo {pages} {38} (\bibinfo {year} {2022})}\BibitemShut {NoStop}%
\bibitem [{\citenamefont {Fedotov}\ \emph {et~al.}(2023)\citenamefont
  {Fedotov}, \citenamefont {Ilderton}, \citenamefont {Karbstein}, \citenamefont
  {King}, \citenamefont {Seipt}, \citenamefont {Taya},\ and\ \citenamefont
  {Torgrimsson}}]{fedetov2023qed}%
  \BibitemOpen
  \bibfield  {author} {\bibinfo {author} {\bibfnamefont {A.}~\bibnamefont
  {Fedotov}}, \bibinfo {author} {\bibfnamefont {A.}~\bibnamefont {Ilderton}},
  \bibinfo {author} {\bibfnamefont {F.}~\bibnamefont {Karbstein}}, \bibinfo
  {author} {\bibfnamefont {B.}~\bibnamefont {King}}, \bibinfo {author}
  {\bibfnamefont {D.}~\bibnamefont {Seipt}}, \bibinfo {author} {\bibfnamefont
  {H.}~\bibnamefont {Taya}},\ and\ \bibinfo {author} {\bibfnamefont
  {G.}~\bibnamefont {Torgrimsson}},\ }\bibfield  {title} {\bibinfo {title}
  {{Advances in QED with intense background fields}},\ }\href
  {https://doi.org/https://doi.org/10.1016/j.physrep.2023.01.003} {\bibfield
  {journal} {\bibinfo  {journal} {Phys. Rep.}\ }\textbf {\bibinfo {volume}
  {1010}},\ \bibinfo {pages} {1} (\bibinfo {year} {2023})}\BibitemShut
  {NoStop}%
\bibitem [{\citenamefont {Popruzhenko}\ and\ \citenamefont
  {Fedotov}(2023)}]{popruzhenko2023dynamics}%
  \BibitemOpen
  \bibfield  {author} {\bibinfo {author} {\bibfnamefont {S.~V.}\ \bibnamefont
  {Popruzhenko}}\ and\ \bibinfo {author} {\bibfnamefont {A.~M.}\ \bibnamefont
  {Fedotov}},\ }\bibfield  {title} {\bibinfo {title} {{Dynamics and radiation
  of charged particles in ultra-intense laser fields}},\ }\href@noop {}
  {\bibfield  {journal} {\bibinfo  {journal} {Phys.-Usp.}\ }\textbf {\bibinfo
  {volume} {66}},\ \bibinfo {pages} {460} (\bibinfo {year} {2023})}\BibitemShut
  {NoStop}%
\bibitem [{\citenamefont {Breit}\ and\ \citenamefont
  {Wheeler}(1934)}]{breit1934collision}%
  \BibitemOpen
  \bibfield  {author} {\bibinfo {author} {\bibfnamefont {G.}~\bibnamefont
  {Breit}}\ and\ \bibinfo {author} {\bibfnamefont {J.~A.}\ \bibnamefont
  {Wheeler}},\ }\bibfield  {title} {\bibinfo {title} {{Collision of two light
  quanta}},\ }\href@noop {} {\bibfield  {journal} {\bibinfo  {journal} {Phys.
  Rev.}\ }\textbf {\bibinfo {volume} {46}},\ \bibinfo {pages} {1087} (\bibinfo
  {year} {1934})}\BibitemShut {NoStop}%
\bibitem [{\citenamefont {Bula}\ \emph {et~al.}(1996)\citenamefont {Bula},
  \citenamefont {McDonald}, \citenamefont {Prebys}, \citenamefont {Bamber},
  \citenamefont {Boege}, \citenamefont {Kotseroglou}, \citenamefont
  {Melissinos}, \citenamefont {Meyerhofer}, \citenamefont {Ragg}, \citenamefont
  {Burke}, \citenamefont {Field}, \citenamefont {Horton-Smith}, \citenamefont
  {Odian}, \citenamefont {Spencer}, \citenamefont {Walz}, \citenamefont
  {Berridge}, \citenamefont {Bugg}, \citenamefont {Shmakov},\ and\
  \citenamefont {Weidemann}}]{PhysRevLett.76.3116}%
  \BibitemOpen
  \bibfield  {author} {\bibinfo {author} {\bibfnamefont {C.}~\bibnamefont
  {Bula}}, \bibinfo {author} {\bibfnamefont {K.~T.}\ \bibnamefont {McDonald}},
  \bibinfo {author} {\bibfnamefont {E.~J.}\ \bibnamefont {Prebys}}, \bibinfo
  {author} {\bibfnamefont {C.}~\bibnamefont {Bamber}}, \bibinfo {author}
  {\bibfnamefont {S.}~\bibnamefont {Boege}}, \bibinfo {author} {\bibfnamefont
  {T.}~\bibnamefont {Kotseroglou}}, \bibinfo {author} {\bibfnamefont {A.~C.}\
  \bibnamefont {Melissinos}}, \bibinfo {author} {\bibfnamefont {D.~D.}\
  \bibnamefont {Meyerhofer}}, \bibinfo {author} {\bibfnamefont
  {W.}~\bibnamefont {Ragg}}, \bibinfo {author} {\bibfnamefont {D.~L.}\
  \bibnamefont {Burke}}, \bibinfo {author} {\bibfnamefont {R.~C.}\ \bibnamefont
  {Field}}, \bibinfo {author} {\bibfnamefont {G.}~\bibnamefont {Horton-Smith}},
  \bibinfo {author} {\bibfnamefont {A.~C.}\ \bibnamefont {Odian}}, \bibinfo
  {author} {\bibfnamefont {J.~E.}\ \bibnamefont {Spencer}}, \bibinfo {author}
  {\bibfnamefont {D.}~\bibnamefont {Walz}}, \bibinfo {author} {\bibfnamefont
  {S.~C.}\ \bibnamefont {Berridge}}, \bibinfo {author} {\bibfnamefont {W.~M.}\
  \bibnamefont {Bugg}}, \bibinfo {author} {\bibfnamefont {K.}~\bibnamefont
  {Shmakov}},\ and\ \bibinfo {author} {\bibfnamefont {A.~W.}\ \bibnamefont
  {Weidemann}},\ }\bibfield  {title} {\bibinfo {title} {{Observation of
  Nonlinear Effects in Compton Scattering}},\ }\href
  {https://doi.org/10.1103/PhysRevLett.76.3116} {\bibfield  {journal} {\bibinfo
   {journal} {Phys. Rev. Lett.}\ }\textbf {\bibinfo {volume} {76}},\ \bibinfo
  {pages} {3116} (\bibinfo {year} {1996})}\BibitemShut {NoStop}%
\bibitem [{\citenamefont {Burke}\ \emph {et~al.}(1997)\citenamefont {Burke},
  \citenamefont {Field}, \citenamefont {Horton-Smith}, \citenamefont {Spencer},
  \citenamefont {Walz}, \citenamefont {Berridge}, \citenamefont {Bugg},
  \citenamefont {Shmakov}, \citenamefont {Weidemann}, \citenamefont {Bula},
  \citenamefont {McDonald}, \citenamefont {Prebys}, \citenamefont {Bamber},
  \citenamefont {Boege}, \citenamefont {Koffas}, \citenamefont {Kotseroglou},
  \citenamefont {Melissinos}, \citenamefont {Meyerhofer}, \citenamefont
  {Reis},\ and\ \citenamefont {Ragg}}]{PhysRevLett.79.1626}%
  \BibitemOpen
  \bibfield  {author} {\bibinfo {author} {\bibfnamefont {D.~L.}\ \bibnamefont
  {Burke}}, \bibinfo {author} {\bibfnamefont {R.~C.}\ \bibnamefont {Field}},
  \bibinfo {author} {\bibfnamefont {G.}~\bibnamefont {Horton-Smith}}, \bibinfo
  {author} {\bibfnamefont {J.~E.}\ \bibnamefont {Spencer}}, \bibinfo {author}
  {\bibfnamefont {D.}~\bibnamefont {Walz}}, \bibinfo {author} {\bibfnamefont
  {S.~C.}\ \bibnamefont {Berridge}}, \bibinfo {author} {\bibfnamefont {W.~M.}\
  \bibnamefont {Bugg}}, \bibinfo {author} {\bibfnamefont {K.}~\bibnamefont
  {Shmakov}}, \bibinfo {author} {\bibfnamefont {A.~W.}\ \bibnamefont
  {Weidemann}}, \bibinfo {author} {\bibfnamefont {C.}~\bibnamefont {Bula}},
  \bibinfo {author} {\bibfnamefont {K.~T.}\ \bibnamefont {McDonald}}, \bibinfo
  {author} {\bibfnamefont {E.~J.}\ \bibnamefont {Prebys}}, \bibinfo {author}
  {\bibfnamefont {C.}~\bibnamefont {Bamber}}, \bibinfo {author} {\bibfnamefont
  {S.~J.}\ \bibnamefont {Boege}}, \bibinfo {author} {\bibfnamefont
  {T.}~\bibnamefont {Koffas}}, \bibinfo {author} {\bibfnamefont
  {T.}~\bibnamefont {Kotseroglou}}, \bibinfo {author} {\bibfnamefont {A.~C.}\
  \bibnamefont {Melissinos}}, \bibinfo {author} {\bibfnamefont {D.~D.}\
  \bibnamefont {Meyerhofer}}, \bibinfo {author} {\bibfnamefont {D.~A.}\
  \bibnamefont {Reis}},\ and\ \bibinfo {author} {\bibfnamefont
  {W.}~\bibnamefont {Ragg}},\ }\bibfield  {title} {\bibinfo {title} {{Positron
  Production in Multiphoton Light-by-Light Scattering}},\ }\href
  {https://doi.org/10.1103/PhysRevLett.79.1626} {\bibfield  {journal} {\bibinfo
   {journal} {Phys. Rev. Lett.}\ }\textbf {\bibinfo {volume} {79}},\ \bibinfo
  {pages} {1626} (\bibinfo {year} {1997})}\BibitemShut {NoStop}%
\bibitem [{\citenamefont {Bamber}\ \emph {et~al.}(1999)\citenamefont {Bamber},
  \citenamefont {Boege}, \citenamefont {Koffas}, \citenamefont {Kotseroglou},
  \citenamefont {Melissinos}, \citenamefont {Meyerhofer}, \citenamefont {Reis},
  \citenamefont {Ragg}, \citenamefont {Bula}, \citenamefont {McDonald},
  \citenamefont {Prebys}, \citenamefont {Burke}, \citenamefont {Field},
  \citenamefont {Horton-Smith}, \citenamefont {Spencer}, \citenamefont {Walz},
  \citenamefont {Berridge}, \citenamefont {Bugg}, \citenamefont {Shmakov},\
  and\ \citenamefont {Weidemann}}]{PhysRevD.60.092004}%
  \BibitemOpen
  \bibfield  {author} {\bibinfo {author} {\bibfnamefont {C.}~\bibnamefont
  {Bamber}}, \bibinfo {author} {\bibfnamefont {S.~J.}\ \bibnamefont {Boege}},
  \bibinfo {author} {\bibfnamefont {T.}~\bibnamefont {Koffas}}, \bibinfo
  {author} {\bibfnamefont {T.}~\bibnamefont {Kotseroglou}}, \bibinfo {author}
  {\bibfnamefont {A.~C.}\ \bibnamefont {Melissinos}}, \bibinfo {author}
  {\bibfnamefont {D.~D.}\ \bibnamefont {Meyerhofer}}, \bibinfo {author}
  {\bibfnamefont {D.~A.}\ \bibnamefont {Reis}}, \bibinfo {author}
  {\bibfnamefont {W.}~\bibnamefont {Ragg}}, \bibinfo {author} {\bibfnamefont
  {C.}~\bibnamefont {Bula}}, \bibinfo {author} {\bibfnamefont {K.~T.}\
  \bibnamefont {McDonald}}, \bibinfo {author} {\bibfnamefont {E.~J.}\
  \bibnamefont {Prebys}}, \bibinfo {author} {\bibfnamefont {D.~L.}\
  \bibnamefont {Burke}}, \bibinfo {author} {\bibfnamefont {R.~C.}\ \bibnamefont
  {Field}}, \bibinfo {author} {\bibfnamefont {G.}~\bibnamefont {Horton-Smith}},
  \bibinfo {author} {\bibfnamefont {J.~E.}\ \bibnamefont {Spencer}}, \bibinfo
  {author} {\bibfnamefont {D.}~\bibnamefont {Walz}}, \bibinfo {author}
  {\bibfnamefont {S.~C.}\ \bibnamefont {Berridge}}, \bibinfo {author}
  {\bibfnamefont {W.~M.}\ \bibnamefont {Bugg}}, \bibinfo {author}
  {\bibfnamefont {K.}~\bibnamefont {Shmakov}},\ and\ \bibinfo {author}
  {\bibfnamefont {A.~W.}\ \bibnamefont {Weidemann}},\ }\bibfield  {title}
  {\bibinfo {title} {{Studies of nonlinear QED in collisions of 46.6 GeV
  electrons with intense laser pulses}},\ }\href
  {https://doi.org/10.1103/PhysRevD.60.092004} {\bibfield  {journal} {\bibinfo
  {journal} {Phys. Rev. D}\ }\textbf {\bibinfo {volume} {60}},\ \bibinfo
  {pages} {092004} (\bibinfo {year} {1999})}\BibitemShut {NoStop}%
\bibitem [{\citenamefont {Neville}\ and\ \citenamefont
  {Rohrlich}(1971)}]{PhysRevD.3.1692}%
  \BibitemOpen
  \bibfield  {author} {\bibinfo {author} {\bibfnamefont {R.~A.}\ \bibnamefont
  {Neville}}\ and\ \bibinfo {author} {\bibfnamefont {F.}~\bibnamefont
  {Rohrlich}},\ }\bibfield  {title} {\bibinfo {title} {{Quantum Electrodynamics
  on Null Planes and Applications to Lasers}},\ }\href
  {https://doi.org/10.1103/PhysRevD.3.1692} {\bibfield  {journal} {\bibinfo
  {journal} {Phys. Rev. D}\ }\textbf {\bibinfo {volume} {3}},\ \bibinfo {pages}
  {1692} (\bibinfo {year} {1971})}\BibitemShut {NoStop}%
\bibitem [{\citenamefont {Boca}\ and\ \citenamefont
  {Florescu}(2009)}]{PhysRevA.80.053403}%
  \BibitemOpen
  \bibfield  {author} {\bibinfo {author} {\bibfnamefont {M.}~\bibnamefont
  {Boca}}\ and\ \bibinfo {author} {\bibfnamefont {V.}~\bibnamefont
  {Florescu}},\ }\bibfield  {title} {\bibinfo {title} {{Nonlinear Compton
  scattering with a laser pulse}},\ }\href
  {https://doi.org/10.1103/PhysRevA.80.053403} {\bibfield  {journal} {\bibinfo
  {journal} {Phys. Rev. A}\ }\textbf {\bibinfo {volume} {80}},\ \bibinfo
  {pages} {053403} (\bibinfo {year} {2009})}\BibitemShut {NoStop}%
\bibitem [{\citenamefont {Seipt}\ and\ \citenamefont
  {K\"ampfer}(2011)}]{PhysRevA.83.022101}%
  \BibitemOpen
  \bibfield  {author} {\bibinfo {author} {\bibfnamefont {D.}~\bibnamefont
  {Seipt}}\ and\ \bibinfo {author} {\bibfnamefont {B.}~\bibnamefont
  {K\"ampfer}},\ }\bibfield  {title} {\bibinfo {title} {{Nonlinear Compton
  scattering of ultrashort intense laser pulses}},\ }\href
  {https://doi.org/10.1103/PhysRevA.83.022101} {\bibfield  {journal} {\bibinfo
  {journal} {Phys. Rev. A}\ }\textbf {\bibinfo {volume} {83}},\ \bibinfo
  {pages} {022101} (\bibinfo {year} {2011})}\BibitemShut {NoStop}%
\bibitem [{\citenamefont {Mackenroth}\ and\ \citenamefont
  {Di~Piazza}(2011)}]{PhysRevA.83.032106}%
  \BibitemOpen
  \bibfield  {author} {\bibinfo {author} {\bibfnamefont {F.}~\bibnamefont
  {Mackenroth}}\ and\ \bibinfo {author} {\bibfnamefont {A.}~\bibnamefont
  {Di~Piazza}},\ }\bibfield  {title} {\bibinfo {title} {{Nonlinear Compton
  scattering in ultrashort laser pulses}},\ }\href
  {https://doi.org/10.1103/PhysRevA.83.032106} {\bibfield  {journal} {\bibinfo
  {journal} {Phys. Rev. A}\ }\textbf {\bibinfo {volume} {83}},\ \bibinfo
  {pages} {032106} (\bibinfo {year} {2011})}\BibitemShut {NoStop}%
\bibitem [{\citenamefont {Krajewska}\ and\ \citenamefont
  {Kami\ifmmode~\acute{n}\else
  \'{n}\fi{}ski}(2012{\natexlab{a}})}]{PhysRevA.85.062102}%
  \BibitemOpen
  \bibfield  {author} {\bibinfo {author} {\bibfnamefont {K.}~\bibnamefont
  {Krajewska}}\ and\ \bibinfo {author} {\bibfnamefont {J.~Z.}\ \bibnamefont
  {Kami\ifmmode~\acute{n}\else \'{n}\fi{}ski}},\ }\bibfield  {title} {\bibinfo
  {title} {{Compton process in intense short laser pulses}},\ }\href
  {https://doi.org/10.1103/PhysRevA.85.062102} {\bibfield  {journal} {\bibinfo
  {journal} {Phys. Rev. A}\ }\textbf {\bibinfo {volume} {85}},\ \bibinfo
  {pages} {062102} (\bibinfo {year} {2012}{\natexlab{a}})}\BibitemShut
  {NoStop}%
\bibitem [{\citenamefont {Heinzl}\ \emph {et~al.}(2010)\citenamefont {Heinzl},
  \citenamefont {Ilderton},\ and\ \citenamefont
  {Marklund}}]{Heinzl2010FiniteSE}%
  \BibitemOpen
  \bibfield  {author} {\bibinfo {author} {\bibfnamefont {T.}~\bibnamefont
  {Heinzl}}, \bibinfo {author} {\bibfnamefont {A.}~\bibnamefont {Ilderton}},\
  and\ \bibinfo {author} {\bibfnamefont {M.}~\bibnamefont {Marklund}},\
  }\bibfield  {title} {\bibinfo {title} {{Finite size effects in stimulated
  laser pair production}},\ }\href
  {https://api.semanticscholar.org/CorpusID:119203153} {\bibfield  {journal}
  {\bibinfo  {journal} {Phys. Lett. B}\ }\textbf {\bibinfo {volume} {692}},\
  \bibinfo {pages} {250} (\bibinfo {year} {2010})}\BibitemShut {NoStop}%
\bibitem [{\citenamefont {Titov}\ \emph {et~al.}(2012)\citenamefont {Titov},
  \citenamefont {Takabe}, \citenamefont {K\"ampfer},\ and\ \citenamefont
  {Hosaka}}]{PhysRevLett.108.240406}%
  \BibitemOpen
  \bibfield  {author} {\bibinfo {author} {\bibfnamefont {A.~I.}\ \bibnamefont
  {Titov}}, \bibinfo {author} {\bibfnamefont {H.}~\bibnamefont {Takabe}},
  \bibinfo {author} {\bibfnamefont {B.}~\bibnamefont {K\"ampfer}},\ and\
  \bibinfo {author} {\bibfnamefont {A.}~\bibnamefont {Hosaka}},\ }\bibfield
  {title} {\bibinfo {title} {{Enhanced Subthreshold
  ${e}^{\mathbf{+}}{e}^{\mathbf{\ensuremath{-}}}$ Production in Short Laser
  Pulses}},\ }\href {https://doi.org/10.1103/PhysRevLett.108.240406} {\bibfield
   {journal} {\bibinfo  {journal} {Phys. Rev. Lett.}\ }\textbf {\bibinfo
  {volume} {108}},\ \bibinfo {pages} {240406} (\bibinfo {year}
  {2012})}\BibitemShut {NoStop}%
\bibitem [{\citenamefont {Krajewska}\ and\ \citenamefont
  {Kami\ifmmode~\acute{n}\else
  \'{n}\fi{}ski}(2012{\natexlab{b}})}]{PhysRevA.86.052104}%
  \BibitemOpen
  \bibfield  {author} {\bibinfo {author} {\bibfnamefont {K.}~\bibnamefont
  {Krajewska}}\ and\ \bibinfo {author} {\bibfnamefont {J.~Z.}\ \bibnamefont
  {Kami\ifmmode~\acute{n}\else \'{n}\fi{}ski}},\ }\bibfield  {title} {\bibinfo
  {title} {{Breit-Wheeler process in intense short laser pulses}},\ }\href
  {https://doi.org/10.1103/PhysRevA.86.052104} {\bibfield  {journal} {\bibinfo
  {journal} {Phys. Rev. A}\ }\textbf {\bibinfo {volume} {86}},\ \bibinfo
  {pages} {052104} (\bibinfo {year} {2012}{\natexlab{b}})}\BibitemShut
  {NoStop}%
\bibitem [{\citenamefont {Meuren}\ \emph {et~al.}(2015)\citenamefont {Meuren},
  \citenamefont {Hatsagortsyan}, \citenamefont {Keitel},\ and\ \citenamefont
  {Di~Piazza}}]{PhysRevD.91.013009}%
  \BibitemOpen
  \bibfield  {author} {\bibinfo {author} {\bibfnamefont {S.}~\bibnamefont
  {Meuren}}, \bibinfo {author} {\bibfnamefont {K.~Z.}\ \bibnamefont
  {Hatsagortsyan}}, \bibinfo {author} {\bibfnamefont {C.~H.}\ \bibnamefont
  {Keitel}},\ and\ \bibinfo {author} {\bibfnamefont {A.}~\bibnamefont
  {Di~Piazza}},\ }\bibfield  {title} {\bibinfo {title} {{Polarization-operator
  approach to pair creation in short laser pulses}},\ }\href
  {https://doi.org/10.1103/PhysRevD.91.013009} {\bibfield  {journal} {\bibinfo
  {journal} {Phys. Rev. D}\ }\textbf {\bibinfo {volume} {91}},\ \bibinfo
  {pages} {013009} (\bibinfo {year} {2015})}\BibitemShut {NoStop}%
\bibitem [{\citenamefont {Meuren}\ \emph {et~al.}(2016)\citenamefont {Meuren},
  \citenamefont {Keitel},\ and\ \citenamefont
  {Di~Piazza}}]{PhysRevD.93.085028}%
  \BibitemOpen
  \bibfield  {author} {\bibinfo {author} {\bibfnamefont {S.}~\bibnamefont
  {Meuren}}, \bibinfo {author} {\bibfnamefont {C.~H.}\ \bibnamefont {Keitel}},\
  and\ \bibinfo {author} {\bibfnamefont {A.}~\bibnamefont {Di~Piazza}},\
  }\bibfield  {title} {\bibinfo {title} {{Semiclassical picture for
  electron-positron photoproduction in strong laser fields}},\ }\href
  {https://doi.org/10.1103/PhysRevD.93.085028} {\bibfield  {journal} {\bibinfo
  {journal} {Phys. Rev. D}\ }\textbf {\bibinfo {volume} {93}},\ \bibinfo
  {pages} {085028} (\bibinfo {year} {2016})}\BibitemShut {NoStop}%
\bibitem [{\citenamefont {Di~Piazza}(2016)}]{PhysRevLett.117.213201}%
  \BibitemOpen
  \bibfield  {author} {\bibinfo {author} {\bibfnamefont {A.}~\bibnamefont
  {Di~Piazza}},\ }\bibfield  {title} {\bibinfo {title} {{Nonlinear
  Breit-Wheeler Pair Production in a Tightly Focused Laser Beam}},\ }\href
  {https://doi.org/10.1103/PhysRevLett.117.213201} {\bibfield  {journal}
  {\bibinfo  {journal} {Phys. Rev. Lett.}\ }\textbf {\bibinfo {volume} {117}},\
  \bibinfo {pages} {213201} (\bibinfo {year} {2016})}\BibitemShut {NoStop}%
\bibitem [{\citenamefont {Gao}\ and\ \citenamefont
  {Tang}(2022)}]{PhysRevD.106.056003}%
  \BibitemOpen
  \bibfield  {author} {\bibinfo {author} {\bibfnamefont {Y.}~\bibnamefont
  {Gao}}\ and\ \bibinfo {author} {\bibfnamefont {S.}~\bibnamefont {Tang}},\
  }\bibfield  {title} {\bibinfo {title} {{Optimal photon polarization toward
  the observation of the nonlinear Breit-Wheeler pair production}},\ }\href
  {https://doi.org/10.1103/PhysRevD.106.056003} {\bibfield  {journal} {\bibinfo
   {journal} {Phys. Rev. D}\ }\textbf {\bibinfo {volume} {106}},\ \bibinfo
  {pages} {056003} (\bibinfo {year} {2022})}\BibitemShut {NoStop}%
\bibitem [{\citenamefont {Tang}(2022)}]{PhysRevD.105.056018}%
  \BibitemOpen
  \bibfield  {author} {\bibinfo {author} {\bibfnamefont {S.}~\bibnamefont
  {Tang}},\ }\bibfield  {title} {\bibinfo {title} {{Fully polarized nonlinear
  Breit-Wheeler pair production in pulsed plane waves}},\ }\href
  {https://doi.org/10.1103/PhysRevD.105.056018} {\bibfield  {journal} {\bibinfo
   {journal} {Phys. Rev. D}\ }\textbf {\bibinfo {volume} {105}},\ \bibinfo
  {pages} {056018} (\bibinfo {year} {2022})}\BibitemShut {NoStop}%
\bibitem [{\citenamefont {Tang}\ and\ \citenamefont
  {King}(2021)}]{PhysRevD.104.096019}%
  \BibitemOpen
  \bibfield  {author} {\bibinfo {author} {\bibfnamefont {S.}~\bibnamefont
  {Tang}}\ and\ \bibinfo {author} {\bibfnamefont {B.}~\bibnamefont {King}},\
  }\bibfield  {title} {\bibinfo {title} {{Pulse envelope effects in nonlinear
  Breit-Wheeler pair creation}},\ }\href
  {https://doi.org/10.1103/PhysRevD.104.096019} {\bibfield  {journal} {\bibinfo
   {journal} {Phys. Rev. D}\ }\textbf {\bibinfo {volume} {104}},\ \bibinfo
  {pages} {096019} (\bibinfo {year} {2021})}\BibitemShut {NoStop}%
\bibitem [{\citenamefont {Jiang}\ \emph {et~al.}(2024)\citenamefont {Jiang},
  \citenamefont {Dai}, \citenamefont {Zhuang}, \citenamefont {Gao},
  \citenamefont {Tang},\ and\ \citenamefont {Chen}}]{PhysRevD.109.036030}%
  \BibitemOpen
  \bibfield  {author} {\bibinfo {author} {\bibfnamefont {J.-J.}\ \bibnamefont
  {Jiang}}, \bibinfo {author} {\bibfnamefont {Y.-N.}\ \bibnamefont {Dai}},
  \bibinfo {author} {\bibfnamefont {K.-H.}\ \bibnamefont {Zhuang}}, \bibinfo
  {author} {\bibfnamefont {Y.}~\bibnamefont {Gao}}, \bibinfo {author}
  {\bibfnamefont {S.}~\bibnamefont {Tang}},\ and\ \bibinfo {author}
  {\bibfnamefont {Y.-Y.}\ \bibnamefont {Chen}},\ }\bibfield  {title} {\bibinfo
  {title} {{Interferences effects in the polarized nonlinear Breit-Wheeler
  process}},\ }\href {https://doi.org/10.1103/PhysRevD.109.036030} {\bibfield
  {journal} {\bibinfo  {journal} {Phys. Rev. D}\ }\textbf {\bibinfo {volume}
  {109}},\ \bibinfo {pages} {036030} (\bibinfo {year} {2024})}\BibitemShut
  {NoStop}%
\bibitem [{\citenamefont {Dirac}(1930{\natexlab{a}})}]{dirac1930theory}%
  \BibitemOpen
  \bibfield  {author} {\bibinfo {author} {\bibfnamefont {P.~A.~M.}\
  \bibnamefont {Dirac}},\ }\bibfield  {title} {\bibinfo {title} {{A theory of
  electrons and protons}},\ }\href@noop {} {\bibfield  {journal} {\bibinfo
  {journal} {Proc. R. Soc. London, Ser. A}\ }\textbf {\bibinfo {volume}
  {126}},\ \bibinfo {pages} {360} (\bibinfo {year}
  {1930}{\natexlab{a}})}\BibitemShut {NoStop}%
\bibitem [{\citenamefont {Dirac}(1930{\natexlab{b}})}]{Dirac1930Annihilation}%
  \BibitemOpen
  \bibfield  {author} {\bibinfo {author} {\bibfnamefont {P.~A.~M.}\
  \bibnamefont {Dirac}},\ }\bibfield  {title} {\bibinfo {title} {{On the
  Annihilation of Electrons and Protons}},\ }\href
  {https://doi.org/10.1017/S0305004100016091} {\bibfield  {journal} {\bibinfo
  {journal} {Math. Proc. Cambridge Phil. Soc.}\ }\textbf {\bibinfo {volume}
  {26}},\ \bibinfo {pages} {361} (\bibinfo {year}
  {1930}{\natexlab{b}})}\BibitemShut {NoStop}%
\bibitem [{\citenamefont {Heisenberg}\ and\ \citenamefont
  {Euler}(1936)}]{heisenberg1936diractheory}%
  \BibitemOpen
  \bibfield  {author} {\bibinfo {author} {\bibfnamefont {W.}~\bibnamefont
  {Heisenberg}}\ and\ \bibinfo {author} {\bibfnamefont {H.}~\bibnamefont
  {Euler}},\ }\bibfield  {title} {\bibinfo {title} {{Consequences of Dirac
  Theory of the Positron}},\ }\href@noop {} {\bibfield  {journal} {\bibinfo
  {journal} {Zeit. Phys.}\ }\textbf {\bibinfo {volume} {98}},\ \bibinfo {pages}
  {714} (\bibinfo {year} {1936})}\BibitemShut {NoStop}%
\bibitem [{\citenamefont {Schwinger}(1951)}]{schwinger1951gaugeinvariance}%
  \BibitemOpen
  \bibfield  {author} {\bibinfo {author} {\bibfnamefont {J.}~\bibnamefont
  {Schwinger}},\ }\bibfield  {title} {\bibinfo {title} {{On Gauge Invariance
  and Vacuum Polarization}},\ }\href {https://doi.org/10.1103/PhysRev.82.664}
  {\bibfield  {journal} {\bibinfo  {journal} {Phys. Rev.}\ }\textbf {\bibinfo
  {volume} {82}},\ \bibinfo {pages} {664} (\bibinfo {year} {1951})}\BibitemShut
  {NoStop}%
\bibitem [{\citenamefont {Sauter}(1931)}]{sauter1931pairdirac}%
  \BibitemOpen
  \bibfield  {author} {\bibinfo {author} {\bibfnamefont {F.}~\bibnamefont
  {Sauter}},\ }\bibfield  {title} {\bibinfo {title} {{{\"U}ber das Verhalten
  eines Elektrons im homogenen elektrischen Feld nach der relativistischen
  Theorie Diracs}},\ }\href@noop {} {\bibfield  {journal} {\bibinfo  {journal}
  {Zeit. Phys.}\ }\textbf {\bibinfo {volume} {69}},\ \bibinfo {pages} {742}
  (\bibinfo {year} {1931})}\BibitemShut {NoStop}%
\bibitem [{\citenamefont {Sch\"utzhold}\ \emph {et~al.}(2008)\citenamefont
  {Sch\"utzhold}, \citenamefont {Gies},\ and\ \citenamefont
  {Dunne}}]{schutzhold2008schwinger}%
  \BibitemOpen
  \bibfield  {author} {\bibinfo {author} {\bibfnamefont {R.}~\bibnamefont
  {Sch\"utzhold}}, \bibinfo {author} {\bibfnamefont {H.}~\bibnamefont {Gies}},\
  and\ \bibinfo {author} {\bibfnamefont {G.}~\bibnamefont {Dunne}},\ }\bibfield
   {title} {\bibinfo {title} {{Dynamically Assisted Schwinger Mechanism}},\
  }\href {https://doi.org/10.1103/PhysRevLett.101.130404} {\bibfield  {journal}
  {\bibinfo  {journal} {Phys. Rev. Lett.}\ }\textbf {\bibinfo {volume} {101}},\
  \bibinfo {pages} {130404} (\bibinfo {year} {2008})}\BibitemShut {NoStop}%
\bibitem [{\citenamefont {Lv}\ \emph {et~al.}(2018)\citenamefont {Lv},
  \citenamefont {Dong}, \citenamefont {Li}, \citenamefont {Sheng},
  \citenamefont {Su},\ and\ \citenamefont {Grobe}}]{PhysRevA.97.022515}%
  \BibitemOpen
  \bibfield  {author} {\bibinfo {author} {\bibfnamefont {Q.~Z.}\ \bibnamefont
  {Lv}}, \bibinfo {author} {\bibfnamefont {S.}~\bibnamefont {Dong}}, \bibinfo
  {author} {\bibfnamefont {Y.~T.}\ \bibnamefont {Li}}, \bibinfo {author}
  {\bibfnamefont {Z.~M.}\ \bibnamefont {Sheng}}, \bibinfo {author}
  {\bibfnamefont {Q.}~\bibnamefont {Su}},\ and\ \bibinfo {author}
  {\bibfnamefont {R.}~\bibnamefont {Grobe}},\ }\bibfield  {title} {\bibinfo
  {title} {{Role of the spatial inhomogeneity on the laser-induced vacuum
  decay}},\ }\href {https://doi.org/10.1103/PhysRevA.97.022515} {\bibfield
  {journal} {\bibinfo  {journal} {Phys. Rev. A}\ }\textbf {\bibinfo {volume}
  {97}},\ \bibinfo {pages} {022515} (\bibinfo {year} {2018})}\BibitemShut
  {NoStop}%
\bibitem [{\citenamefont {Villalba-Ch\'avez}\ and\ \citenamefont
  {M\"uller}(2019{\natexlab{a}})}]{PhysRevD.100.116018}%
  \BibitemOpen
  \bibfield  {author} {\bibinfo {author} {\bibfnamefont {S.}~\bibnamefont
  {Villalba-Ch\'avez}}\ and\ \bibinfo {author} {\bibfnamefont {C.}~\bibnamefont
  {M\"uller}},\ }\bibfield  {title} {\bibinfo {title} {{Signatures of the
  Schwinger mechanism assisted by a fast-oscillating electric field}},\ }\href
  {https://doi.org/10.1103/PhysRevD.100.116018} {\bibfield  {journal} {\bibinfo
   {journal} {Phys. Rev. D}\ }\textbf {\bibinfo {volume} {100}},\ \bibinfo
  {pages} {116018} (\bibinfo {year} {2019}{\natexlab{a}})}\BibitemShut
  {NoStop}%
\bibitem [{\citenamefont {Aleksandrov}\ \emph {et~al.}(2018)\citenamefont
  {Aleksandrov}, \citenamefont {Plunien},\ and\ \citenamefont
  {Shabaev}}]{PhysRevD.97.116001}%
  \BibitemOpen
  \bibfield  {author} {\bibinfo {author} {\bibfnamefont {I.~A.}\ \bibnamefont
  {Aleksandrov}}, \bibinfo {author} {\bibfnamefont {G.}~\bibnamefont
  {Plunien}},\ and\ \bibinfo {author} {\bibfnamefont {V.~M.}\ \bibnamefont
  {Shabaev}},\ }\bibfield  {title} {\bibinfo {title} {{Dynamically assisted
  Schwinger effect beyond the spatially-uniform-field approximation}},\ }\href
  {https://doi.org/10.1103/PhysRevD.97.116001} {\bibfield  {journal} {\bibinfo
  {journal} {Phys. Rev. D}\ }\textbf {\bibinfo {volume} {97}},\ \bibinfo
  {pages} {116001} (\bibinfo {year} {2018})}\BibitemShut {NoStop}%
\bibitem [{\citenamefont {Castro~Neto}\ \emph {et~al.}(2009)\citenamefont
  {Castro~Neto}, \citenamefont {Guinea}, \citenamefont {Peres}, \citenamefont
  {Novoselov},\ and\ \citenamefont {Geim}}]{RevModPhys.81.109}%
  \BibitemOpen
  \bibfield  {author} {\bibinfo {author} {\bibfnamefont {A.~H.}\ \bibnamefont
  {Castro~Neto}}, \bibinfo {author} {\bibfnamefont {F.}~\bibnamefont {Guinea}},
  \bibinfo {author} {\bibfnamefont {N.~M.~R.}\ \bibnamefont {Peres}}, \bibinfo
  {author} {\bibfnamefont {K.~S.}\ \bibnamefont {Novoselov}},\ and\ \bibinfo
  {author} {\bibfnamefont {A.~K.}\ \bibnamefont {Geim}},\ }\bibfield  {title}
  {\bibinfo {title} {{The electronic properties of graphene}},\ }\href
  {https://doi.org/10.1103/RevModPhys.81.109} {\bibfield  {journal} {\bibinfo
  {journal} {Rev. Mod. Phys.}\ }\textbf {\bibinfo {volume} {81}},\ \bibinfo
  {pages} {109} (\bibinfo {year} {2009})}\BibitemShut {NoStop}%
\bibitem [{\citenamefont {Golub}\ \emph {et~al.}(2020)\citenamefont {Golub},
  \citenamefont {Egger}, \citenamefont {M\"uller},\ and\ \citenamefont
  {Villalba-Ch\'avez}}]{PhysRevLett.124.110403}%
  \BibitemOpen
  \bibfield  {author} {\bibinfo {author} {\bibfnamefont {A.}~\bibnamefont
  {Golub}}, \bibinfo {author} {\bibfnamefont {R.}~\bibnamefont {Egger}},
  \bibinfo {author} {\bibfnamefont {C.}~\bibnamefont {M\"uller}},\ and\
  \bibinfo {author} {\bibfnamefont {S.}~\bibnamefont {Villalba-Ch\'avez}},\
  }\bibfield  {title} {\bibinfo {title} {{Dimensionality-Driven Photoproduction
  of Massive Dirac Pairs near Threshold in Gapped Graphene Monolayers}},\
  }\href {https://doi.org/10.1103/PhysRevLett.124.110403} {\bibfield  {journal}
  {\bibinfo  {journal} {Phys. Rev. Lett.}\ }\textbf {\bibinfo {volume} {124}},\
  \bibinfo {pages} {110403} (\bibinfo {year} {2020})}\BibitemShut {NoStop}%
\bibitem [{\citenamefont {Schubert}(2001)}]{schubert2001perturbative}%
  \BibitemOpen
  \bibfield  {author} {\bibinfo {author} {\bibfnamefont {C.}~\bibnamefont
  {Schubert}},\ }\bibfield  {title} {\bibinfo {title} {{Perturbative quantum
  field theory in the string-inspired formalism}},\ }\href@noop {} {\bibfield
  {journal} {\bibinfo  {journal} {Phys. Rep.}\ }\textbf {\bibinfo {volume}
  {355}},\ \bibinfo {pages} {73} (\bibinfo {year} {2001})}\BibitemShut
  {NoStop}%
\bibitem [{\citenamefont {Dunne}\ and\ \citenamefont
  {Schubert}(2005)}]{PhysRevD.72.105004}%
  \BibitemOpen
  \bibfield  {author} {\bibinfo {author} {\bibfnamefont {G.~V.}\ \bibnamefont
  {Dunne}}\ and\ \bibinfo {author} {\bibfnamefont {C.}~\bibnamefont
  {Schubert}},\ }\bibfield  {title} {\bibinfo {title} {{Worldline instantons
  and pair production in inhomogenous fields}},\ }\href
  {https://doi.org/10.1103/PhysRevD.72.105004} {\bibfield  {journal} {\bibinfo
  {journal} {Phys. Rev. D}\ }\textbf {\bibinfo {volume} {72}},\ \bibinfo
  {pages} {105004} (\bibinfo {year} {2005})}\BibitemShut {NoStop}%
\bibitem [{\citenamefont {Degli~Esposti}\ and\ \citenamefont
  {Torgrimsson}(2023{\natexlab{a}})}]{PhysRevD.107.056019}%
  \BibitemOpen
  \bibfield  {author} {\bibinfo {author} {\bibfnamefont {G.}~\bibnamefont
  {Degli~Esposti}}\ and\ \bibinfo {author} {\bibfnamefont {G.}~\bibnamefont
  {Torgrimsson}},\ }\bibfield  {title} {\bibinfo {title} {{Worldline instantons
  for the momentum spectrum of Schwinger pair production in spacetime dependent
  fields}},\ }\href {https://doi.org/10.1103/PhysRevD.107.056019} {\bibfield
  {journal} {\bibinfo  {journal} {Phys. Rev. D}\ }\textbf {\bibinfo {volume}
  {107}},\ \bibinfo {pages} {056019} (\bibinfo {year}
  {2023}{\natexlab{a}})}\BibitemShut {NoStop}%
\bibitem [{\citenamefont {T.~Cheng}\ and\ \citenamefont
  {Grobe}(2010)}]{doi:10.1080/00107510903450559}%
  \BibitemOpen
  \bibfield  {author} {\bibinfo {author} {\bibfnamefont {Q.~S.}\ \bibnamefont
  {T.~Cheng}}\ and\ \bibinfo {author} {\bibfnamefont {R.}~\bibnamefont
  {Grobe}},\ }\bibfield  {title} {\bibinfo {title} {{Introductory review on
  quantum field theory with space–time resolution}},\ }\href
  {https://doi.org/10.1080/00107510903450559} {\bibfield  {journal} {\bibinfo
  {journal} {Contemp. Phys.}\ }\textbf {\bibinfo {volume} {51}},\ \bibinfo
  {pages} {315} (\bibinfo {year} {2010})}\BibitemShut {NoStop}%
\bibitem [{\citenamefont {Li}\ \emph {et~al.}(2023)\citenamefont {Li},
  \citenamefont {Li}, \citenamefont {Su},\ and\ \citenamefont
  {Grobe}}]{PhysRevA.108.033112}%
  \BibitemOpen
  \bibfield  {author} {\bibinfo {author} {\bibfnamefont {C.~K.}\ \bibnamefont
  {Li}}, \bibinfo {author} {\bibfnamefont {Y.~J.}\ \bibnamefont {Li}}, \bibinfo
  {author} {\bibfnamefont {Q.}~\bibnamefont {Su}},\ and\ \bibinfo {author}
  {\bibfnamefont {R.}~\bibnamefont {Grobe}},\ }\bibfield  {title} {\bibinfo
  {title} {{Phase sensitivity of the pair-creation process in colliding laser
  pulses}},\ }\href {https://doi.org/10.1103/PhysRevA.108.033112} {\bibfield
  {journal} {\bibinfo  {journal} {Phys. Rev. A}\ }\textbf {\bibinfo {volume}
  {108}},\ \bibinfo {pages} {033112} (\bibinfo {year} {2023})}\BibitemShut
  {NoStop}%
\bibitem [{\citenamefont {Bia{\l}ynicki-Birula}\ \emph
  {et~al.}(1991)\citenamefont {Bia{\l}ynicki-Birula}, \citenamefont
  {G\'ornicki},\ and\ \citenamefont {Rafelski}}]{bialynicki1991diracvacuum}%
  \BibitemOpen
  \bibfield  {author} {\bibinfo {author} {\bibfnamefont {I.}~\bibnamefont
  {Bia{\l}ynicki-Birula}}, \bibinfo {author} {\bibfnamefont {P.}~\bibnamefont
  {G\'ornicki}},\ and\ \bibinfo {author} {\bibfnamefont {J.}~\bibnamefont
  {Rafelski}},\ }\bibfield  {title} {\bibinfo {title} {{Phase-space structure
  of the Dirac vacuum}},\ }\href {https://doi.org/10.1103/PhysRevD.44.1825}
  {\bibfield  {journal} {\bibinfo  {journal} {Phys. Rev. D}\ }\textbf {\bibinfo
  {volume} {44}},\ \bibinfo {pages} {1825} (\bibinfo {year}
  {1991})}\BibitemShut {NoStop}%
\bibitem [{\citenamefont {Bechler}\ \emph {et~al.}(2023)\citenamefont
  {Bechler}, \citenamefont {V{\'e}lez}, \citenamefont {Krajewska},\ and\
  \citenamefont {Kami{\'n}ski}}]{bechler2023schwinger}%
  \BibitemOpen
  \bibfield  {author} {\bibinfo {author} {\bibfnamefont {A.}~\bibnamefont
  {Bechler}}, \bibinfo {author} {\bibfnamefont {F.~C.}\ \bibnamefont
  {V{\'e}lez}}, \bibinfo {author} {\bibfnamefont {K.}~\bibnamefont
  {Krajewska}},\ and\ \bibinfo {author} {\bibfnamefont {J.~Z.}\ \bibnamefont
  {Kami{\'n}ski}},\ }\bibfield  {title} {\bibinfo {title} {{Vortex Structures
  and Momentum Sharing in Dynamic Sauter–Schwinger Process}},\ }\href
  {https://doi.org/10.12693/APhysPolA.143.S18} {\bibfield  {journal} {\bibinfo
  {journal} {Acta Phys. Pol. A}\ }\textbf {\bibinfo {volume} {143}},\ \bibinfo
  {pages} {S18} (\bibinfo {year} {2023})}\BibitemShut {NoStop}%
\bibitem [{\citenamefont {Kami\ifmmode~\acute{n}\else \'{n}\fi{}ski}\ \emph
  {et~al.}(2018)\citenamefont {Kami\ifmmode~\acute{n}\else \'{n}\fi{}ski},
  \citenamefont {Twardy},\ and\ \citenamefont
  {Krajewska}}]{PhysRevD.98.056009}%
  \BibitemOpen
  \bibfield  {author} {\bibinfo {author} {\bibfnamefont {J.~Z.}\ \bibnamefont
  {Kami\ifmmode~\acute{n}\else \'{n}\fi{}ski}}, \bibinfo {author}
  {\bibfnamefont {M.}~\bibnamefont {Twardy}},\ and\ \bibinfo {author}
  {\bibfnamefont {K.}~\bibnamefont {Krajewska}},\ }\bibfield  {title} {\bibinfo
  {title} {{Diffraction at a time grating in electron-positron pair creation
  from the vacuum}},\ }\href {https://doi.org/10.1103/PhysRevD.98.056009}
  {\bibfield  {journal} {\bibinfo  {journal} {Phys. Rev. D}\ }\textbf {\bibinfo
  {volume} {98}},\ \bibinfo {pages} {056009} (\bibinfo {year}
  {2018})}\BibitemShut {NoStop}%
\bibitem [{\citenamefont {Taya}(2019)}]{PhysRevD.99.056006}%
  \BibitemOpen
  \bibfield  {author} {\bibinfo {author} {\bibfnamefont {H.}~\bibnamefont
  {Taya}},\ }\bibfield  {title} {\bibinfo {title} {{Franz-Keldysh effect in
  strong-field QED}},\ }\href {https://doi.org/10.1103/PhysRevD.99.056006}
  {\bibfield  {journal} {\bibinfo  {journal} {Phys. Rev. D}\ }\textbf {\bibinfo
  {volume} {99}},\ \bibinfo {pages} {056006} (\bibinfo {year}
  {2019})}\BibitemShut {NoStop}%
\bibitem [{\citenamefont {Huang}\ and\ \citenamefont
  {Taya}(2019)}]{PhysRevD.100.016013}%
  \BibitemOpen
  \bibfield  {author} {\bibinfo {author} {\bibfnamefont {X.-G.}\ \bibnamefont
  {Huang}}\ and\ \bibinfo {author} {\bibfnamefont {H.}~\bibnamefont {Taya}},\
  }\bibfield  {title} {\bibinfo {title} {{Spin-dependent dynamically assisted
  Schwinger mechanism}},\ }\href {https://doi.org/10.1103/PhysRevD.100.016013}
  {\bibfield  {journal} {\bibinfo  {journal} {Phys. Rev. D}\ }\textbf {\bibinfo
  {volume} {100}},\ \bibinfo {pages} {016013} (\bibinfo {year}
  {2019})}\BibitemShut {NoStop}%
\bibitem [{\citenamefont {Kohlf\"urst}(2019)}]{PhysRevD.99.096017}%
  \BibitemOpen
  \bibfield  {author} {\bibinfo {author} {\bibfnamefont {C.}~\bibnamefont
  {Kohlf\"urst}},\ }\bibfield  {title} {\bibinfo {title} {{Spin states in
  multiphoton pair production for circularly polarized light}},\ }\href
  {https://doi.org/10.1103/PhysRevD.99.096017} {\bibfield  {journal} {\bibinfo
  {journal} {Phys. Rev. D}\ }\textbf {\bibinfo {volume} {99}},\ \bibinfo
  {pages} {096017} (\bibinfo {year} {2019})}\BibitemShut {NoStop}%
\bibitem [{\citenamefont {Dyson}(1949{\natexlab{a}})}]{dyson1949radiation}%
  \BibitemOpen
  \bibfield  {author} {\bibinfo {author} {\bibfnamefont {F.~J.}\ \bibnamefont
  {Dyson}},\ }\bibfield  {title} {\bibinfo {title} {{The radiation theories of
  Tomonaga, Schwinger, and Feynman}},\ }\href@noop {} {\bibfield  {journal}
  {\bibinfo  {journal} {Phys. Rev.}\ }\textbf {\bibinfo {volume} {75}},\
  \bibinfo {pages} {486} (\bibinfo {year} {1949}{\natexlab{a}})}\BibitemShut
  {NoStop}%
\bibitem [{\citenamefont {Dyson}(1949{\natexlab{b}})}]{dyson1949s}%
  \BibitemOpen
  \bibfield  {author} {\bibinfo {author} {\bibfnamefont {F.~J.}\ \bibnamefont
  {Dyson}},\ }\bibfield  {title} {\bibinfo {title} {{The S matrix in quantum
  electrodynamics}},\ }\href@noop {} {\bibfield  {journal} {\bibinfo  {journal}
  {Phys. Rev.}\ }\textbf {\bibinfo {volume} {75}},\ \bibinfo {pages} {1736}
  (\bibinfo {year} {1949}{\natexlab{b}})}\BibitemShut {NoStop}%
\bibitem [{\citenamefont {Lehmann}\ \emph {et~al.}(1955)\citenamefont
  {Lehmann}, \citenamefont {Symanzik},\ and\ \citenamefont
  {Zimmermann}}]{lehmann1955formulierung}%
  \BibitemOpen
  \bibfield  {author} {\bibinfo {author} {\bibfnamefont {H.}~\bibnamefont
  {Lehmann}}, \bibinfo {author} {\bibfnamefont {K.}~\bibnamefont {Symanzik}},\
  and\ \bibinfo {author} {\bibfnamefont {W.}~\bibnamefont {Zimmermann}},\
  }\bibfield  {title} {\bibinfo {title} {{Zur formulierung quantisierter
  feldtheorien}},\ }\href@noop {} {\bibfield  {journal} {\bibinfo  {journal}
  {Il Nuovo Cim. (1955-1965)}\ }\textbf {\bibinfo {volume} {1}},\ \bibinfo
  {pages} {205} (\bibinfo {year} {1955})}\BibitemShut {NoStop}%
\bibitem [{\citenamefont {Lehmann}\ \emph {et~al.}(1957)\citenamefont
  {Lehmann}, \citenamefont {Symanzik},\ and\ \citenamefont
  {Zimmermann}}]{lehmann1957formulation}%
  \BibitemOpen
  \bibfield  {author} {\bibinfo {author} {\bibfnamefont {H.}~\bibnamefont
  {Lehmann}}, \bibinfo {author} {\bibfnamefont {K.}~\bibnamefont {Symanzik}},\
  and\ \bibinfo {author} {\bibfnamefont {W.}~\bibnamefont {Zimmermann}},\
  }\bibfield  {title} {\bibinfo {title} {{On the formulation of quantized field
  theories}},\ }\href@noop {} {\bibfield  {journal} {\bibinfo  {journal} {Il
  Nuovo Cim. (1955-1965)}\ }\textbf {\bibinfo {volume} {6}},\ \bibinfo {pages}
  {319} (\bibinfo {year} {1957})}\BibitemShut {NoStop}%
\bibitem [{\citenamefont {Itzykson}\ and\ \citenamefont
  {Zuber}(1980)}]{itzykson1980quantumfield}%
  \BibitemOpen
  \bibfield  {author} {\bibinfo {author} {\bibfnamefont {C.}~\bibnamefont
  {Itzykson}}\ and\ \bibinfo {author} {\bibfnamefont {J.-B.}\ \bibnamefont
  {Zuber}},\ }\href@noop {} {\emph {\bibinfo {title} {{Quantum Field
  Theory}}}}\ (\bibinfo  {publisher} {McGraw-Hill},\ \bibinfo {address} {New
  York},\ \bibinfo {year} {1980})\BibitemShut {NoStop}%
\bibitem [{\citenamefont {Degli~Esposti}\ and\ \citenamefont
  {Torgrimsson}(2023{\natexlab{b}})}]{degli2023worldline}%
  \BibitemOpen
  \bibfield  {author} {\bibinfo {author} {\bibfnamefont {G.}~\bibnamefont
  {Degli~Esposti}}\ and\ \bibinfo {author} {\bibfnamefont {G.}~\bibnamefont
  {Torgrimsson}},\ }\bibfield  {title} {\bibinfo {title} {{Worldline instantons
  for the momentum spectrum of Schwinger pair production in spacetime dependent
  fields}},\ }\href@noop {} {\bibfield  {journal} {\bibinfo  {journal} {Phys.
  Rev. D}\ }\textbf {\bibinfo {volume} {107}},\ \bibinfo {pages} {056019}
  (\bibinfo {year} {2023}{\natexlab{b}})}\BibitemShut {NoStop}%
\bibitem [{\citenamefont {Bjorken}\ and\ \citenamefont
  {Drell}(1964)}]{bjorken1964relquantum}%
  \BibitemOpen
  \bibfield  {author} {\bibinfo {author} {\bibfnamefont {J.~D.}\ \bibnamefont
  {Bjorken}}\ and\ \bibinfo {author} {\bibfnamefont {S.~D.}\ \bibnamefont
  {Drell}},\ }\href@noop {} {\emph {\bibinfo {title} {{Relativistic Quantum
  Mechanics}}}}\ (\bibinfo  {publisher} {McGraw-Hill},\ \bibinfo {address} {New
  York},\ \bibinfo {year} {1964})\BibitemShut {NoStop}%
\bibitem [{\citenamefont {Bia{\l}ynicki-Birula}\ and\ \citenamefont
  {Bia{\l}ynicka-Birula}(1975)}]{bialynicki1975quantumelectro}%
  \BibitemOpen
  \bibfield  {author} {\bibinfo {author} {\bibfnamefont {I.}~\bibnamefont
  {Bia{\l}ynicki-Birula}}\ and\ \bibinfo {author} {\bibfnamefont
  {Z.}~\bibnamefont {Bia{\l}ynicka-Birula}},\ }\href@noop {} {\emph {\bibinfo
  {title} {{Quantum Electrodynamics}}}}\ (\bibinfo  {publisher} {Pergamon},\
  \bibinfo {address} {Oxford},\ \bibinfo {year} {1975})\BibitemShut {NoStop}%
\bibitem [{\citenamefont {Bothe}(1926)}]{bothe1926kopplung}%
  \BibitemOpen
  \bibfield  {author} {\bibinfo {author} {\bibfnamefont {W.}~\bibnamefont
  {Bothe}},\ }\bibfield  {title} {\bibinfo {title} {{{\"U}ber die Kopplung
  zwischen elementaren Strahlungsvorg{\"a}ngen}},\ }\href@noop {} {\bibfield
  {journal} {\bibinfo  {journal} {Zeit. Phys.}\ }\textbf {\bibinfo {volume}
  {37}},\ \bibinfo {pages} {547} (\bibinfo {year} {1926})}\BibitemShut
  {NoStop}%
\bibitem [{\citenamefont {Garrison}\ and\ \citenamefont
  {Chiao}(2008)}]{garrison2008quantumopt}%
  \BibitemOpen
  \bibfield  {author} {\bibinfo {author} {\bibfnamefont {J.~C.}\ \bibnamefont
  {Garrison}}\ and\ \bibinfo {author} {\bibfnamefont {R.~Y.}\ \bibnamefont
  {Chiao}},\ }\href@noop {} {\emph {\bibinfo {title} {{Quantum Optics}}}}\
  (\bibinfo  {publisher} {Oxford},\ \bibinfo {address} {Oxford},\ \bibinfo
  {year} {2008})\BibitemShut {NoStop}%
\bibitem [{\citenamefont {Goldberger}\ and\ \citenamefont
  {Watson}(1964)}]{goldberger1964collision}%
  \BibitemOpen
  \bibfield  {author} {\bibinfo {author} {\bibfnamefont {M.~L.}\ \bibnamefont
  {Goldberger}}\ and\ \bibinfo {author} {\bibfnamefont {K.~M.}\ \bibnamefont
  {Watson}},\ }\href@noop {} {\emph {\bibinfo {title} {{Collision Theory}}}}\
  (\bibinfo  {publisher} {John Wiley \& Sons},\ \bibinfo {address} {New York},\
  \bibinfo {year} {1964})\BibitemShut {NoStop}%
\bibitem [{\citenamefont {Newton}(1982)}]{newton1982scattering}%
  \BibitemOpen
  \bibfield  {author} {\bibinfo {author} {\bibfnamefont {R.~G.}\ \bibnamefont
  {Newton}},\ }\href@noop {} {\emph {\bibinfo {title} {{Scattering Theory of
  Waves and Particles}}}}\ (\bibinfo  {publisher} {Springer},\ \bibinfo
  {address} {New York},\ \bibinfo {year} {1982})\BibitemShut {NoStop}%
\bibitem [{\citenamefont {Feynman}(1949{\natexlab{a}})}]{feymann1949qed}%
  \BibitemOpen
  \bibfield  {author} {\bibinfo {author} {\bibfnamefont {R.~P.}\ \bibnamefont
  {Feynman}},\ }\bibfield  {title} {\bibinfo {title} {{Space-Time Approach to
  Quantum Electrodynamics}},\ }\href {https://doi.org/10.1103/PhysRev.76.769}
  {\bibfield  {journal} {\bibinfo  {journal} {Phys. Rev.}\ }\textbf {\bibinfo
  {volume} {76}},\ \bibinfo {pages} {769} (\bibinfo {year}
  {1949}{\natexlab{a}})}\BibitemShut {NoStop}%
\bibitem [{\citenamefont {Feynman}(1949{\natexlab{b}})}]{feymann1949positrons}%
  \BibitemOpen
  \bibfield  {author} {\bibinfo {author} {\bibfnamefont {R.~P.}\ \bibnamefont
  {Feynman}},\ }\bibfield  {title} {\bibinfo {title} {{The Theory of
  Positrons}},\ }\href {https://doi.org/10.1103/PhysRev.76.749} {\bibfield
  {journal} {\bibinfo  {journal} {Phys. Rev.}\ }\textbf {\bibinfo {volume}
  {76}},\ \bibinfo {pages} {749} (\bibinfo {year}
  {1949}{\natexlab{b}})}\BibitemShut {NoStop}%
\bibitem [{\citenamefont {Bia{\l}ynicki-Birula}\ \emph
  {et~al.}(1992)\citenamefont {Bia{\l}ynicki-Birula}, \citenamefont {Cieplak},\
  and\ \citenamefont {Kami{\'n}ski}}]{bialynicki1991TheoryOfQuanta}%
  \BibitemOpen
  \bibfield  {author} {\bibinfo {author} {\bibfnamefont {I.}~\bibnamefont
  {Bia{\l}ynicki-Birula}}, \bibinfo {author} {\bibfnamefont {M.}~\bibnamefont
  {Cieplak}},\ and\ \bibinfo {author} {\bibfnamefont {J.}~\bibnamefont
  {Kami{\'n}ski}},\ }\href@noop {} {\emph {\bibinfo {title} {{Theory of
  Quanta}}}}\ (\bibinfo  {publisher} {Oxford},\ \bibinfo {address} {New York},\
  \bibinfo {year} {1992})\BibitemShut {NoStop}%
\bibitem [{\citenamefont {Bia{\l}ynicki-Birula}\ and\ \citenamefont
  {Rudnicki}(2011)}]{PhysRevD.83.065020}%
  \BibitemOpen
  \bibfield  {author} {\bibinfo {author} {\bibfnamefont {I.}~\bibnamefont
  {Bia{\l}ynicki-Birula}}\ and\ \bibinfo {author} {\bibfnamefont
  {{\L}.}~\bibnamefont {Rudnicki}},\ }\bibfield  {title} {\bibinfo {title}
  {{Time evolution of the QED vacuum in a uniform electric field: Complete
  analytic solution by spinorial decomposition}},\ }\href
  {https://doi.org/10.1103/PhysRevD.83.065020} {\bibfield  {journal} {\bibinfo
  {journal} {Phys. Rev. D}\ }\textbf {\bibinfo {volume} {83}},\ \bibinfo
  {pages} {065020} (\bibinfo {year} {2011})}\BibitemShut {NoStop}%
\bibitem [{\citenamefont {Blaschke}\ \emph {et~al.}(2013)\citenamefont
  {Blaschke}, \citenamefont {K\"ampfer}, \citenamefont {Schmidt}, \citenamefont
  {Panferov}, \citenamefont {Prozorkevich},\ and\ \citenamefont
  {Smolyansky}}]{PhysRevD.88.045017}%
  \BibitemOpen
  \bibfield  {author} {\bibinfo {author} {\bibfnamefont {D.~B.}\ \bibnamefont
  {Blaschke}}, \bibinfo {author} {\bibfnamefont {B.}~\bibnamefont {K\"ampfer}},
  \bibinfo {author} {\bibfnamefont {S.~M.}\ \bibnamefont {Schmidt}}, \bibinfo
  {author} {\bibfnamefont {A.~D.}\ \bibnamefont {Panferov}}, \bibinfo {author}
  {\bibfnamefont {A.~V.}\ \bibnamefont {Prozorkevich}},\ and\ \bibinfo {author}
  {\bibfnamefont {S.~A.}\ \bibnamefont {Smolyansky}},\ }\bibfield  {title}
  {\bibinfo {title} {{Properties of the electron-positron plasma created from a
  vacuum in a strong laser field: Quasiparticle excitations}},\ }\href
  {https://doi.org/10.1103/PhysRevD.88.045017} {\bibfield  {journal} {\bibinfo
  {journal} {Phys. Rev. D}\ }\textbf {\bibinfo {volume} {88}},\ \bibinfo
  {pages} {045017} (\bibinfo {year} {2013})}\BibitemShut {NoStop}%
\bibitem [{\citenamefont {Krajewska}\ and\ \citenamefont
  {Kami\ifmmode~\acute{n}\else
  \'{n}\fi{}ski}(2019{\natexlab{a}})}]{PhysRevA.100.012104}%
  \BibitemOpen
  \bibfield  {author} {\bibinfo {author} {\bibfnamefont {K.}~\bibnamefont
  {Krajewska}}\ and\ \bibinfo {author} {\bibfnamefont {J.~Z.}\ \bibnamefont
  {Kami\ifmmode~\acute{n}\else \'{n}\fi{}ski}},\ }\bibfield  {title} {\bibinfo
  {title} {{Threshold effects in electron-positron pair creation from the
  vacuum: Stabilization and longitudinal versus transverse momentum sharing}},\
  }\href {https://doi.org/10.1103/PhysRevA.100.012104} {\bibfield  {journal}
  {\bibinfo  {journal} {Phys. Rev. A}\ }\textbf {\bibinfo {volume} {100}},\
  \bibinfo {pages} {012104} (\bibinfo {year} {2019}{\natexlab{a}})}\BibitemShut
  {NoStop}%
\bibitem [{\citenamefont {Blinne}\ and\ \citenamefont
  {Gies}(2014)}]{blinne2014pair}%
  \BibitemOpen
  \bibfield  {author} {\bibinfo {author} {\bibfnamefont {A.}~\bibnamefont
  {Blinne}}\ and\ \bibinfo {author} {\bibfnamefont {H.}~\bibnamefont {Gies}},\
  }\bibfield  {title} {\bibinfo {title} {{Pair production in rotating electric
  fields}},\ }\href {https://doi.org/10.1103/PhysRevD.89.085001} {\bibfield
  {journal} {\bibinfo  {journal} {Phys. Rev. D}\ }\textbf {\bibinfo {volume}
  {89}},\ \bibinfo {pages} {085001} (\bibinfo {year} {2014})}\BibitemShut
  {NoStop}%
\bibitem [{\citenamefont {Blinne}\ and\ \citenamefont
  {Strobel}(2016)}]{blinne2016wigner}%
  \BibitemOpen
  \bibfield  {author} {\bibinfo {author} {\bibfnamefont {A.}~\bibnamefont
  {Blinne}}\ and\ \bibinfo {author} {\bibfnamefont {E.}~\bibnamefont
  {Strobel}},\ }\bibfield  {title} {\bibinfo {title} {{Comparison of
  semiclassical and Wigner function methods in pair production in rotating
  fields}},\ }\href {https://doi.org/10.1103/PhysRevD.93.025014} {\bibfield
  {journal} {\bibinfo  {journal} {Phys. Rev. D}\ }\textbf {\bibinfo {volume}
  {93}},\ \bibinfo {pages} {025014} (\bibinfo {year} {2016})}\BibitemShut
  {NoStop}%
\bibitem [{\citenamefont {Li}\ \emph {et~al.}(2017)\citenamefont {Li},
  \citenamefont {Li},\ and\ \citenamefont {Xie}}]{li2017pairvortex}%
  \BibitemOpen
  \bibfield  {author} {\bibinfo {author} {\bibfnamefont {Z.~L.}\ \bibnamefont
  {Li}}, \bibinfo {author} {\bibfnamefont {Y.~J.}\ \bibnamefont {Li}},\ and\
  \bibinfo {author} {\bibfnamefont {B.~S.}\ \bibnamefont {Xie}},\ }\bibfield
  {title} {\bibinfo {title} {{Momentum Vortices on Pairs Production by Two
  Counter-Rotating Fields}},\ }\href
  {https://doi.org/10.1103/PhysRevD.96.076010} {\bibfield  {journal} {\bibinfo
  {journal} {Phys. Rev. D}\ }\textbf {\bibinfo {volume} {96}},\ \bibinfo
  {pages} {076010} (\bibinfo {year} {2017})}\BibitemShut {NoStop}%
\bibitem [{\citenamefont {Hu}\ \emph {et~al.}(2023)\citenamefont {Hu},
  \citenamefont {Amat}, \citenamefont {Wang}, \citenamefont {Sawut},
  \citenamefont {Fan},\ and\ \citenamefont {Xie}}]{li2023pairspiralorginal}%
  \BibitemOpen
  \bibfield  {author} {\bibinfo {author} {\bibfnamefont {L.-N.}\ \bibnamefont
  {Hu}}, \bibinfo {author} {\bibfnamefont {O.}~\bibnamefont {Amat}}, \bibinfo
  {author} {\bibfnamefont {L.}~\bibnamefont {Wang}}, \bibinfo {author}
  {\bibfnamefont {A.}~\bibnamefont {Sawut}}, \bibinfo {author} {\bibfnamefont
  {H.-H.}\ \bibnamefont {Fan}},\ and\ \bibinfo {author} {\bibfnamefont {B.~S.}\
  \bibnamefont {Xie}},\ }\bibfield  {title} {\bibinfo {title} {{Momentum
  spirals in multiphoton pair production revisited}},\ }\href
  {https://doi.org/10.1103/PhysRevD.107.116010} {\bibfield  {journal} {\bibinfo
   {journal} {Phys. Rev. D}\ }\textbf {\bibinfo {volume} {107}},\ \bibinfo
  {pages} {116010} (\bibinfo {year} {2023})}\BibitemShut {NoStop}%
\bibitem [{\citenamefont {Grib}\ \emph {et~al.}(1972)\citenamefont {Grib},
  \citenamefont {Mostepanenko},\ and\ \citenamefont {Frolov}}]{grib1972pair}%
  \BibitemOpen
  \bibfield  {author} {\bibinfo {author} {\bibfnamefont {A.~A.}\ \bibnamefont
  {Grib}}, \bibinfo {author} {\bibfnamefont {V.~M.}\ \bibnamefont
  {Mostepanenko}},\ and\ \bibinfo {author} {\bibfnamefont {V.~M.}\ \bibnamefont
  {Frolov}},\ }\bibfield  {title} {\bibinfo {title} {{Particle creation from
  vacuum by a homogeneous electric field in the canonical formalism}},\
  }\href@noop {} {\bibfield  {journal} {\bibinfo  {journal} {Theor. Math.
  Phys.}\ }\textbf {\bibinfo {volume} {13}},\ \bibinfo {pages} {207} (\bibinfo
  {year} {1972})}\BibitemShut {NoStop}%
\bibitem [{\citenamefont {Mostepanenko}\ and\ \citenamefont
  {Frolov}(1974)}]{mostepanenko1974pair}%
  \BibitemOpen
  \bibfield  {author} {\bibinfo {author} {\bibfnamefont {V.~M.}\ \bibnamefont
  {Mostepanenko}}\ and\ \bibinfo {author} {\bibfnamefont {V.~M.}\ \bibnamefont
  {Frolov}},\ }\bibfield  {title} {\bibinfo {title} {{Particle production from
  vacuum by homogeneous electric field with periodical time dependence}},\
  }\href@noop {} {\bibfield  {journal} {\bibinfo  {journal} {Yad. Fiz.}\
  }\textbf {\bibinfo {volume} {19}},\ \bibinfo {pages} {885} (\bibinfo {year}
  {1974})}\BibitemShut {NoStop}%
\bibitem [{\citenamefont {Bagrov}\ \emph {et~al.}(1975)\citenamefont {Bagrov},
  \citenamefont {Gitman},\ and\ \citenamefont {Shvartsman}}]{bagrov1975pair}%
  \BibitemOpen
  \bibfield  {author} {\bibinfo {author} {\bibfnamefont {V.~G.}\ \bibnamefont
  {Bagrov}}, \bibinfo {author} {\bibfnamefont {D.~M.}\ \bibnamefont {Gitman}},\
  and\ \bibinfo {author} {\bibfnamefont {S.~M.}\ \bibnamefont {Shvartsman}},\
  }\bibfield  {title} {\bibinfo {title} {{Concerning the production of
  electron-positron pairs from vacuum}},\ }\href@noop {} {\bibfield  {journal}
  {\bibinfo  {journal} {Zh. Eksp. Teor. Fiz.}\ }\textbf {\bibinfo {volume}
  {68}},\ \bibinfo {pages} {392} (\bibinfo {year} {1975})}\BibitemShut
  {NoStop}%
\bibitem [{\citenamefont {Grib}\ \emph {et~al.}(1994)\citenamefont {Grib},
  \citenamefont {Mamayev},\ and\ \citenamefont
  {Mostepanenko}}]{grib1994vacuumquantum}%
  \BibitemOpen
  \bibfield  {author} {\bibinfo {author} {\bibfnamefont {A.~A.}\ \bibnamefont
  {Grib}}, \bibinfo {author} {\bibfnamefont {S.~G.}\ \bibnamefont {Mamayev}},\
  and\ \bibinfo {author} {\bibfnamefont {V.~M.}\ \bibnamefont {Mostepanenko}},\
  }\href@noop {} {\emph {\bibinfo {title} {{Vacuum quantum effects in strong
  fields}}}}\ (\bibinfo  {publisher} {Friedmann Laboratory Publishing},\
  \bibinfo {address} {St.Petersburg},\ \bibinfo {year} {1994})\BibitemShut
  {NoStop}%
\bibitem [{\citenamefont {Gavrilov}\ and\ \citenamefont
  {Gitman}(1996)}]{gavrilov1996vacuum}%
  \BibitemOpen
  \bibfield  {author} {\bibinfo {author} {\bibfnamefont {S.~P.}\ \bibnamefont
  {Gavrilov}}\ and\ \bibinfo {author} {\bibfnamefont {D.~M.}\ \bibnamefont
  {Gitman}},\ }\bibfield  {title} {\bibinfo {title} {{Vacuum instability in
  external fields}},\ }\href {https://doi.org/10.1103/PhysRevD.53.7162}
  {\bibfield  {journal} {\bibinfo  {journal} {Phys. Rev. D}\ }\textbf {\bibinfo
  {volume} {53}},\ \bibinfo {pages} {7162} (\bibinfo {year}
  {1996})}\BibitemShut {NoStop}%
\bibitem [{\citenamefont {Krajewska}\ and\ \citenamefont
  {Kami\ifmmode~\acute{n}\else
  \'{n}\fi{}ski}(2019{\natexlab{b}})}]{PhysRevA.100.062116}%
  \BibitemOpen
  \bibfield  {author} {\bibinfo {author} {\bibfnamefont {K.}~\bibnamefont
  {Krajewska}}\ and\ \bibinfo {author} {\bibfnamefont {J.~Z.}\ \bibnamefont
  {Kami\ifmmode~\acute{n}\else \'{n}\fi{}ski}},\ }\bibfield  {title} {\bibinfo
  {title} {{Unitary versus pseudounitary time evolution and statistical effects
  in the dynamical Sauter-Schwinger process}},\ }\href
  {https://doi.org/10.1103/PhysRevA.100.062116} {\bibfield  {journal} {\bibinfo
   {journal} {Phys. Rev. A}\ }\textbf {\bibinfo {volume} {100}},\ \bibinfo
  {pages} {062116} (\bibinfo {year} {2019}{\natexlab{b}})}\BibitemShut
  {NoStop}%
\bibitem [{\citenamefont {Avetissian}\ \emph {et~al.}(2002)\citenamefont
  {Avetissian}, \citenamefont {Avetissian}, \citenamefont {Mkrtchian},\ and\
  \citenamefont {Sedrakian}}]{avetissian2002pair}%
  \BibitemOpen
  \bibfield  {author} {\bibinfo {author} {\bibfnamefont {H.~K.}\ \bibnamefont
  {Avetissian}}, \bibinfo {author} {\bibfnamefont {A.~K.}\ \bibnamefont
  {Avetissian}}, \bibinfo {author} {\bibfnamefont {G.~F.}\ \bibnamefont
  {Mkrtchian}},\ and\ \bibinfo {author} {\bibfnamefont {K.~V.}\ \bibnamefont
  {Sedrakian}},\ }\bibfield  {title} {\bibinfo {title} {{Electron-positron pair
  production in the field of superstrong oppositely directed laser beams}},\
  }\href {https://doi.org/10.1103/PhysRevE.66.016502} {\bibfield  {journal}
  {\bibinfo  {journal} {Phys. Rev. E}\ }\textbf {\bibinfo {volume} {66}},\
  \bibinfo {pages} {016502} (\bibinfo {year} {2002})}\BibitemShut {NoStop}%
\bibitem [{\citenamefont {Jansen}\ and\ \citenamefont
  {M\"uller}(2013)}]{jansen2013strongpair}%
  \BibitemOpen
  \bibfield  {author} {\bibinfo {author} {\bibfnamefont {M.~J.~A.}\
  \bibnamefont {Jansen}}\ and\ \bibinfo {author} {\bibfnamefont
  {C.}~\bibnamefont {M\"uller}},\ }\bibfield  {title} {\bibinfo {title}
  {{Strongly enhanced pair production in combined high- and low-frequency laser
  fields}},\ }\href {https://doi.org/10.1103/PhysRevA.88.052125} {\bibfield
  {journal} {\bibinfo  {journal} {Phys. Rev. A}\ }\textbf {\bibinfo {volume}
  {88}},\ \bibinfo {pages} {052125} (\bibinfo {year} {2013})}\BibitemShut
  {NoStop}%
\bibitem [{\citenamefont {Villalba-Ch\'avez}\ and\ \citenamefont
  {M\"uller}(2019{\natexlab{b}})}]{villalbachavez2019schwinger}%
  \BibitemOpen
  \bibfield  {author} {\bibinfo {author} {\bibfnamefont {S.}~\bibnamefont
  {Villalba-Ch\'avez}}\ and\ \bibinfo {author} {\bibfnamefont {C.}~\bibnamefont
  {M\"uller}},\ }\bibfield  {title} {\bibinfo {title} {{Signatures of the
  Schwinger mechanism assisted by a fast-oscillating electric field}},\ }\href
  {https://doi.org/10.1103/PhysRevD.100.116018} {\bibfield  {journal} {\bibinfo
   {journal} {Phys. Rev. D}\ }\textbf {\bibinfo {volume} {100}},\ \bibinfo
  {pages} {116018} (\bibinfo {year} {2019}{\natexlab{b}})}\BibitemShut
  {NoStop}%
\bibitem [{\citenamefont {Folkerts}\ \emph {et~al.}(2023)\citenamefont
  {Folkerts}, \citenamefont {Putzer}, \citenamefont {Villalba-Ch\'avez},\ and\
  \citenamefont {M\"uller}}]{folkerts2023pair}%
  \BibitemOpen
  \bibfield  {author} {\bibinfo {author} {\bibfnamefont {N.}~\bibnamefont
  {Folkerts}}, \bibinfo {author} {\bibfnamefont {J.}~\bibnamefont {Putzer}},
  \bibinfo {author} {\bibfnamefont {S.}~\bibnamefont {Villalba-Ch\'avez}},\
  and\ \bibinfo {author} {\bibfnamefont {C.}~\bibnamefont {M\"uller}},\
  }\bibfield  {title} {\bibinfo {title} {{Electron-positron-pair creation in
  the superposition of two oscillating electric-field pulses with largely
  different frequency, duration, and relative positioning}},\ }\href
  {https://doi.org/10.1103/PhysRevA.107.052210} {\bibfield  {journal} {\bibinfo
   {journal} {Phys. Rev. A}\ }\textbf {\bibinfo {volume} {107}},\ \bibinfo
  {pages} {052210} (\bibinfo {year} {2023})}\BibitemShut {NoStop}%
\bibitem [{\citenamefont {Brezin}\ and\ \citenamefont
  {Itzykson}(1970)}]{brezin1970pair}%
  \BibitemOpen
  \bibfield  {author} {\bibinfo {author} {\bibfnamefont {E.}~\bibnamefont
  {Brezin}}\ and\ \bibinfo {author} {\bibfnamefont {C.}~\bibnamefont
  {Itzykson}},\ }\bibfield  {title} {\bibinfo {title} {{Pair Production in
  Vacuum by an Alternating Field}},\ }\href
  {https://doi.org/10.1103/PhysRevD.2.1191} {\bibfield  {journal} {\bibinfo
  {journal} {Phys. Rev. D}\ }\textbf {\bibinfo {volume} {2}},\ \bibinfo {pages}
  {1191} (\bibinfo {year} {1970})}\BibitemShut {NoStop}%
\bibitem [{\citenamefont {Popov}(2004)}]{VladimirSPopov2004}%
  \BibitemOpen
  \bibfield  {author} {\bibinfo {author} {\bibfnamefont {V.~S.}\ \bibnamefont
  {Popov}},\ }\bibfield  {title} {\bibinfo {title} {{Tunnel and multiphoton
  ionization of atoms and ions in a strong laser field (Keldysh theory)}},\
  }\href {https://doi.org/10.1070/PU2004v047n09ABEH001812} {\bibfield
  {journal} {\bibinfo  {journal} {Phys.-Usp.}\ }\textbf {\bibinfo {volume}
  {47}},\ \bibinfo {pages} {855} (\bibinfo {year} {2004})}\BibitemShut
  {NoStop}%
\bibitem [{\citenamefont {Dirac}(1931)}]{dirac1931quantised}%
  \BibitemOpen
  \bibfield  {author} {\bibinfo {author} {\bibfnamefont {P.~A.~M.}\
  \bibnamefont {Dirac}},\ }\bibfield  {title} {\bibinfo {title} {{Quantised
  singularities in the electromagnetic field}},\ }\href@noop {} {\bibfield
  {journal} {\bibinfo  {journal} {Proc. R. Soc. London, Ser. A}\ }\textbf
  {\bibinfo {volume} {133}},\ \bibinfo {pages} {60} (\bibinfo {year}
  {1931})}\BibitemShut {NoStop}%
\bibitem [{\citenamefont {Bia{\l}ynicki-Birula}\ \emph
  {et~al.}(2000)\citenamefont {Bia{\l}ynicki-Birula}, \citenamefont
  {Bia{\l}ynicka-Birula},\ and\ \citenamefont {\ifmmode~\acute{S}\else
  \'{S}\fi{}liwa}}]{PhysRevA.61.032110}%
  \BibitemOpen
  \bibfield  {author} {\bibinfo {author} {\bibfnamefont {I.}~\bibnamefont
  {Bia{\l}ynicki-Birula}}, \bibinfo {author} {\bibfnamefont {Z.}~\bibnamefont
  {Bia{\l}ynicka-Birula}},\ and\ \bibinfo {author} {\bibfnamefont
  {C.}~\bibnamefont {\ifmmode~\acute{S}\else \'{S}\fi{}liwa}},\ }\bibfield
  {title} {\bibinfo {title} {{Motion of vortex lines in quantum mechanics}},\
  }\href {https://doi.org/10.1103/PhysRevA.61.032110} {\bibfield  {journal}
  {\bibinfo  {journal} {Phys. Rev. A}\ }\textbf {\bibinfo {volume} {61}},\
  \bibinfo {pages} {032110} (\bibinfo {year} {2000})}\BibitemShut {NoStop}%
\bibitem [{\citenamefont {Larionov}\ \emph {et~al.}(2018)\citenamefont
  {Larionov}, \citenamefont {Ovchinnikov}, \citenamefont {Smirnovsky},\ and\
  \citenamefont {Schmidt}}]{Larionov2018Perturbation}%
  \BibitemOpen
  \bibfield  {author} {\bibinfo {author} {\bibfnamefont {N.~V.}\ \bibnamefont
  {Larionov}}, \bibinfo {author} {\bibfnamefont {S.~Y.}\ \bibnamefont
  {Ovchinnikov}}, \bibinfo {author} {\bibfnamefont {A.~A.}\ \bibnamefont
  {Smirnovsky}},\ and\ \bibinfo {author} {\bibfnamefont {A.~A.}\ \bibnamefont
  {Schmidt}},\ }\bibfield  {title} {\bibinfo {title} {{Perturbation Theory in
  the Analysis of Quantum Vortices Formed by Impact of Ultrashort
  Electromagnetic Pulse on Atom}},\ }\href
  {https://doi.org/10.1134/S1063784218110166} {\bibfield  {journal} {\bibinfo
  {journal} {Tech. Phys.}\ }\textbf {\bibinfo {volume} {63}},\ \bibinfo {pages}
  {1569 } (\bibinfo {year} {2018})}\BibitemShut {NoStop}%
\bibitem [{\citenamefont {Cajiao~V\'elez}\ \emph {et~al.}(2020)\citenamefont
  {Cajiao~V\'elez}, \citenamefont {Geng}, \citenamefont
  {Kami\ifmmode~\acute{n}\else \'{n}\fi{}ski}, \citenamefont {Peng},\ and\
  \citenamefont {Krajewska}}]{PhysRevA.102.043102}%
  \BibitemOpen
  \bibfield  {author} {\bibinfo {author} {\bibfnamefont {F.}~\bibnamefont
  {Cajiao~V\'elez}}, \bibinfo {author} {\bibfnamefont {L.}~\bibnamefont
  {Geng}}, \bibinfo {author} {\bibfnamefont {J.~Z.}\ \bibnamefont
  {Kami\ifmmode~\acute{n}\else \'{n}\fi{}ski}}, \bibinfo {author}
  {\bibfnamefont {L.-Y.}\ \bibnamefont {Peng}},\ and\ \bibinfo {author}
  {\bibfnamefont {K.}~\bibnamefont {Krajewska}},\ }\bibfield  {title} {\bibinfo
  {title} {{Vortex streets and honeycomb structures in photodetachment driven
  by linearly polarized few-cycle laser pulses}},\ }\href
  {https://doi.org/10.1103/PhysRevA.102.043102} {\bibfield  {journal} {\bibinfo
   {journal} {Phys. Rev. A}\ }\textbf {\bibinfo {volume} {102}},\ \bibinfo
  {pages} {043102} (\bibinfo {year} {2020})}\BibitemShut {NoStop}%
\bibitem [{\citenamefont {Geng}\ \emph {et~al.}(2020)\citenamefont {Geng},
  \citenamefont {Cajiao~V\'elez}, \citenamefont {Kami\ifmmode~\acute{n}\else
  \'{n}\fi{}ski}, \citenamefont {Peng},\ and\ \citenamefont
  {Krajewska}}]{PhysRevA.102.043117}%
  \BibitemOpen
  \bibfield  {author} {\bibinfo {author} {\bibfnamefont {L.}~\bibnamefont
  {Geng}}, \bibinfo {author} {\bibfnamefont {F.}~\bibnamefont
  {Cajiao~V\'elez}}, \bibinfo {author} {\bibfnamefont {J.~Z.}\ \bibnamefont
  {Kami\ifmmode~\acute{n}\else \'{n}\fi{}ski}}, \bibinfo {author}
  {\bibfnamefont {L.-Y.}\ \bibnamefont {Peng}},\ and\ \bibinfo {author}
  {\bibfnamefont {K.}~\bibnamefont {Krajewska}},\ }\bibfield  {title} {\bibinfo
  {title} {{Vortex structures in photodetachment by few-cycle circularly
  polarized pulses}},\ }\href {https://doi.org/10.1103/PhysRevA.102.043117}
  {\bibfield  {journal} {\bibinfo  {journal} {Phys. Rev. A}\ }\textbf {\bibinfo
  {volume} {102}},\ \bibinfo {pages} {043117} (\bibinfo {year}
  {2020})}\BibitemShut {NoStop}%
\bibitem [{\citenamefont {Geng}\ \emph {et~al.}(2021)\citenamefont {Geng},
  \citenamefont {Cajiao~V\'elez}, \citenamefont {Kami\ifmmode~\acute{n}\else
  \'{n}\fi{}ski}, \citenamefont {Peng},\ and\ \citenamefont
  {Krajewska}}]{PhysRevA.104.033111}%
  \BibitemOpen
  \bibfield  {author} {\bibinfo {author} {\bibfnamefont {L.}~\bibnamefont
  {Geng}}, \bibinfo {author} {\bibfnamefont {F.}~\bibnamefont
  {Cajiao~V\'elez}}, \bibinfo {author} {\bibfnamefont {J.~Z.}\ \bibnamefont
  {Kami\ifmmode~\acute{n}\else \'{n}\fi{}ski}}, \bibinfo {author}
  {\bibfnamefont {L.-Y.}\ \bibnamefont {Peng}},\ and\ \bibinfo {author}
  {\bibfnamefont {K.}~\bibnamefont {Krajewska}},\ }\bibfield  {title} {\bibinfo
  {title} {{Structured photoelectron distributions in photodetachment induced
  by trains of laser pulses: Vortices versus spirals}},\ }\href
  {https://doi.org/10.1103/PhysRevA.104.033111} {\bibfield  {journal} {\bibinfo
   {journal} {Phys. Rev. A}\ }\textbf {\bibinfo {volume} {104}},\ \bibinfo
  {pages} {033111} (\bibinfo {year} {2021})}\BibitemShut {NoStop}%
\bibitem [{\citenamefont {Cajiao~V\'elez}(2021)}]{PhysRevA.104.043116}%
  \BibitemOpen
  \bibfield  {author} {\bibinfo {author} {\bibfnamefont {F.}~\bibnamefont
  {Cajiao~V\'elez}},\ }\bibfield  {title} {\bibinfo {title} {{Generation of
  quantum vortices in photodetachment: The role of the ground-state wave
  function}},\ }\href {https://doi.org/10.1103/PhysRevA.104.043116} {\bibfield
  {journal} {\bibinfo  {journal} {Phys. Rev. A}\ }\textbf {\bibinfo {volume}
  {104}},\ \bibinfo {pages} {043116} (\bibinfo {year} {2021})}\BibitemShut
  {NoStop}%
\bibitem [{\citenamefont {Majczak}\ \emph {et~al.}(2022)\citenamefont
  {Majczak}, \citenamefont {V\'{e}lez}, \citenamefont {Kami\'{n}ski},\ and\
  \citenamefont {Krajewska}}]{majczak2022vorticesphotodetachment}%
  \BibitemOpen
  \bibfield  {author} {\bibinfo {author} {\bibfnamefont {M.~M.}\ \bibnamefont
  {Majczak}}, \bibinfo {author} {\bibfnamefont {F.~C.}\ \bibnamefont
  {V\'{e}lez}}, \bibinfo {author} {\bibfnamefont {J.~Z.}\ \bibnamefont
  {Kami\'{n}ski}},\ and\ \bibinfo {author} {\bibfnamefont {K.}~\bibnamefont
  {Krajewska}},\ }\bibfield  {title} {\bibinfo {title} {{Carrier-envelope-phase
  and helicity control of electron vortices and spirals in photodetachment}},\
  }\href {https://doi.org/10.1364/OE.473929} {\bibfield  {journal} {\bibinfo
  {journal} {Opt. Express}\ }\textbf {\bibinfo {volume} {30}},\ \bibinfo
  {pages} {43330} (\bibinfo {year} {2022})}\BibitemShut {NoStop}%
\bibitem [{\citenamefont {Hebenstreit}\ \emph {et~al.}(2010)\citenamefont
  {Hebenstreit}, \citenamefont {Alkofer},\ and\ \citenamefont
  {Gies}}]{hebenstreit2010schwinger}%
  \BibitemOpen
  \bibfield  {author} {\bibinfo {author} {\bibfnamefont {F.}~\bibnamefont
  {Hebenstreit}}, \bibinfo {author} {\bibfnamefont {R.}~\bibnamefont
  {Alkofer}},\ and\ \bibinfo {author} {\bibfnamefont {H.}~\bibnamefont
  {Gies}},\ }\bibfield  {title} {\bibinfo {title} {{Schwinger pair production
  in space- and time-dependent electric fields: Relating the Wigner formalism
  to quantum kinetic theory}},\ }\href
  {https://doi.org/10.1103/PhysRevD.82.105026} {\bibfield  {journal} {\bibinfo
  {journal} {Phys. Rev. D}\ }\textbf {\bibinfo {volume} {82}},\ \bibinfo
  {pages} {105026} (\bibinfo {year} {2010})}\BibitemShut {NoStop}%
\bibitem [{\citenamefont {Kohlf\"urst}(2020)}]{kohlfurst2020magneticpair}%
  \BibitemOpen
  \bibfield  {author} {\bibinfo {author} {\bibfnamefont {C.}~\bibnamefont
  {Kohlf\"urst}},\ }\bibfield  {title} {\bibinfo {title} {{Effect of
  time-dependent inhomogeneous magnetic fields on the particle momentum
  spectrum in electron-positron pair production}},\ }\href
  {https://doi.org/10.1103/PhysRevD.101.096003} {\bibfield  {journal} {\bibinfo
   {journal} {Phys. Rev. D}\ }\textbf {\bibinfo {volume} {101}},\ \bibinfo
  {pages} {096003} (\bibinfo {year} {2020})}\BibitemShut {NoStop}%
\bibitem [{\citenamefont {Al-Naseri}\ \emph {et~al.}(2021)\citenamefont
  {Al-Naseri}, \citenamefont {Zamanian},\ and\ \citenamefont
  {Brodin}}]{brodin2021plasmadynamics}%
  \BibitemOpen
  \bibfield  {author} {\bibinfo {author} {\bibfnamefont {H.}~\bibnamefont
  {Al-Naseri}}, \bibinfo {author} {\bibfnamefont {J.}~\bibnamefont
  {Zamanian}},\ and\ \bibinfo {author} {\bibfnamefont {G.}~\bibnamefont
  {Brodin}},\ }\bibfield  {title} {\bibinfo {title} {{Plasma dynamics and
  vacuum pair creation using the Dirac-Heisenberg-Wigner formalism}},\ }\href
  {https://doi.org/10.1103/PhysRevE.104.015207} {\bibfield  {journal} {\bibinfo
   {journal} {Phys. Rev. E}\ }\textbf {\bibinfo {volume} {104}},\ \bibinfo
  {pages} {015207} (\bibinfo {year} {2021})}\BibitemShut {NoStop}%
\bibitem [{\citenamefont {Kohlf\"urst}\ \emph {et~al.}(2022)\citenamefont
  {Kohlf\"urst}, \citenamefont {Ahmadiniaz}, \citenamefont {Oertel},\ and\
  \citenamefont {Sch\"utzhold}}]{kohlfurst2022collidinglaser}%
  \BibitemOpen
  \bibfield  {author} {\bibinfo {author} {\bibfnamefont {C.}~\bibnamefont
  {Kohlf\"urst}}, \bibinfo {author} {\bibfnamefont {N.}~\bibnamefont
  {Ahmadiniaz}}, \bibinfo {author} {\bibfnamefont {J.}~\bibnamefont {Oertel}},\
  and\ \bibinfo {author} {\bibfnamefont {R.}~\bibnamefont {Sch\"utzhold}},\
  }\bibfield  {title} {\bibinfo {title} {{Sauter-Schwinger Effect for Colliding
  Laser Pulses}},\ }\href {https://doi.org/10.1103/PhysRevLett.129.241801}
  {\bibfield  {journal} {\bibinfo  {journal} {Phys. Rev. Lett.}\ }\textbf
  {\bibinfo {volume} {129}},\ \bibinfo {pages} {241801} (\bibinfo {year}
  {2022})}\BibitemShut {NoStop}%
\end{thebibliography}%

\end{document}